A data-driven prospective study of incident dementia among older adults in the United States

Jordan Weiss[1#a*¶&‡], Eli Puterman[2&‡], Aric A. Prather[3&‡], Erin B. Ware[4‡], David H. Rehkopf[5*¶&‡]

[1] Population Studies Center and the Leonard Davis Institute of Health Economics, University of Pennsylvania, Philadelphia, PA, USA

[2] School of Kinesiology, The University of British Columbia, Vancouver BC, Canada

[3] Department of Psychiatry, University of California, San Francisco, San Francisco, CA, USA

[4] Survey Research Center, Institute for Social Research, University of Michigan, Ann Arbor, MI, USA

[5] School of Medicine, Stanford University, Palo Alto, CA, USA

[#a]Current Address: Department of Demography, UC Berkeley, Berkeley, CA, USA

[*]Correspondence to: drehkopf@stanford.edu (DHR), jordanmnw@gmail.com (JW)

[¶]Contributed equally to data curation and formal analysis
[&]Contributed equally to conceptualization, methodology, and original draft preparation
[‡]Contributed equally to interpretation of results and review and editing of manuscript

**Abbreviations**: ADAMS, Aging, Demographics, and Memory Study; CI, Confidence interval; HR, hazard ratio; HRS, Health and Retirement Study; NIA, National Institute on Aging; PGS, Polygenic score; US, United States.

**Abstract**

We conducted a prospective analysis of incident dementia and its association with 65 sociodemographic, early-life, economic, health and behavioral, social, and genetic risk factors in a sample of 7,908 adults over the age of 50 from the nationally representative US-based Health and Retirement Study. We used traditional survival analysis methods (Fine-Gray models) and a data-driven approach (random survival forests for competing risks) which allowed us to account for the competing risk of death with up to 14 years of follow-up. Overall, the top five predictors across all groups were lower education, loneliness, lower wealth and income, and lower self-reported health. However, we observed variation in the leading predictors of dementia across racial/ethnic and gender groups. Our ranked lists may be useful for guiding future observational and quasi-experimental research that investigates understudied domains of risk and emphasizes life course economic and health conditions as well as disparities therein.

**Introduction**

In 2017, the Lancet Commission on Dementia Prevention and Care published a report to consolidate the state of knowledge on preventive and management strategies for cognitive dementia [1]. The Commission reviewed evidence from over 500 scientific peer-reviewed articles, systematic reviews, and meta-analyses and calculated that nearly one third of dementia cases may be preventable. A wide range of factors may contribute to the ability to prevent one third of these cases including educational attainment, social engagement, physical activity, and management of comorbidities.

However, a majority of the studies reviewed in the Commission's report examined risk factors independent of one another to test *a priori* hypotheses about how they may be associated with dementia. Few studies have jointly and comparatively analyzed risk factors for dementia across domains of early-life, sociodemographic, economic, health and behavioral, social, and genetic characteristics while also systematically examining whether these factors differ among racial/ethnic and gender groups. A study by Lourida and colleagues [2] reported that a favorable lifestyle (e.g., not being an active smoker, engaging in regular physical activity, and maintaining a healthy diet) was associated with a lower dementia risk irrespective of genetic risk for dementia in a sample of more than 190,000 participants of European ancestry. Another recent study [3] reported that sociodemographic characteristics (e.g., lower educational attainment, Hispanic origin) and measures of health (e.g., lower rated subjective health, higher levels of body mass index [BMI]) were comparatively better predictors of incident dementia than genetic risk of dementia assessed through polygenic scores. Together, these studies suggest that a healthy lifestyle may help offset the genetic risk of dementia.

Despite these promising findings, there has been limited work integrating risk factors across multiple domains to understand their relative importance for predicting dementia and how these rankings may vary across racial/ethnic and gender groups. An analytic framework that allows for a comprehensive investigation of dementia risk factors may be useful for hypothesis generation and prioritizing group-specific intervention targets to prevent or delay the onset of dementia [4, 5]. Further, this may help shape our understanding of how intervening on specific risk factors may eradicate or exacerbate documented disparities in dementia risk [6, 7].

Prior studies in which investigators used more contemporary statistical approaches to examine dementia have focused primarily on medical risk factors [8] and neuroimaging biomarkers [9] as well as resilience to genetic predispositions to dementia [10]. Although these characteristics are important to studying the onset of dementia, little work [e.g., 3, 11] has combined genetic and life-course environmental risk factors in pursuit of a more comprehensive prediction model. A study by Casanova and colleagues [11] used data from the Health and Retirement Study (HRS) combined with a data-driven approach to predict cognitive impairment using sociodemographic, health, and genetic data. These researchers found that education, age, gender, and history of stroke were among the leading characteristics predicting cognitive impairment. Despite the novelty and innovation of their approach, the authors did not account for the semi-competing risk of death nor did they examine differential rankings of predictors by race/ethnic and gender. Failure to account for the semi-competing risk of mortality when studying older populations can bias results and overestimate the risk of disease [12]. In addition, documented differences in longevity and dementia incidence among race/ethnicity and gender groups could bias results as they absorb much of the variation in this prior study. Due to data limitations, this prior study also examined a smaller list of predictors, for example, using neighborhood socioeconomic status rather than a range of social and economic factors at the individual level. More recently, Aschwanden and colleagues [3] conducted a similar study using the HRS but did not did not account for the semi-competing risk of death nor did they investigate variation across racial/ethnic and gender groups.

Understanding the relative importance and predictive power of these factors remains understudied and is critical for planning group-specific treatment strategies for those who may be at greater risk of dementia. We build on this emerging literature by investigating the relative contribution of 65 early-life, sociodemographic, health, behavioral, social, and genetic characteristics across the life course to dementia in the nationally representative and longitudinal US-based Health and Retirement Study (HRS). We estimated cause-specific hazard ratios of each characteristic for incident dementia while accounting for the semi-competing risk of death. We then compared these results to those obtained within a data-driven framework by utilizing random survival forests for competing risks. All models were stratified by race/ethnicity and gender to examine the differential ranking of each predictor across these demographic strata.

**Results**

The analytic sample for the primary analysis was comprised of 7,908 respondents with an average age at baseline of 65.6 years (standard error = 0.15). Overall, 37.4% of respondents were non-Hispanic white men; 49.9% of respondents were non-Hispanic white women; 4.5% of respondents were non-Hispanic black men; and non-Hispanic black women comprised 8.2% of the sample. Summary characteristics for the analytic HRS sample at baseline are shown in Table S1. Correlations between all predictors by race/ethnicity and gender are presented in Fig S1.

Fig 1 and Supplemental Table S2 present the hazard ratios (HRs) and 95% confidence intervals (CIs) for each risk factor on dementia examined independently in Fine and Gray models stratified by race/ethnicity and gender. Risk factors are categorized by domain and ranked from largest increase in risk of dementia (top of figure) to largest decrease in risk of dementia (bottom of figure). Risk factors with CIs that cross one are not considered statistically significant at the P value = 0.05 level.

**Fig 1. Hazard ratios (HRs) and 95% confidence intervals (CI) of each predictor for incident dementia obtained from Fine-Gray regression models stratified by race and gender.** Predictors with HRs equal to zero are excluded from the figure but retained in Table S3. Models use full analytic sample and classify dementia using the Langa-Weir classification scheme.

Three of the top 10 characteristics for non-Hispanic white men and women were consistent (lower education, lower neighborhood safety, received food stamps) whereas four were consistent for non-Hispanic white and black men (lower education, received food stamps, reported pain, reported Medicaid) and four of the top 10 characteristics were consistent for non-Hispanic white and black women (lower education, received food stamps, heavy alcohol use, and reported headaches). Across racial/ethnic and gender groups, lower education and receipt of food stamps were the only characteristics that consistently ranked in the top 10 of predictors; receipt of food stamps was the only characteristic that consistently ranked in the top five of predictors. These results were consistent when comparing HRs obtained from cause-specific hazard models (Supplemental Table S3) with the exception of lower education, which did not rank in the top 10 of predictors for non-Hispanic white women. However, it is evident that the point estimates and CIs for most predictors overlap across racial/ethnic and gender groups as shown in Fig S2 suggesting that the differences in the associated hazards across racial/ethnic groups are not statistically significant at the P value = 0.05 level.

Results from the random survival forests analysis for competing risks are shown in Fig 2. Blue bars indicate predictors with positive variable importance values; red bars indicate predictors with negative variable importance values, which in this context can be considered statistically insignificant. The positive length of the bar indicates the importance of each predictor. The top predictors across racial/ethnic and gender groups differed from those obtained in the Fine and Gray models, and were more consistent across racial/ethnic groups in the random survival forests analysis. Whereas lower education and lower neighborhood safety were the only consistently ranked predictors in the top 10 across race/ethnicity and gender groups in the Fine and Gray models, lower education and loneliness were consistently ranked in the top 10 in the random survival forests analysis. Lower age at first birth and lower levels of respondent's mother's education appeared in the top 10 for non-Hispanic white and black women. Lower income and self-reported health ranked in the top 10 for all groups with the exception of non-Hispanic black men whereas lower wealth ranked in the top 10 for all groups with the exception of non-Hispanic white women. Fig 3 shows the rank order for predictors, overall, and for each race/ethnicity and gender group. The overall rank order was determined by calculating the unweighted mean of predictor rank orders within each of the four demographic strata, and sorting from lowest to highest mean (i.e., highest rank to lowest rank). The values of the overall rank order in Fig 3 do not represent the means themselves but instead correspond to these rankings. These results illustrate the variation across and within racial/ethnic groups. For example, whereas lower wealth is respectively ranked as the second and fifth leading predictor for non-Hispanic white men and non-Hispanic black women, lower wealth ranked 13[th] for non-Hispanic white women and 10[th] for non-Hispanic black men. Food insecurity, which ranked 10[th] overall, was ranked 20[th] and 23[rd] for non-Hispanic white men and women, but 11[th] and 7[th] for non-Hispanic black men and women.

**Fig 2. Variable importance plot for 65 characteristics predicting dementia obtained from random survival forests for competing risks stratified by race and gender.** Model uses full analytic sample and classifies dementia using the Langa-Weir classification scheme.

**Fig 3. Rank order of predictors obtained from random survival forests for competing risks stratified by race and gender.** Model uses full analytic sample and classifies dementia using the Langa-Weir classification scheme.

In sensitivity analyses among a subsample of 6,746 respondents who were 70+ years of age at baseline and had at least one measure of all four classification schemes for dementia, we observed that the consistency of predictors with the highest HRs from the Fine-Gray models varied by race/ethnicity and gender group. For non-Hispanic white men, four of the top five predictors were consistent across all four classification schemes (lower education, lower neighborhood safety, receipt of food stamps, Medicaid, and psychiatric illness; Figs S3-S6 and Tables S4-S7). Among non-Hispanic white women, only headaches were among the top five predictors with the highest HRs across all four classification schemes (Fig S3-S6). For non-Hispanic black men, only Medicaid was consistently ranked in the top five predictors with the highest HRs and among non-Hispanic black women, self-reported persistent dizziness was the only predictor consistently ranked in the top five (Fig S3-S6). For the random survival forests analysis, three of the top five predictors (lower education, lower income, loneliness) were consistent overall when using the four different classification schemes (Figs S7-S13). When used as the outcome in the Fine-Gray and random survival forests analyses, the Langa-Weir classifier, which does not account for race/ethnicity, gender, or education, resulted in more socioeconomic risk factors being ranked in the top five. Models which used the Hurd, Expert, and LASSO classifiers, which do account for respondent demographics, were more likely to produce health, behavioral, and genetic risk factors as top predictors.

**Discussion**

In this 14-year population-based study of older adults in the US with available polygenic score data, we found that the relative importance of risk factors for predicting dementia varied across racial/ethnic and gender groups using two distinct methodologies. We also found the predictor rankings to vary based on the type of dementia classification used. Although not all observed differences were statistically significant, our stratified models may offer insight into substantive differences in the relative importance of risk factors across racial/ethnic and gender groups.

We observed variation in the rank order of predictors across and within racial/ethnic and gender groups in both the Fine-Gray models and random survival forests. The consistency of our primary results across both analyses suggests that our findings are robust to these two distinct approaches. However, in our sensitivity analyses in which we compared four alternative classification schemes for dementia, we found the results to vary which may be due to the different criteria used in each of these classification schemes for dementia.

The low ranking of genetic predictors was apparent across our analyses, whereas we saw stronger associations between characteristics measured in mid- or later-life particularly those in the economic domain. Lower levels of education and receipt of food stamps were the only characteristics that consistently ranked in the top 10 of predictors for all groups in the random survival forests analysis. There was heterogeneity within and outside of the top 10 predictors highlighting the importance of identifying effective methods to promote health and mitigate dementia burden and its risk factors within racial/ethnic and gender groups. The results obtained in our analyses were similar to those reported by Casanova and colleagues [11] and Aschwanden et al. [3] but our account of the competing risk of mortality and stratification by race/ethnicity and gender may offer more accurate estimates and provide additional insight into the differential ranking of risk factors within these groups.

In their recent study, Aschwanden and colleagues [3] used Cox proportional hazard models and a machine-learning approach to predict cognitive impairment and dementia in the HRS over a 10-year period. The authors used 52 multi-domain risk factors in their random survival forests analysis and found that increases in body mass index, higher levels of emotional distress, diabetes, self-reporting race as black, and higher reports of childhood trauma were the top five predictors of dementia over the study period. The authors incorporated two polygenic scores—one for Alzheimer's disease which included the apolipoprotein e4 allele and one which did not. The authors then examined a subset of top predictors from their random survival forests model in a semi-parametric survival analysis framework using Cox proportional hazards model. Of the six predictors the authors examined, only emotional distress was significantly associated with incident dementia.

Despite similar methodologies and data, none of the top five predictors from Aschwanden and colleagues' study appeared in the overall top 10 from our random survival forests model and only two risk factors (education and self-reported childhood health) ranked in both top 10 lists. In our random survival forests analysis which accounted for the competing risk of death, we found that psychiatric illness—which is different from but most comparable to the author's emotional distress measure—ranked fourth among non-Hispanic white men, 12th among non-Hispanic white women, 62nd among non-Hispanic black men, and 41st among non-Hispanic black women. Interestingly, however, the HR for emotional distress obtained from Aschwanden and colleagues' study (HR: 1.85; 95%CI: 1.41, 2.44) overlapped with the HR for psychiatric illness among non-Hispanic white men in our independent Fine-Gray model (HR: 1.61; 95%CI: 1.13, 2.31) whereas the HR for psychiatric illness was not statistically significant for any other subgroup in our analysis. Moreover, the prior study reported increasing trajectories of body mass index as the top predictor of dementia in their random survival forests analysis whereas in our study, which used baseline body mass index in the year 2000, body mass index was ranked overall as the 59th predictor out of 65 and at best, ranked 29th for non-Hispanic black women. The relationship between body mass index and dementia is complex [13-15], with investigators reporting in one study that midlife obesity was associated with an increased risk of dementia compared to those with normal body mass index whereas this association reversed in later-life [14]. It is possible that, by not accounting for the competing risk of mortality, the top predictors reported in Aschwanden and colleagues' study are driven by their association with mortality although we did not test this directly.

    As noted above, a major strength of this study is our account of the competing risk of death. Recent studies have reported that studying age-related conditions, including dementia, while not accounting for the competing risk of death may produce biased or misleading results [16, 17]. An additional strength of our study is the inclusion of 65 risk factors spanning sociodemographic, early-life, economic, health, behavioral, social, and genetic risk factor

domains across the life course. Our inclusion of genetic risk factors was in the form of polygenic scores which are able to capture genotypic variation across multiple genetic loci compared to individual genotypes. In addition, we compared our primary results using the Langa-Weir classification scheme to more recent approaches which may be better suited to studying disparities in dementia across racial/ethnic groups as well as among adults who vary with respect to socioeconomic status [18, 19].

This study also had several limitations. First, there were several risk factors in the report by the Lancet Commission on Dementia Prevention and Care that are not available in the HRS. These measures include for example, dietary quality, exposure to environmental contaminants, and questions about cognitive training and stimulation. Second, although we used clinically validated cut points derived from the ADAMS for assessing dementia in the HRS cohort, these classifications may be subject to measurement error. This measurement error could affect the coefficient estimates and rankings if a large enough sample of respondents were misclassified with respect to their cognitive status. We conducted sensitivity analyses using three alternative classification schemes for dementia which may alleviate some of this concern. Third, as with any tree-based approach such as the random forest algorithm, variables with a wider range and therefore more points at which they can be split, tend to have higher predictive power [20]. We addressed this limitation by standardizing continuous measures to make them as equivalent as possible across our study.

We identified heterogeneity in the association between dementia and its risk factors within racial/ethnic and gender groups using a more traditional approach to account for competing risks (the Fine-Gray model) as well as a more contemporary data-driven approach (random survival forests for competing risks). These results may be useful for understanding and further exploring recent reports documenting disparity trends in dementia across racial/ethnic and gender groups [7, 21, 22]. We advise caution in treating these results with a causal interpretation and instead suggest that these results can be used for hypothesis generation and to inform future observational and clinical studies to identify the multiple pathways through which these risk factors may be differentially associated with the risk of dementia across demographic strata.

**Materials and Methods**

**Study population**

The HRS is a nationally representative and longitudinal study of more than 30,000 community-dwelling US adults aged 50+ years and their spouses of any age. Since 1992, the HRS has biennially collected economic, social, and health information from respondents who undergo detailed telephone or in-person interviews. Respondents who are unable or unwilling to participate may be surveyed by a proxy respondent, typically a spouse or adult child, who completes the survey on their behalf. The HRS is under current IRB approval at the University of Michigan and the National Institute on Aging (NIA) with support from the NIA (NIA; U01AG009740) and the Social Security Administration [23]. Polygenic score data were available for HRS respondents who consented and provided salivary DNA from 2006 through 2012 [24].

We used a base year of 2000 with follow-up through 2014 during which time cognitive information was consistently ascertained for community-dwelling and nursing home residents. We restricted the analytic sample to non-Hispanic men and women aged 51 years and older who were dementia-free at baseline in 2000 who had polygenic score data, a valid sampling weight, and at least one measure of cognitive function over the study period (2000 to 2014). We further excluded respondents who self-reported their race as "Other Race" due to low sample sizes.

**Measures**

**Outcome.** Different protocols were used to assess cognitive function among self- and proxy-respondents in the HRS [25]. Among self-respondents, cognitive function was determined through a series of cognitive tests which included immediate and delayed 10-noun free recall tests (range: 0-10 points each), a serial 7s subtraction test (range: 0-5 points), and a backwards counting test (range: 0-2 points). Scores ranged from 0 to 27, with higher scores reflecting better cognitive performance. Among respondents surveyed through a proxy, cognitive scores were based on the proxy's assessment of the respondent's memory (range: 0-4; excellent, very good, good, fair, poor), limitations in five instrumental activities of daily living (range: 0-5; managing money, taking medication, preparing hot meals, using phones and doing groceries), and the interviewer's assessment of the respondent's difficulty completing the interview due to cognitive limitations (range: 0-2; none, some, prevents completion) to produce a score ranging from 0 to 11, with higher scores reflecting a higher degree of impairment.

Cut points for dementia using these scales in the HRS were validated against the Aging, Demographics, and Memory Study (ADAMS). The ADAMS is a clinical substudy of 856 HRS respondents who underwent extensive in-home neuropsychological and clinical assessments [26]. We used the Langa-Weir approach [27] in our primary analysis to classify respondents with dementia (self-respondent: 0-6 out of 27; proxy: 6-11 out of 11).

In sensitivity analyses, we used three additional classification schemes for dementia which are reported to have greater sensitivity to racial/ethnic and sociodemographic disparities [18, 19]. These alternative schemes, referred to as the Hurd Model, the Expert Model, and the LASSO Model were also validated against the ADAMS as described elsewhere [19]. Our sensitivity analyses were conducted in a subsample of respondents who, in addition to the inclusion criteria for the analytic sample, were 70 years or older at baseline and had available information on all four dementia classification methods (which were estimated among HRS respondents aged 70+ years).

**Risk factors**

We conducted a thorough review of the articles cited in the Lancet Commission's report and selected 65 risk factors that were available in the HRS. We classified risk factors into seven domains: sociodemographic (1), early-life (2), economic (3), health (4), behavioral (5), social ties (7), and genetic markers (8). A complete list of risk factors, their definition, and coding is provided in the supplemental S1 Appendix. All risk factors measured on a continuous scale were standardized to a normal distribution (mean = 0, standard deviation = 1). Risk factors were coded such that higher scores reflected a higher degree of risk. All risk factors were measured in 1998 or 2000.

**Statistical Analysis**

All statistical analysis was performed in R version 3.6.1 [28]. All analyses used respondent-level sampling weights and, where appropriate, included robust standard errors to account for the clustering of individuals within households in the HRS. In preparing the data file, we excluded risk factors that were missing among 20% or more of respondents (see Table S1). Missing data values for the remaining predictors were imputed using a non-parametric approach implemented with the R package 'missForest' with five iterations each fit with 500 trees [29]. We examined associations between all predictors by creating correlation matrices for all risk factors across racial/ethnic and gender groups. The distribution of all 65 risk factors at baseline was examined by computing the prevalence or mean and standard deviation of each risk factor after imputation.

We used inverse probability weighting to account for selection into the HRS genetic sample. This process upweighted respondents with a lower propensity for providing genetic data, creating a pseudo population which more closely reflects the representativeness of the HRS sample [30, 31]. The respondent-level sampling weights used to generate new base weights for our analysis were calculated and provided by the HRS investigators. Specifically, we used the respondent-level sampling weights that account for both community-dwelling respondents and those residing in nursing homes. All analysis used these

We examined bivariate associations between each predictor and dementia using the method proposed by Fine and Gray [32] to account for the semi-competing risk of death. This approach accounts for the fact that respondents who die prior to incident dementia are no longer considered at risk for dementia, as opposed to methods such as the Kaplan-Meier estimator which treats the semi-competing risk of death as noninformative censoring which may bias results and overestimate associations between risk factors and dementia [12]. Respondents were observed from baseline until incident dementia, death, or censoring. We used age as the timescale (i.e., age at first dementia diagnosis, age at death, age at last visit) due to its strong association with dementia. All models were stratified by race/ethnicity and gender due to known differences in longevity and dementia risk across these strata [33, 34]. We separately examined bivariate associations between each predictor and dementia following the same procedures as described above but using cause-specific hazard models as recommended by Latouche and colleagues [35]. More details on the Fine-Gray model are provided in the supplemental S2 Appendix

We then used random survival forests for competing risks [36] to simultaneously investigate the relative importance of each predictor for dementia across racial/ethnic and gender groups while accounting for the semi-competing risk of death. Age was used as the timescale. Random survival forests are an extension of the random forest algorithm, an ensemble-based classification method which fits a series of classification and regression trees and then pools results across the trees [37]. We implemented this approach using the R package 'randomForestSRC' with 1,000 trees [38]. More details on the random survival forests procedure is provided in the supplemental S2 Appendix

Fig 1

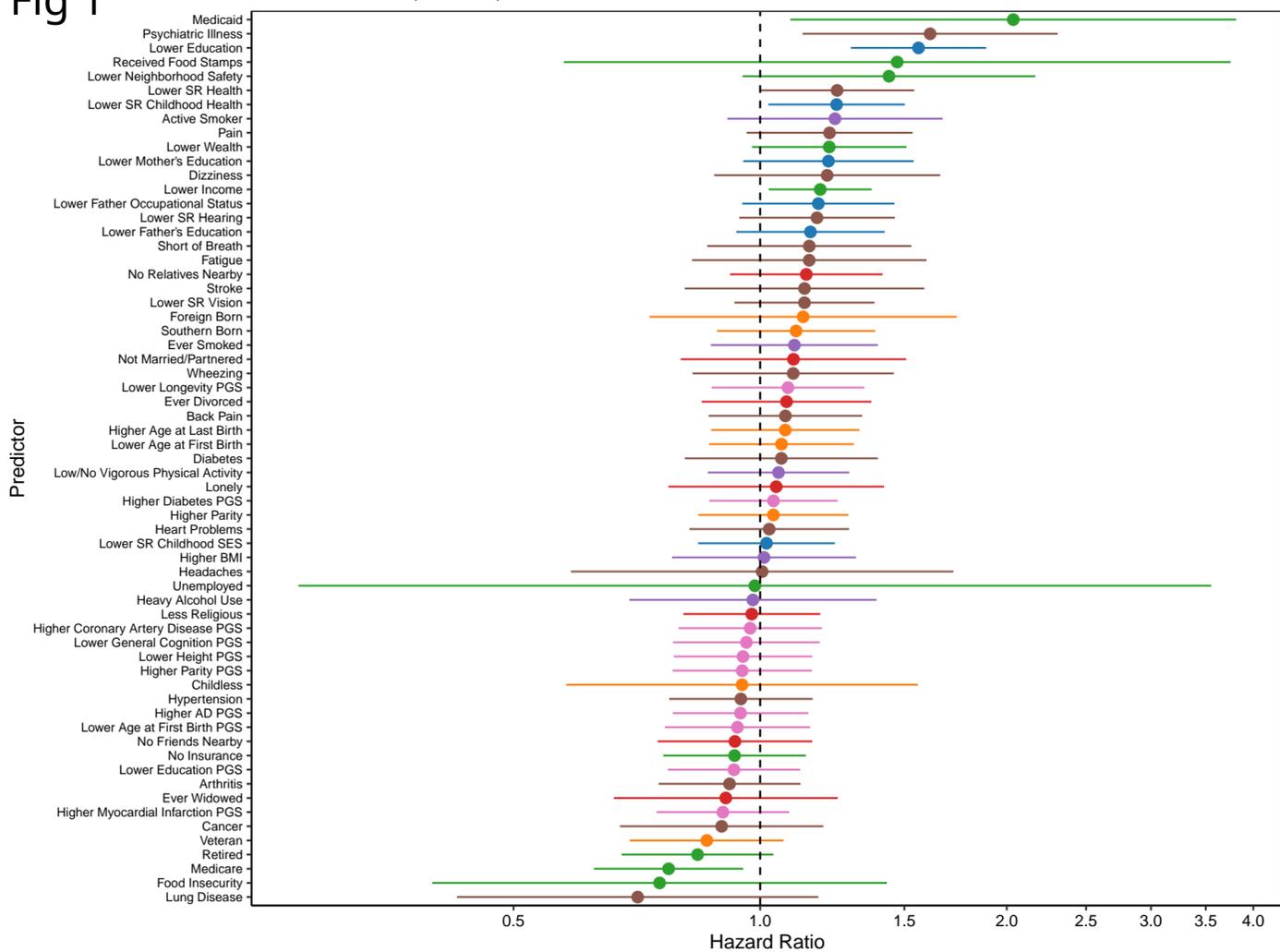

NH White Men (n=2960)

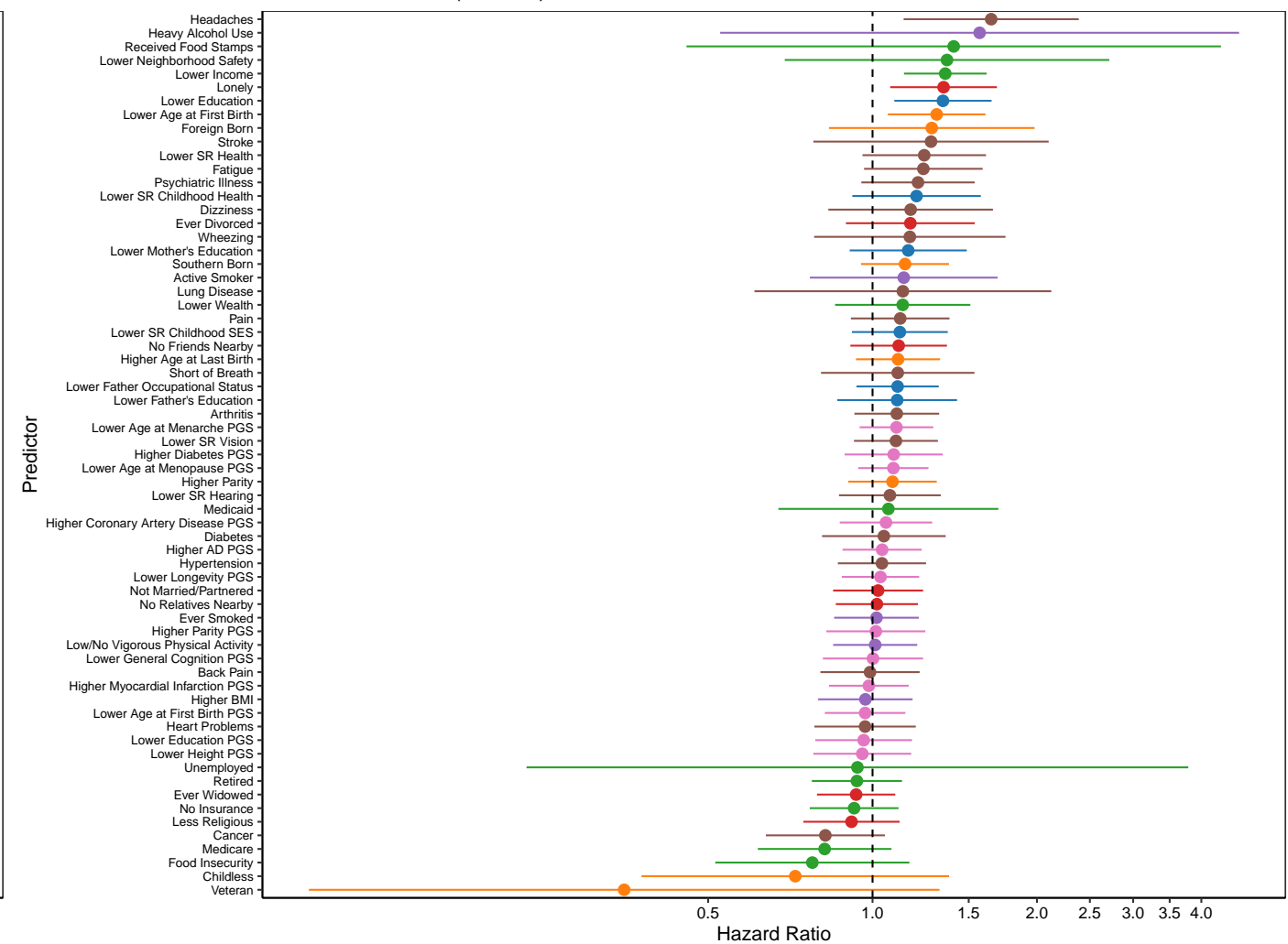

NH White Women (n=3946)

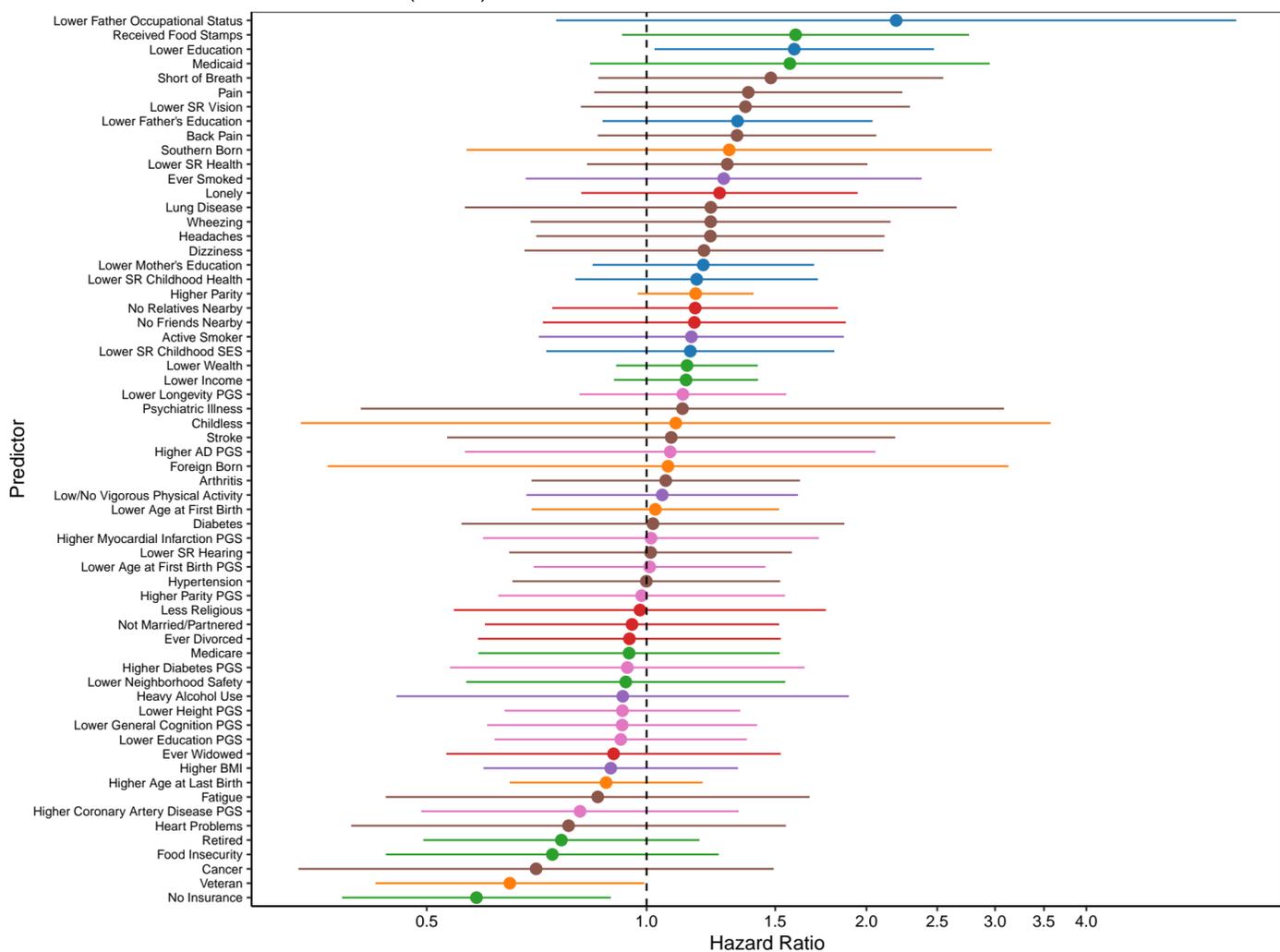

NH Black Men (n=353)

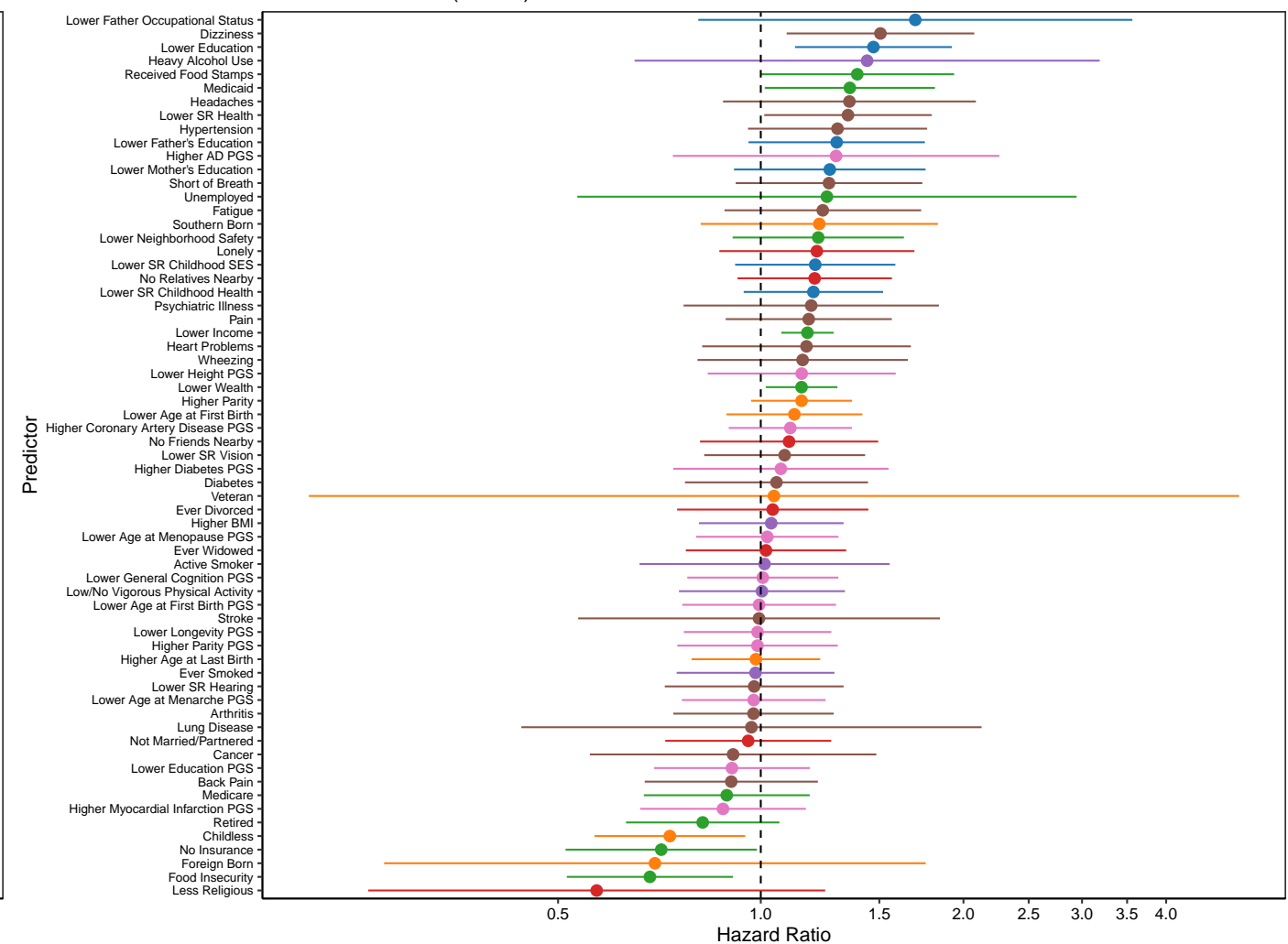

NH Black Women (n=649)

Early-Life   Economic   Behaviors   Genetic
Sociodemographic   Social Ties   Health

Fig 2

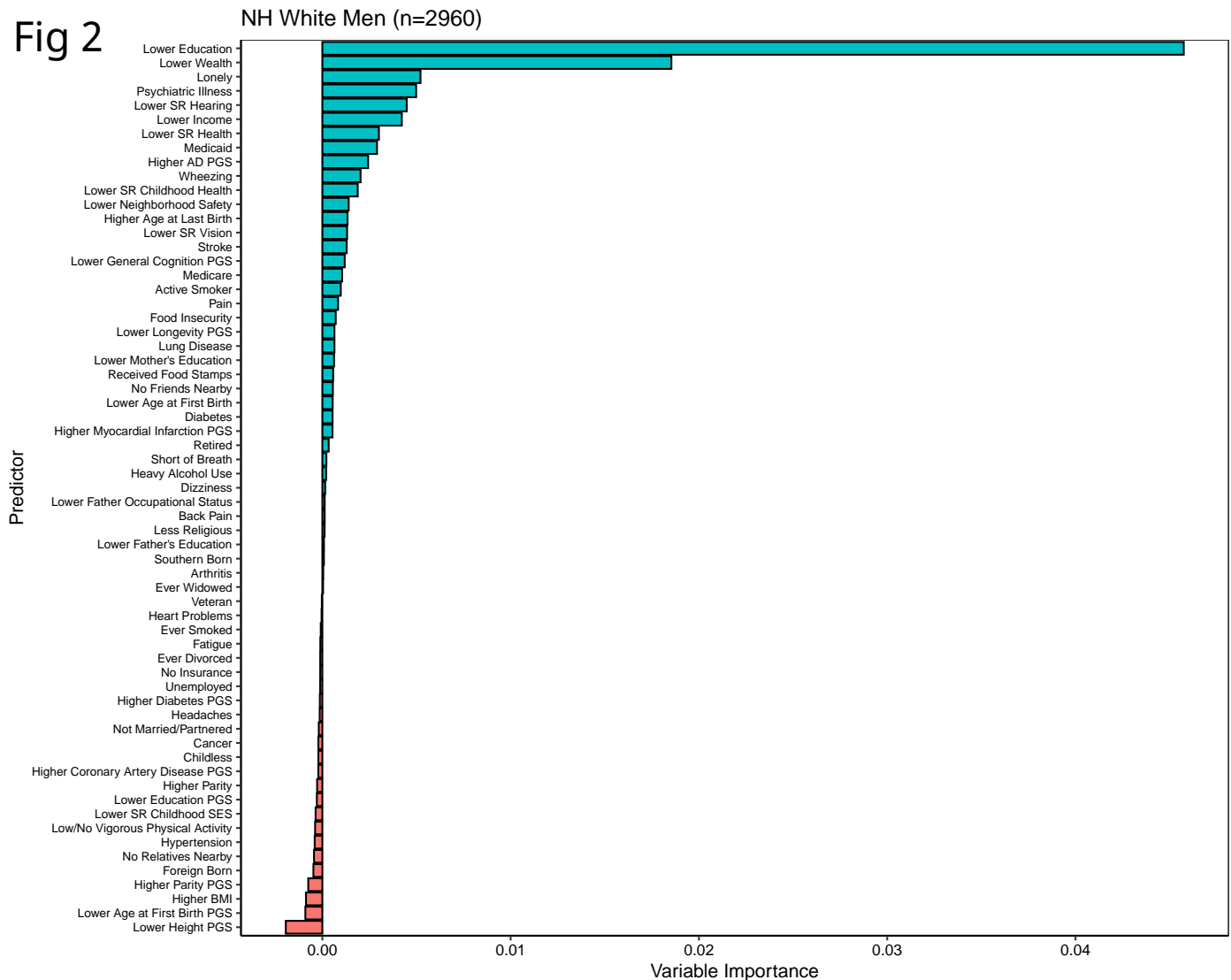

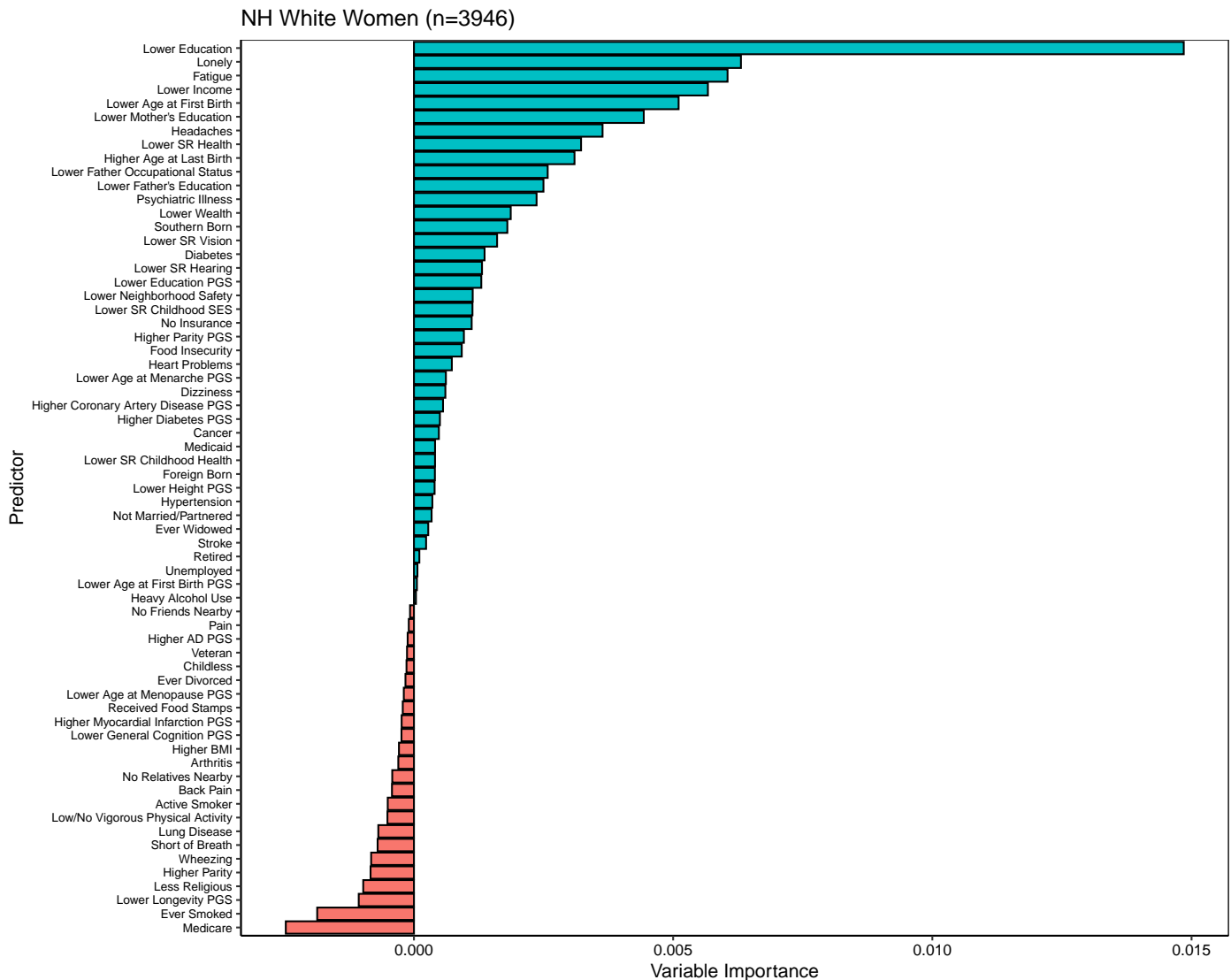

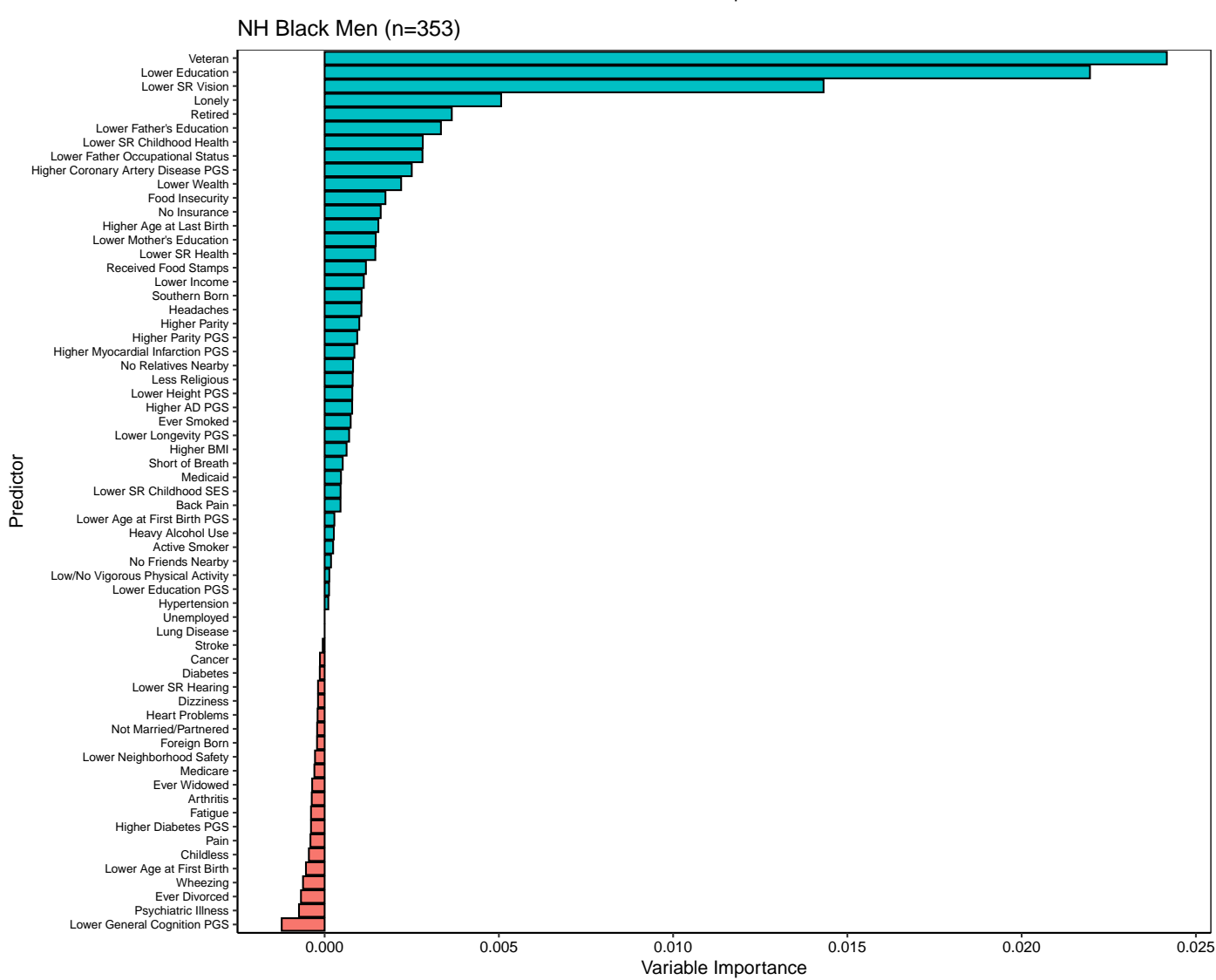

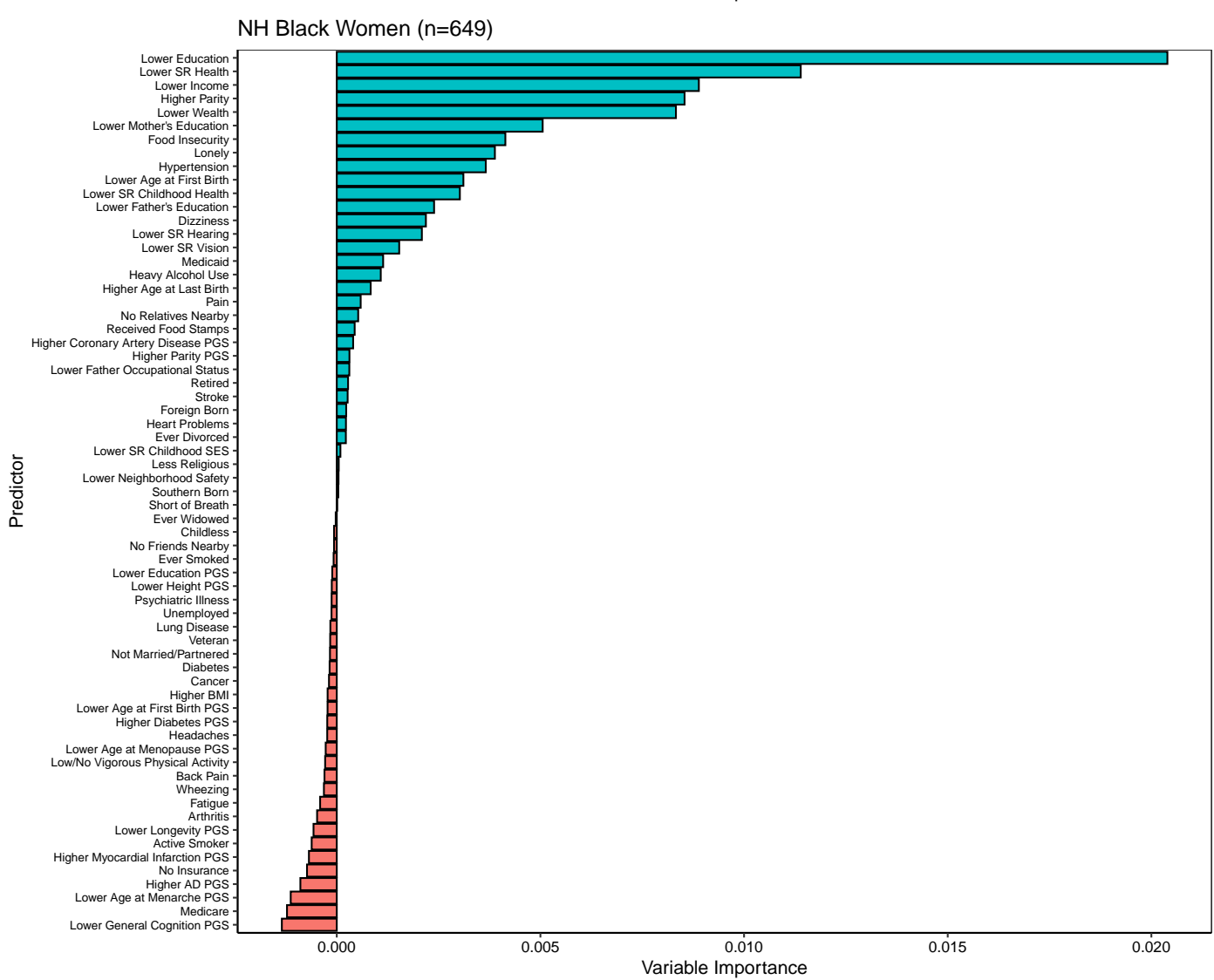

Fig 3

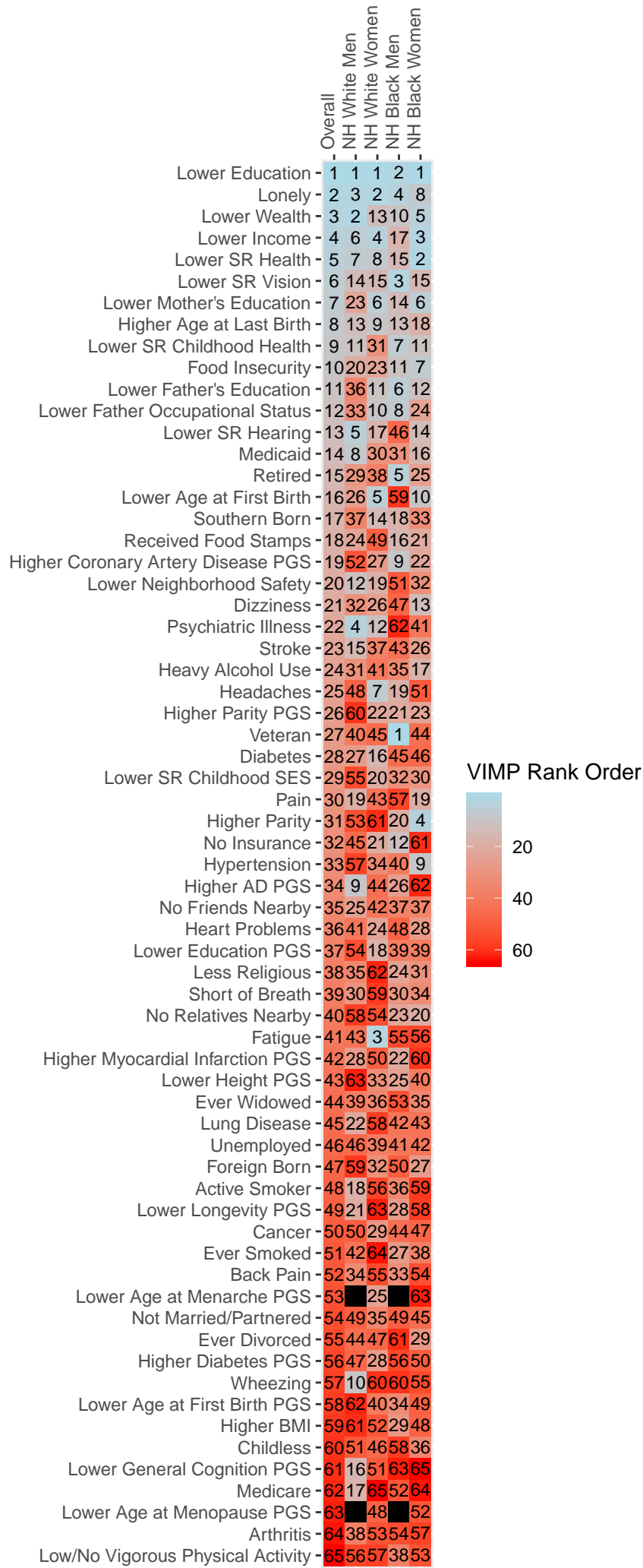

## Supporting information

**S1 Fig. Correlation matrices for 65** predictors **stratified by race and gender.**

**S2 Fig. Comparison of hazard ratios (HRs) and 95% confidence intervals (CI) of each predictor for incident dementia obtained from Fine-Gray regression models stratified by race and gender.** Model uses full analytic sample and classifies dementia using the Langa-Weir classification scheme.

**S3 Fig. Hazard ratios (HRs) and 95% confidence intervals (CI) of each predictor for incident dementia obtained from cause-specific regression models stratified by race and gender.** Models use restricted analytic sample and classify dementia using the Langa-Weir classification scheme. Predictors with HRs equal to zero are excluded from the figure but retained in Table S5.

**S4 Fig. Hazard ratios (HRs) and 95% confidence intervals (CI) of each predictor for incident dementia obtained from Fine-Gray regression models stratified by race and gender.** Models use restricted analytic sample and classify dementia using the Hurd classification scheme. Predictors with HRs equal to zero are excluded from the figure but retained in Table S6.

**S5 Fig. Hazard ratios (HRs) and 95% confidence intervals (CI) of each predictor for incident dementia obtained from Fine-Gray regression models stratified by race and gender.** Models use restricted analytic sample and classify dementia using the Expert

classification scheme. Predictors with HRs equal to zero are excluded from the figure but retained in Table S7.

**S6 Fig. Hazard ratios (HRs) and 95% confidence intervals (CI) of each predictor for incident dementia obtained from Fine-Gray regression models stratified by race and gender.** Models use restricted analytic sample and classify dementia using the LASSO classification scheme. Predictors with HRs equal to zero are excluded from the figure but retained in Table S8.

**S7 Fig. Variable importance plot for 65 characteristics predicting dementia obtained from random survival forests for competing risks stratified by race and gender.** Model uses restricted analytic sample and classifies dementia using the Langa-Weir classification scheme.

**S8 Fig. Variable importance plot for 65 characteristics predicting dementia obtained from random survival forests for competing risks stratified by race and gender.** Model uses restricted analytic sample and classifies dementia using the Hurd classification scheme.

**S9 Fig. Variable importance plot for 65 characteristics predicting dementia obtained from random survival forests for competing risks stratified by race and gender.** Model uses restricted analytic sample and classifies dementia using the Expert classification scheme.

**S10 Fig. Variable importance plot for 65 characteristics predicting dementia obtained from random survival forests for competing risks stratified by race and gender.** Model uses restricted analytic sample and classifies dementia using the LASSO classification scheme.

**S11 Fig. Rank order of predictors obtained from random survival forests for competing risks stratified by race and gender.** Model uses restricted analytic sample and classifies dementia using the Langa-Weir classification scheme.

**S12 Fig. Rank order of predictors obtained from random survival forests for competing risks stratified by race and gender.** Model uses restricted analytic sample and classifies dementia using the Hurd classification scheme.

**S13 Fig. Rank order of predictors obtained from random survival forests for competing risks stratified by race and gender.** Model uses restricted analytic sample and classifies dementia using the Expert classification scheme.

**S14 Fig. Rank order of predictors obtained from random survival forests for competing risks stratified by race and gender.** Model uses restricted analytic sample and classifies dementia using the LASSO classification scheme.

**S1 Table. Descriptive characteristics of the full analytic sample, HRS 2000.**

| Characteristic | Statistic[*] | % Missing |
|---|---|---|
| *Sociodemographic* | | |
| Baseline Age | 65.59 (0.15) | 0.000 |
| Female, % | 0.56 (0.01) | 0.000 |
| Non-Hispanic Black, % | 0.10 (0.00) | 0.000 |
| Foreign Born, % | 0.08 (0.01) | 0.001 |
| Southern Born, % | 0.31 (0.01) | 0.001 |
| Veteran, % | 0.28 (0.01) | 0.001 |
| Childless, % | 0.03 (0.00) | 0.001 |
| Parity | 2.94 (0.02) | 0.001 |
| Age at First Birth | 24.90 (0.08) | 0.000 |
| Age at Last Birth | 31.98 (0.10) | 0.000 |
| *Early-Life* | | |
| Mother's Education, years | 9.64 (0.05) | 0.08 |
| Father's Education, years | 9.26 (0.05) | 0.123 |
| Father's Occupational Status, unit | 2.15 (0.03) | 0.132 |
| SR Childhood Health | 4.27 (0.01) | 0.013 |
| SR Childhood SES | 1.80 (0.01) | 0.013 |
| Education, years | 3.32 (0.02) | 0.000 |
| *Economic* | | |
| Food Insecurity, % | 0.04 (0.00) | 0.005 |
| Income, dollars | 66270.17 (2039.69) | 0.000 |
| Neighborhood Safety | 1.91 (0.01) | 0.006 |
| Wealth, dollars | 371402.5 (13790.59) | 0.000 |
| Medicaid, % | 0.04 (0.00) | 0.001 |
| Medicare, % | 0.45 (0.01) | 0.001 |
| No Insurance, % | 0.37 (0.01) | 0.021 |
| Received Food Stamps, % | 0.03 (0.00) | 0.006 |
| Retired, % | 0.52 (0.01) | 0.000 |
| Unemployed, % | 0.01 (0.00) | 0.000 |
| *Health* | | |
| Arthritis, % | 0.49 (0.01) | 0.001 |
| Back Pain, % | 0.31 (0.01) | 0.000 |
| Cancer, % | 0.11 (0.00) | 0.001 |
| Diabetes, % | 0.11 (0.00) | 0.000 |
| Dizziness, % | 0.09 (0.00) | 0.000 |
| Fatigue, % | 0.14 (0.01) | 0.000 |
| Headaches, % | 0.08 (0.00) | 0.000 |
| Heart Problems, % | 0.17 (0.01) | 0.000 |
| Hypertension, % | 0.42 (0.01) | 0.001 |
| SR Health | 3.53 (0.02) | 0.000 |
| SR Hearing | 4.47 (0.02) | 0.001 |
| SR Vision | 4.36 (0.01) | 0.000 |
| Lung Disease, % | 0.05 (0.00) | 0.000 |
| Pain, % | 0.26 (0.01) | 0.000 |
| Psychiatric Illness, % | 0.08 (0.00) | 0.000 |
| Short of Breath, % | 0.13 (0.00) | 0.000 |
| Stroke, % | 0.05 (0.00) | 0.000 |
| Wheezing, % | 0.10 (0.00) | 0.000 |
| Cataracts [Excluded] | | 0.545 |
| Falls [Excluded] | | 0.522 |
| Glaucoma [Excluded] | | 0.541 |
| Swelling [Excluded] | | 0.448 |
| *Behaviors* | | |
| Active Smoker, % | 0.14 (0.01) | 0.000 |
| Ever Smoked, % | 0.59 (0.01) | 0.000 |
| Heavy Alcohol Use, % | 0.06 (0.00) | 0.001 |
| BMI | 27.27 (0.07) | 0.01 |
| Low/No Vigorous Physical Activity, % | 0.5 (0.01) | 0.000 |
| *Social Connections* | | |
| Ever Divorced, % | 0.29 (0.01) | 0.000 |
| Ever Widowed, % | 0.23 (0.01) | 0.000 |
| Religious | 2.52 (0.01) | 0.001 |
| Lonely, % | 0.15 (0.01) | 0.038 |
| No Friends Nearby, % | 0.69 (0.01) | 0.005 |
| No Relatives Nearby, % | 0.30 (0.01) | 0.006 |
| Not Married/Partnered, % | 0.33 (0.01) | 0.001 |
| *Genetic* | | |
| AD PGS | 0.00 (0.01) | 0.000 |
| Coronary Artery Disease PGS | -0.01 (0.01) | 0.000 |
| Diabetes PGS | 0.01 (0.01) | 0.000 |
| Myocardial Infarction PGS | 0.01 (0.01) | 0.000 |
| Parity PGS | -0.01 (0.01) | 0.000 |
| Age at First Birth PGS | 0.00 (0.01) | 0.000 |
| Age at Menarche PGS | 0.00 (0.01) | 0.000 |
| Age at Menopause PGS | 0.00 (0.01) | 0.000 |
| Education PGS | 0.00 (0.01) | 0.000 |
| General Cognition PGS | 0.00 (0.01) | 0.000 |
| Height PGS | 0.00 (0.01) | 0.000 |
| Longevity PGS | 0.00 (0.01) | 0.000 |

[*] Weighted means or proportions with standard errors listed in parentheses.

**S2 Table. Hazard ratios (HRs) and 95% confidence intervals (CI) of each predictor for incident dementia obtained from Fine-Gray regression models stratified by race and gender.** Models use full analytic sample and classify dementia using the Langa-Weir classification scheme.

| | NH White | | NH Black | |
|---|---|---|---|---|
| Characteristic | Men | Women | Men | Women |
| *Sociodemographic* | | | | |
| Foreign Born | 1.13 (0.73, 1.74) | 1.28 (0.83, 1.98) | 1.07 (0.37, 3.13) | 0.70 (0.28, 1.76) |
| Southern Born | 1.11 (0.89, 1.38) | 1.15 (0.95, 1.38) | 1.30 (0.57, 2.97) | 1.22 (0.81, 1.83) |
| Veteran | 0.86 (0.69, 1.07) | 0.35 (0.09, 1.33) | 0.65 (0.43, 0.99) | 1.05 (0.21, 5.13) |
| Childless | 0.95 (0.58, 1.56) | 0.72 (0.38, 1.38) | 1.10 (0.34, 3.58) | 0.73 (0.57, 0.95) |
| Higher Parity | 1.04 (0.84, 1.28) | 1.09 (0.90, 1.31) | 1.17 (0.97, 1.40) | 1.15 (0.97, 1.37) |
| Lower Age at First Birth | 1.06 (0.87, 1.30) | 1.31 (1.07, 1.61) | 1.03 (0.70, 1.52) | 1.12 (0.89, 1.42) |
| Higher Age at Last Birth | 1.07 (0.87, 1.32) | 1.11 (0.93, 1.33) | 0.88 (0.65, 1.19) | 0.98 (0.79, 1.23) |
| *Early-Life* | | | | |
| Lower Mother's Education | 1.21 (0.95, 1.54) | 1.16 (0.91, 1.49) | 1.20 (0.84, 1.70) | 1.27 (0.91, 1.76) |
| Lower Father's Education | 1.15 (0.94, 1.42) | 1.11 (0.86, 1.43) | 1.33 (0.87, 2.04) | 1.30 (0.96, 1.75) |
| Lower Father's Occupational Status | 1.18 (0.95, 1.46) | 1.11 (0.93, 1.32) | 2.20 (0.75, 6.42) | 1.70 (0.81, 3.56) |
| Lower SR Childhood Health | 1.24 (1.02, 1.50) | 1.20 (0.92, 1.58) | 1.17 (0.80, 1.72) | 1.20 (0.94, 1.52) |
| Lower SR Childhood SES | 1.02 (0.84, 1.23) | 1.12 (0.92, 1.37) | 1.15 (0.73, 1.81) | 1.20 (0.92, 1.58) |
| Lower Education | 1.56 (1.29, 1.89) | 1.35 (1.10, 1.65) | 1.59 (1.03, 2.47) | 1.47 (1.12, 1.92) |
| *Economic* | | | | |
| Food Insecurity | 0.75 (0.40, 1.43) | 0.78 (0.52, 1.17) | 0.74 (0.44, 1.26) | 0.68 (0.52, 0.91) |
| Lower Income | 1.18 (1.02, 1.37) | 1.36 (1.14, 1.62) | 1.13 (0.90, 1.42) | 1.17 (1.07, 1.28) |
| Lower Neighborhood Safety | 1.44 (0.95, 2.17) | 1.37 (0.69, 2.71) | 0.94 (0.57, 1.55) | 1.22 (0.91, 1.63) |
| Lower Wealth | 1.21 (0.98, 1.51) | 1.14 (0.85, 1.51) | 1.14 (0.91, 1.42) | 1.15 (1.02, 1.30) |
| Medicaid | 2.04 (1.09, 3.81) | 1.07 (0.67, 1.70) | 1.57 (0.84, 2.95) | 1.36 (1.01, 1.81) |
| Medicare | 0.77 (0.63, 0.95) | 0.82 (0.62, 1.08) | 0.95 (0.59, 1.52) | 0.89 (0.67, 1.18) |
| No Insurance | 0.93 (0.76, 1.14) | 0.93 (0.77, 1.12) | 0.58 (0.38, 0.89) | 0.71 (0.51, 0.99) |
| Received Food Stamps | 1.47 (0.58, 3.75) | 1.41 (0.46, 4.34) | 1.60 (0.93, 2.76) | 1.39 (1.00, 1.94) |
| Retired | 0.84 (0.68, 1.04) | 0.94 (0.77, 1.13) | 0.76 (0.49, 1.18) | 0.82 (0.63, 1.07) |
| Unemployed | 0.99 (0.27, 3.55) | 0.94 (0.23, 3.79) | 0.00 (0.00, 0.00) | 1.25 (0.53, 2.94) |
| *Health* | | | | |
| Arthritis | 0.92 (0.75, 1.12) | 1.11 (0.93, 1.32) | 1.06 (0.70, 1.62) | 0.98 (0.74, 1.28) |
| Back Pain | 1.07 (0.87, 1.33) | 0.99 (0.80, 1.22) | 1.33 (0.86, 2.06) | 0.90 (0.67, 1.22) |
| Cancer | 0.90 (0.67, 1.19) | 0.82 (0.64, 1.05) | 0.71 (0.33, 1.49) | 0.91 (0.56, 1.48) |
| Diabetes | 1.06 (0.81, 1.39) | 1.05 (0.81, 1.36) | 1.02 (0.56, 1.87) | 1.06 (0.77, 1.44) |
| Dizziness | 1.21 (0.88, 1.66) | 1.17 (0.83, 1.66) | 1.20 (0.68, 2.11) | 1.51 (1.09, 2.08) |
| Fatigue | 1.15 (0.83, 1.60) | 1.24 (0.96, 1.59) | 0.86 (0.44, 1.67) | 1.24 (0.88, 1.73) |
| Headaches | 1.01 (0.59, 1.72) | 1.65 (1.14, 2.39) | 1.22 (0.71, 2.12) | 1.35 (0.88, 2.09) |
| Heart Problems | 1.03 (0.82, 1.28) | 0.97 (0.78, 1.20) | 0.78 (0.39, 1.55) | 1.17 (0.82, 1.67) |
| Hypertension | 0.95 (0.77, 1.16) | 1.04 (0.86, 1.25) | 1.00 (0.66, 1.52) | 1.30 (0.96, 1.77) |
| Lower SR Health | 1.24 (1.00, 1.54) | 1.24 (0.96, 1.61) | 1.29 (0.83, 2.01) | 1.35 (1.01, 1.79) |
| Lower SR Hearing | 1.17 (0.94, 1.46) | 1.08 (0.87, 1.33) | 1.01 (0.65, 1.58) | 0.98 (0.72, 1.33) |
| Lower SR Vision | 1.13 (0.93, 1.38) | 1.10 (0.92, 1.32) | 1.37 (0.81, 2.29) | 1.09 (0.82, 1.43) |
| Lung Disease | 0.71 (0.43, 1.18) | 1.14 (0.61, 2.12) | 1.22 (0.56, 2.66) | 0.97 (0.44, 2.13) |
| Pain | 1.22 (0.96, 1.54) | 1.12 (0.91, 1.38) | 1.38 (0.85, 2.24) | 1.18 (0.89, 1.57) |
| Psychiatric Illness | 1.61 (1.13, 2.31) | 1.21 (0.95, 1.54) | 1.12 (0.41, 3.09) | 1.19 (0.77, 1.84) |
| Short of Breath | 1.15 (0.86, 1.53) | 1.11 (0.80, 1.54) | 1.48 (0.86, 2.55) | 1.26 (0.92, 1.74) |
| Stroke | 1.13 (0.81, 1.59) | 1.28 (0.78, 2.10) | 1.08 (0.53, 2.19) | 0.99 (0.54, 1.85) |
| Wheezing | 1.10 (0.83, 1.46) | 1.17 (0.78, 1.75) | 1.22 (0.69, 2.16) | 1.15 (0.81, 1.65) |
| *Behaviors* | | | | |
| Active Smoker | 1.23 (0.91, 1.67) | 1.15 (0.71, 1.86) | 1.14 (0.77, 1.69) | 1.01 (0.66, 1.55) |
| Ever Smoked | 1.10 (0.87, 1.39) | 1.27 (0.68, 2.38) | 1.02 (0.85, 1.22) | 0.98 (0.75, 1.29) |
| Heavy Alcohol Use | 0.98 (0.69, 1.39) | 0.93 (0.45, 1.89) | 1.57 (0.53, 4.69) | 1.44 (0.65, 3.19) |
| Higher BMI | 1.01 (0.78, 1.31) | 0.89 (0.60, 1.33) | 0.97 (0.80, 1.18) | 1.04 (0.81, 1.33) |
| Low/No Vigorous Physical Activity | 1.05 (0.86, 1.29) | 1.05 (0.68, 1.61) | 1.01 (0.85, 1.21) | 1.00 (0.76, 1.33) |
| *Social Connections* | | | | |
| Ever Divorced | 1.08 (0.85, 1.37) | 1.17 (0.89, 1.54) | 0.95 (0.59, 1.53) | 1.04 (0.75, 1.44) |
| Ever Widowed | 0.91 (0.66, 1.24) | 0.93 (0.79, 1.10) | 0.90 (0.53, 1.53) | 1.02 (0.77, 1.34) |
| Less Religious | 0.98 (0.81, 1.18) | 0.92 (0.75, 1.12) | 0.98 (0.54, 1.76) | 0.57 (0.26, 1.25) |
| Lonely | 1.05 (0.77, 1.42) | 1.35 (1.08, 1.69) | 1.26 (0.81, 1.95) | 1.21 (0.87, 1.69) |
| No Friends Nearby | 0.93 (0.75, 1.16) | 1.12 (0.91, 1.37) | 1.16 (0.72, 1.87) | 1.10 (0.81, 1.49) |
| No Relatives Nearby | 1.14 (0.92, 1.41) | 1.02 (0.86, 1.21) | 1.17 (0.74, 1.83) | 1.20 (0.92, 1.57) |
| Not Married/Partnered | 1.10 (0.80, 1.51) | 1.02 (0.85, 1.24) | 0.96 (0.60, 1.52) | 0.96 (0.72, 1.27) |

| *Genetic* | | | | |
|---|---|---|---|---|
| Higher AD PGS | 0.95 (0.78, 1.15) | 1.04 (0.88, 1.23) | 1.08 (0.56, 2.06) | 1.29 (0.74, 2.26) |
| Higher Coronary Artery Disease PGS | 0.97 (0.79, 1.19) | 1.06 (0.87, 1.29) | 0.81 (0.49, 1.34) | 1.11 (0.90, 1.37) |
| Higher Diabetes PGS | 1.04 (0.87, 1.24) | 1.09 (0.89, 1.34) | 0.94 (0.54, 1.65) | 1.07 (0.74, 1.55) |
| Higher Myocardial Infarction PGS | 0.90 (0.75, 1.09) | 0.98 (0.83, 1.16) | 1.01 (0.60, 1.72) | 0.88 (0.66, 1.17) |
| Higher Parity PGS | 0.95 (0.78, 1.16) | 1.01 (0.82, 1.25) | 0.98 (0.63, 1.55) | 0.99 (0.75, 1.30) |
| Lower Age at First Birth PGS | 0.94 (0.76, 1.15) | 0.97 (0.82, 1.15) | 1.01 (0.70, 1.45) | 0.99 (0.76, 1.29) |
| Lower Age at Menarche PGS | — | 1.11 (0.95, 1.29) | — | 0.98 (0.76, 1.25) |
| Lower Age at Menopause PGS | — | 1.09 (0.94, 1.27) | — | 1.02 (0.80, 1.30) |
| Lower Education PGS | 0.93 (0.77, 1.12) | 0.96 (0.79, 1.18) | 0.92 (0.62, 1.37) | 0.91 (0.69, 1.18) |
| Lower General Cognition PGS | 0.96 (0.78, 1.18) | 1.00 (0.81, 1.24) | 0.93 (0.60, 1.42) | 1.01 (0.78, 1.30) |
| Lower Height PGS | 0.95 (0.78, 1.16) | 0.96 (0.78, 1.18) | 0.93 (0.64, 1.34) | 1.15 (0.83, 1.59) |
| Lower Longevity PGS | 1.08 (0.87, 1.34) | 1.03 (0.88, 1.22) | 1.12 (0.81, 1.55) | 0.99 (0.77, 1.27) |

**S3 Table. Cause-specific hazard ratios (HRs) and 95% confidence intervals (CI) of each predictor for incident dementia obtained from independent Cox models stratified by race and gender.** Models use full analytic sample and classify dementia using the Langa-Weir classification scheme.

| | NH White | | NH Black | |
|---|---|---|---|---|
| Characteristic | Men | Women | Men | Women |
| *Sociodemographic* | | | | |
| Foreign Born | 1.07 (0.68, 1.67) | 1.23 (0.79, 1.91) | 1.12 (0.40, 3.19) | 0.69 (0.26, 1.83) |
| Southern Born | 1.22 (0.97, 1.53) | 1.21 (0.99, 1.47) | 1.27 (0.56, 2.86) | 1.23 (0.80, 1.87) |
| Veteran | 0.87 (0.69, 1.09) | 0.36 (0.09, 1.38) | 0.58 (0.36, 0.92) | 1.03 (0.21, 5.07) |
| Childless | 0.96 (0.60, 1.55) | 0.80 (0.43, 1.50) | 1.07 (0.33, 3.49) | 0.29 (0.04, 2.17) |
| Higher Parity | 1.07 (0.87, 1.33) | 1.12 (0.93, 1.35) | 1.14 (0.96, 1.37) | 1.17 (0.98, 1.39) |
| Lower Age at First Birth | 1.20 (0.96, 1.51) | 1.46 (1.18, 1.81) | 1.00 (0.67, 1.50) | 1.20 (0.95, 1.52) |
| Higher Age at Last Birth | 1.17 (0.93, 1.47) | 1.16 (0.97, 1.39) | 0.92 (0.67, 1.25) | 1.01 (0.80, 1.27) |
| *Early-Life* | | | | |
| Lower Mother's Education | 1.13 (0.86, 1.48) | 1.08 (0.80, 1.45) | 1.17 (0.81, 1.69) | 1.27 (0.90, 1.80) |
| Lower Father's Education | 1.12 (0.88, 1.42) | 1.13 (0.83, 1.52) | 1.27 (0.82, 1.98) | 1.28 (0.94, 1.74) |
| Lower Father's Occupational Status | 1.23 (0.98, 1.53) | 1.13 (0.95, 1.35) | 2.11 (0.78, 5.71) | 1.78 (0.80, 3.96) |
| Lower SR Childhood Health | 1.22 (1.00, 1.48) | 1.18 (0.89, 1.57) | 1.17 (0.80, 1.70) | 1.20 (0.94, 1.53) |
| Lower SR Childhood SES | 1.01 (0.83, 1.23) | 1.16 (0.94, 1.43) | 1.09 (0.69, 1.72) | 1.14 (0.85, 1.53) |
| Lower Education | 1.62 (1.34, 1.97) | 1.32 (1.07, 1.64) | 1.58 (1.02, 2.44) | 1.46 (1.11, 1.92) |
| *Economic* | | | | |
| Food Insecurity | 0.66 (0.34, 1.27) | 0.75 (0.49, 1.16) | 0.71 (0.44, 1.17) | 0.67 (0.50, 0.91) |
| Lower Income | 1.24 (1.06, 1.44) | 1.35 (1.13, 1.62) | 1.13 (0.90, 1.42) | 1.17 (1.07, 1.28) |
| Lower Neighborhood Safety | 1.47 (0.95, 2.27) | 1.45 (0.73, 2.90) | 0.95 (0.59, 1.56) | 1.26 (0.93, 1.70) |
| Lower Wealth | 1.31 (1.05, 1.65) | 1.18 (0.89, 1.58) | 1.11 (0.87, 1.42) | 1.16 (1.02, 1.32) |
| Medicaid | 2.31 (1.20, 4.47) | 1.14 (0.71, 1.84) | 1.63 (0.86, 3.10) | 1.44 (1.05, 1.97) |
| Medicare | 0.55 (0.42, 0.71) | 0.57 (0.42, 0.78) | 0.81 (0.48, 1.36) | 0.69 (0.51, 0.92) |
| No Insurance | 0.96 (0.78, 1.18) | 0.97 (0.80, 1.17) | 0.61 (0.40, 0.94) | 0.73 (0.52, 1.03) |
| Received Food Stamps | 1.80 (0.60, 5.38) | 1.73 (0.57, 5.23) | 1.63 (0.98, 2.72) | 1.63 (1.22, 2.18) |
| Retired | 0.70 (0.56, 0.89) | 0.93 (0.76, 1.14) | 0.62 (0.38, 1.01) | 0.75 (0.57, 1.00) |
| Unemployed | 1.03 (0.28, 3.81) | 1.20 (0.31, 4.70) | 0.00 (0.00, 0.00) | 1.44 (0.77, 2.69) |
| *Health* | | | | |

| Arthritis | 0.93 (0.76, 1.13) | 1.09 (0.91, 1.31) | 1.02 (0.66, 1.57) | 1.00 (0.75, 1.33) |
|---|---|---|---|---|
| Back Pain | 1.09 (0.88, 1.36) | 1.02 (0.82, 1.26) | 1.37 (0.88, 2.12) | 0.93 (0.69, 1.26) |
| Cancer | 0.87 (0.66, 1.17) | 0.79 (0.60, 1.04) | 0.75 (0.36, 1.55) | 1.01 (0.66, 1.54) |
| Diabetes | 1.10 (0.83, 1.46) | 1.20 (0.92, 1.56) | 1.09 (0.60, 1.96) | 1.17 (0.86, 1.60) |
| Dizziness | 1.16 (0.83, 1.63) | 1.19 (0.84, 1.70) | 1.15 (0.62, 2.14) | 1.55 (1.11, 2.16) |
| Fatigue | 1.23 (0.87, 1.75) | 1.29 (1.00, 1.67) | 0.93 (0.50, 1.72) | 1.35 (0.96, 1.89) |
| Headaches | 1.14 (0.67, 1.97) | 1.76 (1.20, 2.58) | 1.22 (0.69, 2.15) | 1.56 (1.05, 2.31) |
| Heart Problems | 1.01 (0.80, 1.27) | 1.00 (0.80, 1.25) | 0.81 (0.41, 1.60) | 1.22 (0.83, 1.78) |
| Hypertension | 0.95 (0.77, 1.16) | 1.04 (0.86, 1.26) | 1.02 (0.67, 1.57) | 1.31 (0.96, 1.78) |
| Lower SR Health | 1.32 (1.05, 1.65) | 1.33 (1.02, 1.74) | 1.31 (0.84, 2.04) | 1.48 (1.10, 2.00) |
| Lower SR Hearing | 1.10 (0.88, 1.38) | 1.03 (0.82, 1.28) | 1.04 (0.67, 1.62) | 0.97 (0.71, 1.34) |
| Lower SR Vision | 1.11 (0.91, 1.36) | 1.08 (0.90, 1.30) | 1.51 (0.90, 2.51) | 1.06 (0.80, 1.41) |
| Lung Disease | 0.83 (0.50, 1.37) | 1.34 (0.72, 2.47) | 1.23 (0.52, 2.90) | 1.13 (0.56, 2.28) |
| Pain | 1.27 (0.99, 1.62) | 1.16 (0.94, 1.44) | 1.42 (0.87, 2.33) | 1.24 (0.93, 1.66) |
| Psychiatric Illness | 1.79 (1.25, 2.57) | 1.24 (0.96, 1.61) | 1.18 (0.45, 3.06) | 1.34 (0.90, 2.02) |
| Short of Breath | 1.28 (0.95, 1.72) | 1.21 (0.87, 1.66) | 1.45 (0.81, 2.60) | 1.36 (0.98, 1.88) |
| Stroke | 1.18 (0.84, 1.65) | 1.28 (0.77, 2.12) | 1.10 (0.57, 2.13) | 1.05 (0.60, 1.85) |
| Wheezing | 1.15 (0.85, 1.55) | 1.33 (0.90, 1.98) | 1.22 (0.67, 2.21) | 1.13 (0.78, 1.65) |
| *Behaviors* | | | | |
| Active Smoker | 1.56 (1.15, 2.13) | 1.50 (1.02, 2.20) | 1.30 (0.81, 2.08) | 1.20 (0.80, 1.79) |
| Ever Smoked | 1.17 (0.92, 1.50) | 1.14 (0.95, 1.36) | 1.24 (0.67, 2.30) | 1.04 (0.79, 1.38) |
| Heavy Alcohol Use | 1.07 (0.76, 1.51) | 1.67 (0.56, 4.99) | 0.93 (0.44, 1.99) | 1.50 (0.66, 3.42) |
| Higher BMI | 1.13 (0.86, 1.49) | 1.04 (0.83, 1.31) | 0.92 (0.60, 1.39) | 1.10 (0.86, 1.40) |
| Low/No Vigorous Physical Activity | 1.11 (0.90, 1.35) | 1.02 (0.85, 1.22) | 1.05 (0.68, 1.61) | 1.02 (0.76, 1.37) |
| *Social Connections* | | | | |
| Ever Divorced | 1.25 (0.98, 1.60) | 1.37 (1.04, 1.80) | 0.94 (0.58, 1.52) | 1.11 (0.78, 1.58) |
| Ever Widowed | 0.84 (0.60, 1.16) | 0.86 (0.72, 1.02) | 0.86 (0.51, 1.45) | 0.91 (0.67, 1.23) |
| Less Religious | 0.98 (0.81, 1.19) | 0.91 (0.74, 1.11) | 0.96 (0.53, 1.75) | 0.62 (0.26, 1.48) |
| Lonely | 1.05 (0.76, 1.44) | 1.36 (1.08, 1.72) | 1.25 (0.81, 1.95) | 1.29 (0.92, 1.81) |
| No Friends Nearby | 0.90 (0.72, 1.13) | 1.08 (0.87, 1.33) | 1.14 (0.71, 1.83) | 1.02 (0.76, 1.38) |
| No Relatives Nearby | 1.13 (0.91, 1.41) | 0.98 (0.82, 1.17) | 1.26 (0.82, 1.96) | 1.25 (0.96, 1.65) |
| Not Married/Partnered | 1.14 (0.82, 1.59) | 1.09 (0.89, 1.33) | 1.03 (0.63, 1.67) | 1.02 (0.76, 1.36) |

| *Genetic* | | | | |
|---|---|---|---|---|
| Higher AD PGS | 0.95 (0.78, 1.15) | 1.04 (0.88, 1.23) | 1.32 (0.67, 2.61) | 1.32 (0.74, 2.38) |
| Higher Coronary Artery Disease PGS | 0.98 (0.80, 1.21) | 1.07 (0.87, 1.31) | 0.78 (0.47, 1.30) | 1.14 (0.92, 1.42) |
| Higher Diabetes PGS | 1.04 (0.86, 1.26) | 1.09 (0.88, 1.34) | 0.94 (0.54, 1.64) | 1.12 (0.77, 1.63) |
| Higher Myocardial Infarction PGS | 0.93 (0.77, 1.12) | 0.98 (0.83, 1.17) | 1.01 (0.59, 1.75) | 0.85 (0.64, 1.14) |
| Higher Parity PGS | 0.95 (0.78, 1.16) | 1.01 (0.81, 1.25) | 0.98 (0.62, 1.56) | 1.00 (0.75, 1.33) |
| Lower Age at First Birth PGS | 0.92 (0.75, 1.13) | 0.97 (0.82, 1.16) | 1.03 (0.71, 1.48) | 1.01 (0.76, 1.33) |
| Lower Age at Menarche PGS | — | 1.08 (0.92, 1.27) | | 0.97 (0.76, 1.23) |
| Lower Age at Menopause PGS | — | 1.08 (0.93, 1.26) | — | 1.09 (0.84, 1.41) |
| Lower Education PGS | 0.93 (0.77, 1.12) | 0.97 (0.79, 1.20) | 0.96 (0.64, 1.42) | 0.88 (0.67, 1.16) |
| Lower General Cognition PGS | 0.97 (0.78, 1.19) | 1.01 (0.81, 1.26) | 0.98 (0.65, 1.48) | 1.01 (0.76, 1.34) |
| Lower Height PGS | 0.97 (0.79, 1.19) | 0.99 (0.80, 1.22) | 0.92 (0.64, 1.33) | 1.13 (0.80, 1.59) |
| Lower Longevity PGS | 1.09 (0.88, 1.35) | 1.03 (0.88, 1.22) | 1.13 (0.82, 1.56) | 0.98 (0.77, 1.26) |

**S4 Table. Hazard ratios (HRs) and 95% confidence intervals (CI) of each predictor for incident dementia obtained from Fine-Gray regression models stratified by race and gender.** Models use restricted analytic sample and classify dementia using the Langa-Weir classification scheme.

| Characteristic | NH White | | NH Black | |
|---|---|---|---|---|
| | Men | Women | Men | Women |
| *Sociodemographic* | | | | |
| Foreign Born | 1.14 (0.73, 1.77) | 1.26 (0.81, 1.97) | 1.20 (0.39, 3.69) | 0.65 (0.23, 1.78) |
| Southern Born | 1.09 (0.87, 1.37) | 1.15 (0.95, 1.39) | 1.18 (0.50, 2.82) | 1.28 (0.83, 1.98) |
| Veteran | 0.85 (0.68, 1.06) | 0.36 (0.09, 1.35) | 0.69 (0.44, 1.09) | 1.13 (0.22, 5.67) |
| Childless | 0.98 (0.60, 1.60) | 0.73 (0.38, 1.41) | 1.17 (0.34, 3.95) | 0.72 (0.56, 0.94) |
| Higher Parity | 1.03 (0.83, 1.29) | 1.08 (0.90, 1.31) | 1.13 (0.92, 1.40) | 1.15 (0.96, 1.38) |
| Lower Age at First Birth | 1.06 (0.85, 1.33) | 1.30 (1.05, 1.60) | 1.02 (0.68, 1.52) | 1.11 (0.87, 1.41) |
| Higher Age at Last Birth | 1.07 (0.86, 1.33) | 1.12 (0.94, 1.34) | 0.88 (0.63, 1.22) | 0.98 (0.79, 1.22) |
| *Early-Life* | | | | |
| Lower Mother's Education | 1.19 (0.92, 1.55) | 1.13 (0.88, 1.46) | 1.26 (0.86, 1.85) | 1.25 (0.90, 1.74) |
| Lower Father's Education | 1.15 (0.93, 1.44) | 1.09 (0.84, 1.42) | 1.39 (0.91, 2.12) | 1.25 (0.91, 1.72) |
| Lower Father's Occupational Status | 1.15 (0.92, 1.43) | 1.12 (0.94, 1.33) | 2.67 (0.61, 11.66) | 1.72 (0.77, 3.87) |
| Lower SR Childhood Health | 1.23 (1.01, 1.50) | 1.21 (0.92, 1.59) | 1.16 (0.76, 1.77) | 1.18 (0.91, 1.51) |
| Lower SR Childhood SES | 1.02 (0.84, 1.24) | 1.10 (0.90, 1.35) | 1.19 (0.72, 1.96) | 1.17 (0.88, 1.57) |
| Lower Education | 1.55 (1.28, 1.87) | 1.34 (1.09, 1.65) | 1.61 (1.01, 2.57) | 1.47 (1.11, 1.94) |
| *Economic* | | | | |
| Food Insecurity | 0.73 (0.36, 1.45) | 0.74 (0.49, 1.13) | 0.77 (0.44, 1.37) | 0.65 (0.48, 0.87) |
| Lower Income | 1.22 (1.03, 1.43) | 1.38 (1.16, 1.63) | 1.07 (0.83, 1.39) | 1.19 (1.08, 1.31) |
| Lower Neighborhood Safety | 1.44 (0.93, 2.23) | 1.33 (0.64, 2.77) | 0.89 (0.52, 1.54) | 1.24 (0.92, 1.67) |
| Lower Wealth | 1.28 (1.00, 1.63) | 1.14 (0.84, 1.53) | 1.12 (0.88, 1.42) | 1.13 (1.00, 1.28) |
| Medicaid | 1.94 (0.97, 3.88) | 1.03 (0.64, 1.67) | 1.44 (0.73, 2.82) | 1.36 (1.00, 1.85) |
| Medicare | 0.71 (0.56, 0.91) | 0.75 (0.53, 1.06) | 0.73 (0.45, 1.19) | 0.83 (0.60, 1.15) |
| No Insurance | 0.94 (0.76, 1.15) | 0.94 (0.78, 1.13) | 0.67 (0.42, 1.06) | 0.73 (0.51, 1.03) |
| Received Food Stamps | 1.89 (0.64, 5.57) | 1.44 (0.47, 4.48) | 1.42 (0.87, 2.33) | 1.49 (1.05, 2.12) |
| Retired | 0.83 (0.65, 1.05) | 0.92 (0.77, 1.12) | 0.72 (0.43, 1.21) | 0.83 (0.62, 1.09) |
| Unemployed | 1.08 (0.29, 4.06) | 0.98 (0.24, 4.03) | 0.00 (0.00, 0.00) | 1.61 (0.89, 2.93) |
| *Health* | | | | |
| Arthritis | 0.89 (0.73, 1.09) | 1.11 (0.93, 1.33) | 0.99 (0.63, 1.56) | 0.93 (0.70, 1.24) |
| Back Pain | 1.09 (0.87, 1.36) | 0.99 (0.80, 1.23) | 1.18 (0.73, 1.92) | 0.85 (0.62, 1.15) |
| Cancer | 0.89 (0.66, 1.18) | 0.81 (0.63, 1.05) | 0.62 (0.30, 1.27) | 0.97 (0.59, 1.59) |
| Diabetes | 1.04 (0.79, 1.38) | 1.07 (0.82, 1.39) | 1.18 (0.63, 2.22) | 1.13 (0.81, 1.56) |
| Dizziness | 1.22 (0.88, 1.70) | 1.18 (0.83, 1.67) | 1.12 (0.60, 2.07) | 1.51 (1.07, 2.12) |
| Fatigue | 1.20 (0.85, 1.69) | 1.25 (0.96, 1.61) | 0.88 (0.42, 1.83) | 1.25 (0.88, 1.79) |
| Headaches | 1.14 (0.66, 1.96) | 1.69 (1.15, 2.49) | 1.12 (0.64, 1.99) | 1.36 (0.85, 2.19) |
| Heart Problems | 1.00 (0.80, 1.26) | 0.97 (0.78, 1.20) | 0.69 (0.31, 1.52) | 1.18 (0.81, 1.73) |
| Hypertension | 0.93 (0.75, 1.14) | 1.03 (0.86, 1.25) | 0.94 (0.59, 1.49) | 1.27 (0.93, 1.74) |
| Lower SR Health | 1.29 (1.03, 1.62) | 1.26 (0.96, 1.65) | 1.13 (0.72, 1.76) | 1.38 (1.01, 1.89) |
| Lower SR Hearing | 1.15 (0.92, 1.44) | 1.07 (0.86, 1.33) | 1.07 (0.69, 1.67) | 0.93 (0.68, 1.29) |
| Lower SR Vision | 1.11 (0.90, 1.36) | 1.10 (0.92, 1.32) | 1.27 (0.74, 2.17) | 1.04 (0.79, 1.38) |
| Lung Disease | 0.75 (0.45, 1.25) | 1.15 (0.60, 2.20) | 1.36 (0.58, 3.15) | 1.14 (0.51, 2.56) |
| Pain | 1.23 (0.97, 1.57) | 1.13 (0.91, 1.39) | 1.16 (0.67, 2.01) | 1.12 (0.83, 1.51) |
| Psychiatric Illness | 1.57 (1.07, 2.31) | 1.20 (0.94, 1.53) | 0.79 (0.23, 2.67) | 1.37 (0.89, 2.11) |
| Short of Breath | 1.18 (0.88, 1.59) | 1.11 (0.80, 1.55) | 1.39 (0.77, 2.51) | 1.17 (0.83, 1.66) |
| Stroke | 1.13 (0.80, 1.59) | 1.27 (0.77, 2.10) | 1.00 (0.49, 2.05) | 0.89 (0.47, 1.70) |
| Wheezing | 1.13 (0.84, 1.51) | 1.17 (0.77, 1.77) | 1.09 (0.59, 2.05) | 1.13 (0.77, 1.66) |
| *Behaviors* | | | | |
| Active Smoker | 1.24 (0.90, 1.72) | 1.15 (0.75, 1.78) | 1.11 (0.64, 1.93) | 1.04 (0.65, 1.66) |
| Ever Smoked | 1.13 (0.89, 1.44) | 1.02 (0.85, 1.22) | 1.27 (0.65, 2.51) | 1.00 (0.75, 1.32) |
| Heavy Alcohol Use | 1.02 (0.71, 1.45) | 1.62 (0.54, 4.86) | 0.86 (0.37, 2.02) | 1.19 (0.56, 2.53) |
| Higher BMI | 1.01 (0.77, 1.32) | 0.99 (0.81, 1.21) | 0.92 (0.56, 1.52) | 1.00 (0.78, 1.28) |
| Low/No Vigorous Physical Activity | 1.06 (0.87, 1.30) | 1.02 (0.85, 1.22) | 0.99 (0.62, 1.58) | 0.98 (0.73, 1.33) |
| *Social Connections* | | | | |
| Ever Divorced | 1.10 (0.86, 1.41) | 1.21 (0.91, 1.60) | 0.92 (0.54, 1.56) | 1.07 (0.75, 1.54) |
| Ever Widowed | 0.90 (0.66, 1.24) | 0.91 (0.77, 1.08) | 0.86 (0.50, 1.46) | 0.98 (0.73, 1.31) |
| Less Religious | 1.00 (0.82, 1.21) | 0.92 (0.75, 1.13) | 1.04 (0.55, 1.94) | 0.61 (0.28, 1.34) |
| Lonely | 1.00 (0.73, 1.38) | 1.35 (1.07, 1.70) | 1.30 (0.83, 2.05) | 1.26 (0.88, 1.81) |
| No Friends Nearby | 0.93 (0.74, 1.17) | 1.11 (0.90, 1.37) | 1.32 (0.79, 2.20) | 1.06 (0.77, 1.46) |
| No Relatives Nearby | 1.13 (0.91, 1.41) | 1.02 (0.85, 1.21) | 1.12 (0.69, 1.82) | 1.22 (0.93, 1.61) |
| Not Married/Partnered | 1.11 (0.80, 1.54) | 1.03 (0.86, 1.25) | 1.01 (0.60, 1.68) | 1.02 (0.76, 1.37) |

| Genetic | | | | |
|---|---|---|---|---|
| Higher AD PGS | 0.95 (0.79, 1.16) | 1.05 (0.88, 1.24) | 1.00 (0.49, 2.02) | 1.42 (0.78, 2.58) |
| Higher Coronary Artery Disease PGS | 0.95 (0.77, 1.17) | 1.06 (0.87, 1.30) | 0.78 (0.43, 1.44) | 1.14 (0.91, 1.43) |
| Higher Diabetes PGS | 1.05 (0.87, 1.27) | 1.11 (0.89, 1.37) | 0.96 (0.53, 1.72) | 1.07 (0.73, 1.57) |
| Higher Myocardial Infarction PGS | 0.91 (0.75, 1.10) | 0.99 (0.83, 1.17) | 0.96 (0.53, 1.73) | 0.85 (0.63, 1.14) |
| Higher Parity PGS | 0.98 (0.80, 1.20) | 1.03 (0.83, 1.28) | 1.02 (0.61, 1.71) | 0.99 (0.74, 1.33) |
| Lower Age at First Birth PGS | 0.93 (0.75, 1.14) | 0.98 (0.82, 1.16) | 1.02 (0.67, 1.54) | 1.01 (0.76, 1.35) |
| Lower Age at Menarche PGS | | 1.11 (0.95, 1.31) | | 0.97 (0.75, 1.25) |
| Lower Age at Menopause PGS | — | 1.09 (0.94, 1.27) | — | 1.02 (0.79, 1.31) |
| Lower Education PGS | 0.93 (0.77, 1.13) | 0.97 (0.79, 1.20) | 0.96 (0.62, 1.48) | 0.91 (0.70, 1.18) |
| Lower General Cognition PGS | 0.96 (0.77, 1.18) | 1.01 (0.81, 1.25) | 0.96 (0.60, 1.54) | 1.07 (0.81, 1.42) |
| Lower Height PGS | 0.97 (0.79, 1.19) | 0.97 (0.79, 1.20) | 0.92 (0.61, 1.39) | 1.11 (0.79, 1.56) |
| Lower Longevity PGS | 1.07 (0.86, 1.34) | 1.04 (0.88, 1.23) | 1.15 (0.83, 1.61) | 1.01 (0.78, 1.30) |

**S5 Table. Hazard ratios (HRs) and 95% confidence intervals (CI) of each predictor for incident dementia obtained from Fine-Gray regression models stratified by race and gender.** Models use restricted analytic sample and classify dementia using the Hurd classification scheme.

| Characteristic | NH White | | NH Black | |
| --- | --- | --- | --- | --- |
| | Men | Women | Men | Women |
| *Sociodemographic* | | | | |
| Foreign Born | 1.06 (0.78, 1.45) | 1.03 (0.80, 1.34) | 0.76 (0.20, 2.95) | 0.76 (0.29, 2.00) |
| Southern Born | 1.12 (0.94, 1.33) | 1.04 (0.91, 1.20) | 1.37 (0.68, 2.75) | 1.19 (0.75, 1.90) |
| Veteran | 0.97 (0.81, 1.15) | 0.79 (0.39, 1.59) | 0.88 (0.55, 1.41) | 0.00 (0.00, 0.00) |
| Childless | 0.95 (0.63, 1.45) | 0.81 (0.49, 1.34) | 1.40 (0.40, 4.92) | 0.71 (0.54, 0.93) |
| Higher Parity | 0.98 (0.83, 1.16) | 0.97 (0.85, 1.11) | 1.17 (0.93, 1.47) | 1.15 (0.95, 1.40) |
| Lower Age at First Birth | 1.01 (0.87, 1.18) | 1.15 (1.00, 1.32) | 1.40 (0.89, 2.18) | 0.95 (0.75, 1.22) |
| Higher Age at Last Birth | 1.03 (0.88, 1.20) | 1.14 (1.00, 1.31) | 1.00 (0.68, 1.47) | 0.85 (0.67, 1.09) |
| *Early-Life* | | | | |
| Lower Mother's Education | 1.23 (1.02, 1.48) | 1.06 (0.88, 1.28) | 1.07 (0.67, 1.72) | 1.16 (0.81, 1.67) |
| Lower Father's Education | 1.19 (1.00, 1.42) | 1.01 (0.84, 1.22) | 1.06 (0.57, 1.98) | 1.19 (0.88, 1.62) |
| Lower Father's Occupational Status | 1.16 (0.99, 1.36) | 1.10 (0.96, 1.26) | 1.60 (0.56, 4.54) | 0.90 (0.45, 1.83) |
| Lower SR Childhood Health | 1.10 (0.94, 1.29) | 1.07 (0.95, 1.21) | 1.14 (0.74, 1.74) | 1.08 (0.82, 1.43) |
| Lower SR Childhood SES | 1.06 (0.92, 1.22) | 1.11 (0.98, 1.26) | 0.91 (0.55, 1.51) | 1.18 (0.88, 1.60) |
| Lower Education | 1.29 (1.12, 1.47) | 1.10 (0.97, 1.25) | 1.22 (0.80, 1.86) | 1.27 (0.97, 1.67) |
| *Economic* | | | | |
| Food Insecurity | 0.76 (0.40, 1.45) | 0.82 (0.57, 1.16) | 1.06 (0.47, 2.37) | 0.80 (0.53, 1.21) |
| Lower Income | 1.14 (0.96, 1.36) | 1.11 (0.98, 1.27) | 1.12 (0.85, 1.46) | 1.03 (0.83, 1.28) |
| Lower Neighborhood Safety | 1.27 (0.88, 1.83) | 1.04 (0.74, 1.45) | 0.72 (0.40, 1.27) | 0.98 (0.69, 1.39) |
| Lower Wealth | 1.21 (0.95, 1.54) | 1.04 (0.89, 1.22) | 1.05 (0.77, 1.42) | 1.13 (0.98, 1.30) |
| Medicaid | 1.72 (0.95, 3.11) | 1.07 (0.73, 1.59) | 1.54 (0.63, 3.72) | 1.23 (0.88, 1.73) |
| Medicare | 0.97 (0.76, 1.23) | 0.94 (0.74, 1.18) | 1.03 (0.60, 1.80) | 0.99 (0.66, 1.50) |
| No Insurance | 0.95 (0.82, 1.09) | 1.00 (0.88, 1.13) | 0.69 (0.43, 1.11) | 0.78 (0.54, 1.12) |
| Received Food Stamps | 2.62 (1.46, 4.68) | 0.92 (0.53, 1.59) | 0.91 (0.29, 2.84) | 1.28 (0.80, 2.06) |
| Retired | 0.95 (0.77, 1.17) | 0.92 (0.81, 1.04) | 0.90 (0.49, 1.67) | 0.96 (0.70, 1.33) |
| Unemployed | 0.50 (0.11, 2.25) | 1.07 (0.39, 2.93) | 0.00 (0.00, 0.00) | 1.70 (0.85, 3.40) |
| *Health* | | | | |
| Arthritis | 1.00 (0.86, 1.15) | 1.05 (0.93, 1.19) | 0.98 (0.62, 1.56) | 0.95 (0.71, 1.27) |
| Back Pain | 1.07 (0.91, 1.26) | 0.96 (0.84, 1.09) | 1.24 (0.75, 2.07) | 0.94 (0.68, 1.30) |
| Cancer | 0.95 (0.78, 1.15) | 0.87 (0.72, 1.05) | 0.84 (0.43, 1.64) | 0.96 (0.55, 1.67) |
| Diabetes | 1.11 (0.91, 1.35) | 1.04 (0.84, 1.29) | 1.32 (0.85, 2.03) | 1.05 (0.75, 1.47) |
| Dizziness | 1.26 (1.01, 1.59) | 1.02 (0.84, 1.24) | 1.13 (0.54, 2.38) | 1.43 (0.91, 2.22) |
| Fatigue | 1.23 (0.95, 1.58) | 1.08 (0.92, 1.27) | 1.06 (0.47, 2.41) | 1.30 (0.88, 1.92) |
| Headaches | 1.14 (0.78, 1.66) | 1.30 (1.03, 1.63) | 1.24 (0.68, 2.26) | 1.62 (0.99, 2.67) |
| Heart Problems | 1.02 (0.87, 1.20) | 1.02 (0.87, 1.21) | 0.61 (0.30, 1.25) | 1.15 (0.74, 1.78) |
| Hypertension | 0.94 (0.81, 1.09) | 1.05 (0.93, 1.18) | 0.81 (0.51, 1.29) | 1.28 (0.93, 1.76) |
| Lower SR Health | 1.21 (1.02, 1.43) | 1.12 (0.97, 1.29) | 0.99 (0.62, 1.58) | 1.32 (0.93, 1.87) |
| Lower SR Hearing | 1.16 (0.99, 1.36) | 1.04 (0.91, 1.17) | 1.07 (0.65, 1.79) | 0.99 (0.69, 1.43) |
| Lower SR Vision | 1.12 (0.97, 1.30) | 1.07 (0.95, 1.21) | 1.24 (0.71, 2.17) | 0.96 (0.72, 1.28) |
| Lung Disease | 0.82 (0.58, 1.15) | 0.96 (0.73, 1.27) | 1.27 (0.44, 3.67) | 1.35 (0.71, 2.56) |
| Pain | 1.17 (0.98, 1.40) | 1.05 (0.92, 1.20) | 1.13 (0.64, 1.98) | 1.17 (0.84, 1.63) |
| Psychiatric Illness | 1.32 (0.91, 1.90) | 1.19 (0.97, 1.47) | 0.59 (0.12, 2.86) | 1.09 (0.62, 1.92) |
| Short of Breath | 1.18 (0.95, 1.47) | 1.02 (0.85, 1.22) | 1.38 (0.78, 2.46) | 1.38 (0.94, 2.01) |
| Stroke | 1.15 (0.87, 1.52) | 1.09 (0.86, 1.38) | 1.17 (0.52, 2.63) | 1.09 (0.65, 1.84) |
| Wheezing | 1.11 (0.90, 1.38) | 1.04 (0.85, 1.28) | 0.77 (0.33, 1.81) | 1.18 (0.77, 1.80) |
| *Behaviors* | | | | |
| Active Smoker | 1.02 (0.76, 1.36) | 0.99 (0.79, 1.25) | 1.06 (0.59, 1.90) | 1.05 (0.62, 1.78) |
| Ever Smoked | 1.05 (0.89, 1.23) | 1.03 (0.91, 1.16) | 1.13 (0.64, 2.00) | 1.02 (0.75, 1.40) |
| Heavy Alcohol Use | 0.98 (0.76, 1.27) | 1.15 (0.80, 1.64) | 0.88 (0.37, 2.13) | 1.03 (0.35, 3.05) |
| Higher BMI | 1.09 (0.89, 1.34) | 1.01 (0.88, 1.16) | 0.95 (0.53, 1.68) | 1.03 (0.80, 1.33) |
| Low/No Vigorous Physical Activity | 1.09 (0.94, 1.26) | 1.03 (0.91, 1.17) | 0.83 (0.52, 1.32) | 0.96 (0.70, 1.31) |
| *Social Connections* | | | | |
| Ever Divorced | 0.98 (0.79, 1.21) | 1.11 (0.94, 1.32) | 0.83 (0.48, 1.45) | 0.94 (0.61, 1.44) |
| Ever Widowed | 0.95 (0.77, 1.17) | 0.98 (0.87, 1.11) | 0.96 (0.59, 1.57) | 1.06 (0.78, 1.44) |
| Less Religious | 0.94 (0.82, 1.07) | 0.94 (0.81, 1.08) | 0.69 (0.32, 1.47) | 0.58 (0.26, 1.30) |
| Lonely | 1.06 (0.84, 1.34) | 1.13 (0.98, 1.31) | 1.61 (1.04, 2.50) | 1.43 (1.03, 1.98) |
| No Friends Nearby | 1.04 (0.87, 1.24) | 0.97 (0.84, 1.12) | 0.99 (0.59, 1.64) | 1.23 (0.87, 1.76) |
| No Relatives Nearby | 1.08 (0.92, 1.27) | 1.05 (0.93, 1.19) | 0.93 (0.53, 1.62) | 1.18 (0.87, 1.59) |
| Not Married/Partnered | 1.03 (0.83, 1.28) | 1.00 (0.89, 1.12) | 1.04 (0.64, 1.69) | 1.02 (0.75, 1.39) |

| *Genetic* | | | | |
|---|---|---|---|---|
| Higher AD PGS | 1.02 (0.89, 1.18) | 1.04 (0.93, 1.16) | 0.74 (0.33, 1.69) | 1.41 (0.74, 2.68) |
| Higher Coronary Artery Disease PGS | 0.99 (0.86, 1.14) | 0.99 (0.87, 1.12) | 0.94 (0.65, 1.35) | 1.10 (0.86, 1.42) |
| Higher Diabetes PGS | 1.03 (0.90, 1.19) | 1.04 (0.93, 1.16) | 0.83 (0.47, 1.48) | 0.90 (0.58, 1.39) |
| Higher Myocardial Infarction PGS | 0.96 (0.84, 1.10) | 1.00 (0.88, 1.14) | 0.79 (0.48, 1.28) | 0.86 (0.62, 1.18) |
| Higher Parity PGS | 0.94 (0.82, 1.07) | 1.06 (0.94, 1.20) | 0.85 (0.58, 1.24) | 0.80 (0.59, 1.09) |
| Lower Age at First Birth PGS | 1.00 (0.86, 1.16) | 1.02 (0.90, 1.16) | 1.09 (0.69, 1.72) | 1.13 (0.84, 1.52) |
| Lower Age at Menarche PGS | — | 1.05 (0.94, 1.18) | — | 0.95 (0.70, 1.28) |
| Lower Age at Menopause PGS | — | 0.99 (0.88, 1.11) | — | 1.01 (0.76, 1.34) |
| Lower Education PGS | 0.98 (0.85, 1.13) | 1.02 (0.91, 1.15) | 0.79 (0.51, 1.21) | 0.96 (0.74, 1.23) |
| Lower General Cognition PGS | 0.97 (0.83, 1.12) | 1.08 (0.95, 1.21) | 0.87 (0.58, 1.31) | 0.99 (0.73, 1.34) |
| Lower Height PGS | 0.96 (0.83, 1.11) | 1.04 (0.93, 1.16) | 0.79 (0.48, 1.28) | 1.13 (0.78, 1.63) |
| Lower Longevity PGS | 1.02 (0.88, 1.19) | 1.00 (0.89, 1.12) | 1.11 (0.82, 1.51) | 0.91 (0.70, 1.17) |

**S6 Table. Hazard ratios (HRs) and 95% confidence intervals (CI) of each predictor for incident dementia obtained from Fine-Gray regression models stratified by race and gender.** Models use restricted analytic sample and classify dementia using the Expert classification scheme.

| Characteristic | NH White | | NH Black | |
|---|---|---|---|---|
| | Men | Women | Men | Women |
| *Sociodemographic* | | | | |
| Foreign Born | 1.07 (0.79, 1.45) | 1.09 (0.85, 1.39) | 0.86 (0.26, 2.85) | 1.02 (0.68, 1.52) |
| Southern Born | 1.10 (0.94, 1.29) | 1.02 (0.90, 1.16) | 1.37 (0.71, 2.63) | 1.04 (0.78, 1.38) |
| Veteran | 0.88 (0.75, 1.04) | 0.87 (0.49, 1.53) | 0.99 (0.63, 1.54) | 0.94 (0.35, 2.53) |
| Childless | 0.78 (0.51, 1.19) | 0.82 (0.51, 1.31) | 1.93 (0.80, 4.61) | 0.29 (0.04, 2.00) |
| Higher Parity | 0.96 (0.81, 1.15) | 1.04 (0.93, 1.17) | 1.08 (0.85, 1.37) | 1.19 (1.01, 1.40) |
| Lower Age at First Birth | 1.00 (0.86, 1.16) | 1.17 (1.03, 1.33) | 1.37 (0.85, 2.20) | 1.08 (0.87, 1.35) |
| Higher Age at Last Birth | 0.99 (0.86, 1.14) | 1.08 (0.95, 1.22) | 1.05 (0.74, 1.50) | 0.93 (0.76, 1.14) |
| *Early-Life* | | | | |
| Lower Mother's Education | 1.31 (1.08, 1.59) | 1.17 (0.97, 1.41) | 1.05 (0.64, 1.73) | 1.13 (0.83, 1.54) |
| Lower Father's Education | 1.23 (1.02, 1.48) | 1.15 (0.95, 1.38) | 1.05 (0.61, 1.82) | 1.07 (0.82, 1.41) |
| Lower Father's Occupational Status | 1.20 (1.03, 1.39) | 1.17 (1.03, 1.32) | 1.34 (0.51, 3.53) | 0.96 (0.52, 1.77) |
| Lower SR Childhood Health | 1.09 (0.95, 1.26) | 1.08 (0.96, 1.21) | 1.12 (0.74, 1.68) | 1.10 (0.87, 1.40) |
| Lower SR Childhood SES | 1.07 (0.94, 1.22) | 1.13 (1.00, 1.27) | 1.06 (0.64, 1.76) | 1.03 (0.83, 1.28) |
| Lower Education | 1.35 (1.19, 1.53) | 1.17 (1.03, 1.32) | 1.03 (0.71, 1.50) | 1.08 (0.86, 1.35) |
| *Economic* | | | | |
| Food Insecurity | 0.93 (0.49, 1.78) | 0.85 (0.62, 1.17) | 1.02 (0.48, 2.17) | 0.75 (0.55, 1.00) |
| Lower Income | 1.17 (1.01, 1.35) | 1.15 (1.01, 1.31) | 1.09 (0.83, 1.42) | 1.01 (0.83, 1.23) |
| Lower Neighborhood Safety | 1.27 (0.92, 1.75) | 1.17 (0.85, 1.61) | 0.73 (0.41, 1.27) | 1.04 (0.78, 1.40) |
| Lower Wealth | 1.16 (0.92, 1.46) | 1.05 (0.90, 1.22) | 1.13 (0.87, 1.46) | 1.07 (0.94, 1.22) |
| Medicaid | 1.42 (0.77, 2.61) | 1.15 (0.80, 1.65) | 1.57 (0.71, 3.47) | 1.17 (0.86, 1.59) |
| Medicare | 0.97 (0.78, 1.20) | 0.94 (0.76, 1.16) | 0.86 (0.53, 1.42) | 1.00 (0.70, 1.44) |
| No Insurance | 0.94 (0.82, 1.07) | 0.99 (0.89, 1.11) | 0.71 (0.45, 1.13) | 0.83 (0.64, 1.08) |
| Received Food Stamps | 2.04 (1.19, 3.50) | 0.98 (0.60, 1.60) | 0.95 (0.33, 2.77) | 1.54 (1.11, 2.14) |
| Retired | 0.96 (0.79, 1.16) | 0.93 (0.83, 1.04) | 0.86 (0.49, 1.52) | 0.91 (0.69, 1.20) |
| Unemployed | 0.63 (0.24, 1.68) | 1.23 (0.52, 2.91) | 0.00 (0.00, 0.00) | 1.35 (0.73, 2.52) |
| *Health* | | | | |
| Arthritis | 1.00 (0.88, 1.15) | 1.04 (0.92, 1.16) | 1.04 (0.66, 1.63) | 1.06 (0.81, 1.37) |
| Back Pain | 1.07 (0.92, 1.25) | 0.97 (0.85, 1.09) | 1.37 (0.85, 2.21) | 1.04 (0.80, 1.34) |
| Cancer | 0.91 (0.76, 1.10) | 0.93 (0.79, 1.10) | 0.81 (0.42, 1.55) | 0.93 (0.59, 1.49) |
| Diabetes | 1.24 (1.03, 1.50) | 1.19 (0.98, 1.44) | 1.28 (0.85, 1.93) | 1.21 (0.86, 1.71) |
| Dizziness | 1.24 (1.01, 1.52) | 1.03 (0.85, 1.25) | 1.03 (0.55, 1.92) | 1.42 (0.98, 2.07) |
| Fatigue | 1.27 (0.98, 1.65) | 1.06 (0.91, 1.23) | 0.94 (0.42, 2.09) | 1.26 (0.90, 1.75) |
| Headaches | 1.08 (0.75, 1.57) | 1.26 (1.02, 1.56) | 1.30 (0.79, 2.14) | 1.34 (0.86, 2.10) |
| Heart Problems | 1.03 (0.88, 1.21) | 1.01 (0.87, 1.17) | 0.56 (0.26, 1.20) | 1.23 (0.93, 1.62) |
| Hypertension | 0.98 (0.85, 1.12) | 1.02 (0.92, 1.15) | 0.87 (0.55, 1.37) | 1.25 (0.96, 1.63) |
| Lower SR Health | 1.23 (1.05, 1.44) | 1.15 (1.02, 1.30) | 0.96 (0.63, 1.47) | 1.42 (1.08, 1.86) |
| Lower SR Hearing | 1.18 (1.01, 1.38) | 1.07 (0.95, 1.20) | 0.96 (0.61, 1.51) | 0.92 (0.67, 1.25) |
| Lower SR Vision | 1.15 (0.99, 1.33) | 1.10 (0.99, 1.23) | 1.23 (0.73, 2.07) | 1.03 (0.78, 1.35) |
| Lung Disease | 0.92 (0.69, 1.23) | 1.04 (0.81, 1.34) | 1.43 (0.61, 3.36) | 1.40 (0.77, 2.55) |
| Pain | 1.14 (0.95, 1.37) | 1.05 (0.93, 1.19) | 1.17 (0.68, 2.01) | 1.16 (0.87, 1.53) |
| Psychiatric Illness | 1.32 (0.95, 1.83) | 1.22 (0.99, 1.51) | 0.73 (0.18, 2.93) | 1.42 (0.88, 2.30) |
| Short of Breath | 1.12 (0.90, 1.40) | 0.98 (0.83, 1.17) | 1.41 (0.88, 2.26) | 1.33 (0.97, 1.83) |
| Stroke | 1.16 (0.89, 1.52) | 1.09 (0.88, 1.35) | 1.05 (0.50, 2.20) | 1.24 (0.85, 1.81) |
| Wheezing | 1.12 (0.91, 1.39) | 1.00 (0.82, 1.23) | 0.76 (0.36, 1.61) | 1.04 (0.70, 1.55) |
| *Behaviors* | | | | |
| Active Smoker | 1.13 (0.88, 1.45) | 1.02 (0.83, 1.27) | 1.06 (0.60, 1.86) | 1.05 (0.71, 1.57) |
| Ever Smoked | 1.11 (0.95, 1.29) | 1.01 (0.90, 1.13) | 1.27 (0.71, 2.28) | 1.02 (0.79, 1.31) |
| Heavy Alcohol Use | 0.98 (0.78, 1.24) | 1.09 (0.76, 1.56) | 1.01 (0.46, 2.25) | 0.81 (0.31, 2.12) |
| Higher BMI | 1.14 (0.95, 1.37) | 1.06 (0.94, 1.19) | 0.96 (0.58, 1.59) | 1.05 (0.87, 1.27) |
| Low/No Vigorous Physical Activity | 1.10 (0.96, 1.26) | 1.02 (0.91, 1.15) | 0.94 (0.60, 1.48) | 0.90 (0.71, 1.14) |
| *Social Connections* | | | | |
| Ever Divorced | 1.04 (0.86, 1.27) | 1.12 (0.95, 1.32) | 1.05 (0.64, 1.73) | 1.26 (0.98, 1.62) |
| Ever Widowed | 1.01 (0.84, 1.23) | 0.97 (0.87, 1.08) | 0.93 (0.58, 1.48) | 0.94 (0.73, 1.22) |
| Less Religious | 0.95 (0.84, 1.07) | 0.95 (0.84, 1.08) | 0.85 (0.42, 1.70) | 0.83 (0.54, 1.27) |
| Lonely | 1.11 (0.86, 1.42) | 1.14 (0.99, 1.31) | 1.65 (1.14, 2.38) | 1.33 (0.97, 1.81) |
| No Friends Nearby | 1.00 (0.85, 1.19) | 0.94 (0.83, 1.07) | 1.12 (0.68, 1.86) | 1.07 (0.80, 1.44) |
| No Relatives Nearby | 1.12 (0.96, 1.30) | 1.01 (0.90, 1.14) | 0.95 (0.57, 1.60) | 1.14 (0.88, 1.47) |
| Not Married/Partnered | 0.92 (0.75, 1.13) | 1.01 (0.90, 1.13) | 0.93 (0.59, 1.47) | 0.97 (0.73, 1.29) |

| *Genetic* | | | | |
|---|---|---|---|---|
| Higher AD PGS | 1.04 (0.91, 1.18) | 1.02 (0.92, 1.14) | 1.01 (0.44, 2.33) | 1.45 (0.83, 2.51) |
| Higher Coronary Artery Disease PGS | 0.99 (0.88, 1.13) | 0.99 (0.88, 1.11) | 0.86 (0.58, 1.25) | 1.12 (0.89, 1.41) |
| Higher Diabetes PGS | 1.04 (0.91, 1.19) | 1.03 (0.92, 1.14) | 0.83 (0.45, 1.52) | 0.84 (0.59, 1.19) |
| Higher Myocardial Infarction PGS | 1.01 (0.88, 1.15) | 0.99 (0.88, 1.11) | 0.80 (0.49, 1.29) | 0.88 (0.69, 1.14) |
| Higher Parity PGS | 1.03 (0.90, 1.18) | 1.02 (0.91, 1.15) | 0.85 (0.57, 1.27) | 0.95 (0.76, 1.21) |
| Lower Age at First Birth PGS | 0.99 (0.86, 1.14) | 1.00 (0.89, 1.12) | 1.10 (0.70, 1.74) | 1.05 (0.83, 1.34) |
| Lower Age at Menarche PGS | — | 1.00 (0.90, 1.11) | — | 0.96 (0.74, 1.24) |
| Lower Age at Menopause PGS | — | 1.01 (0.91, 1.13) | — | 0.99 (0.78, 1.25) |
| Lower Education PGS | 1.03 (0.90, 1.18) | 0.99 (0.89, 1.10) | 0.79 (0.52, 1.20) | 0.96 (0.76, 1.21) |
| Lower General Cognition PGS | 0.96 (0.84, 1.10) | 1.05 (0.94, 1.17) | 0.80 (0.54, 1.19) | 0.97 (0.77, 1.23) |
| Lower Height PGS | 0.98 (0.86, 1.12) | 1.05 (0.94, 1.17) | 0.81 (0.51, 1.30) | 1.11 (0.82, 1.50) |
| Lower Longevity PGS | 0.97 (0.84, 1.13) | 1.01 (0.91, 1.13) | 1.07 (0.80, 1.43) | 0.92 (0.74, 1.16) |

**S7 Table. Hazard ratios (HRs) and 95% confidence intervals (CI) of each predictor for incident dementia obtained from Fine-Gray regression models stratified by race and gender.** Models use restricted analytic sample and classify dementia using the LASSO classification scheme.

| | NH White | | NH Black | |
|---|---|---|---|---|
| Characteristic | Men | Women | Men | Women |
| *Sociodemographic* | | | | |
| Foreign Born | 1.09 (0.81, 1.47) | 0.96 (0.73, 1.27) | 1.04 (0.33, 3.23) | 1.04 (0.71, 1.52) |
| Southern Born | 1.17 (0.99, 1.38) | 1.09 (0.96, 1.25) | 1.23 (0.64, 2.38) | 1.09 (0.80, 1.47) |
| Veteran | 0.86 (0.73, 1.02) | 0.98 (0.57, 1.68) | 1.01 (0.65, 1.58) | 0.00 (0.00, 0.00) |
| Childless | 0.98 (0.67, 1.43) | 0.89 (0.56, 1.41) | 1.07 (0.25, 4.49) | 0.73 (0.47, 1.16) |
| Higher Parity | 0.98 (0.84, 1.15) | 0.98 (0.86, 1.12) | 1.21 (0.98, 1.49) | 1.12 (0.93, 1.35) |
| Lower Age at First Birth | 0.98 (0.85, 1.13) | 1.14 (0.99, 1.31) | 1.42 (0.91, 2.22) | 1.07 (0.85, 1.36) |
| Higher Age at Last Birth | 0.98 (0.85, 1.13) | 1.08 (0.95, 1.23) | 0.97 (0.68, 1.39) | 0.98 (0.80, 1.21) |
| *Early-Life* | | | | |
| Lower Mother's Education | 1.26 (1.05, 1.51) | 1.11 (0.92, 1.33) | 1.11 (0.66, 1.87) | 1.07 (0.79, 1.45) |
| Lower Father's Education | 1.16 (0.96, 1.41) | 1.12 (0.93, 1.35) | 1.17 (0.62, 2.20) | 0.97 (0.72, 1.30) |
| Lower Father's Occupational Status | 1.14 (0.98, 1.33) | 1.17 (1.02, 1.34) | 1.15 (0.61, 2.18) | 0.91 (0.57, 1.45) |
| Lower SR Childhood Health | 1.13 (0.98, 1.30) | 1.09 (0.97, 1.22) | 1.17 (0.77, 1.77) | 1.05 (0.80, 1.37) |
| Lower SR Childhood SES | 1.05 (0.92, 1.21) | 1.15 (1.02, 1.30) | 0.87 (0.51, 1.49) | 1.13 (0.89, 1.44) |
| Lower Education | 1.39 (1.21, 1.59) | 1.17 (1.03, 1.32) | 1.16 (0.80, 1.70) | 1.12 (0.88, 1.41) |
| *Economic* | | | | |
| Food Insecurity | 0.77 (0.42, 1.43) | 0.84 (0.59, 1.18) | 1.07 (0.49, 2.37) | 0.81 (0.58, 1.14) |
| Lower Income | 1.11 (0.93, 1.31) | 1.16 (1.01, 1.32) | 1.08 (0.83, 1.39) | 1.20 (1.03, 1.39) |
| Lower Neighborhood Safety | 1.32 (0.93, 1.86) | 1.05 (0.78, 1.41) | 0.66 (0.39, 1.13) | 1.18 (0.88, 1.59) |
| Lower Wealth | 1.16 (0.92, 1.47) | 1.06 (0.92, 1.22) | 1.07 (0.85, 1.35) | 1.10 (0.96, 1.26) |
| Medicaid | 1.66 (0.95, 2.90) | 1.06 (0.75, 1.51) | 1.43 (0.65, 3.17) | 1.28 (0.92, 1.78) |
| Medicare | 0.99 (0.79, 1.25) | 0.87 (0.71, 1.07) | 0.95 (0.56, 1.61) | 0.98 (0.67, 1.42) |
| No Insurance | 0.98 (0.85, 1.13) | 0.99 (0.88, 1.12) | 0.92 (0.59, 1.44) | 0.80 (0.61, 1.05) |
| Received Food Stamps | 2.23 (1.28, 3.87) | 0.87 (0.50, 1.52) | 0.90 (0.31, 2.64) | 1.48 (1.02, 2.16) |
| Retired | 0.92 (0.75, 1.11) | 0.99 (0.87, 1.12) | 0.85 (0.48, 1.49) | 0.92 (0.70, 1.21) |
| Unemployed | 0.48 (0.11, 2.06) | 0.93 (0.34, 2.55) | 0.00 (0.00, 0.00) | 1.30 (0.66, 2.54) |
| *Health* | | | | |
| Arthritis | 0.98 (0.85, 1.13) | 1.01 (0.90, 1.14) | 1.02 (0.66, 1.58) | 1.04 (0.78, 1.38) |
| Back Pain | 1.05 (0.90, 1.24) | 0.96 (0.85, 1.09) | 1.36 (0.85, 2.17) | 0.91 (0.69, 1.21) |
| Cancer | 0.90 (0.74, 1.09) | 0.87 (0.73, 1.05) | 0.85 (0.42, 1.74) | 0.93 (0.58, 1.47) |
| Diabetes | 1.10 (0.91, 1.34) | 1.06 (0.86, 1.29) | 0.98 (0.62, 1.56) | 1.09 (0.77, 1.55) |
| Dizziness | 1.23 (0.97, 1.55) | 0.98 (0.82, 1.18) | 1.18 (0.70, 1.99) | 1.43 (1.00, 2.04) |
| Fatigue | 1.25 (0.96, 1.63) | 1.05 (0.90, 1.23) | 0.91 (0.41, 2.02) | 1.32 (0.96, 1.81) |
| Headaches | 1.09 (0.74, 1.61) | 1.24 (0.99, 1.55) | 1.13 (0.64, 2.00) | 1.38 (0.90, 2.12) |
| Heart Problems | 1.03 (0.87, 1.22) | 1.03 (0.88, 1.20) | 0.60 (0.31, 1.15) | 1.14 (0.83, 1.56) |
| Hypertension | 0.98 (0.85, 1.14) | 1.03 (0.92, 1.16) | 0.88 (0.56, 1.36) | 1.30 (0.98, 1.73) |
| Lower SR Health | 1.15 (0.97, 1.36) | 1.09 (0.95, 1.25) | 1.01 (0.66, 1.53) | 1.30 (0.96, 1.74) |
| Lower SR Hearing | 1.21 (1.03, 1.43) | 1.06 (0.94, 1.19) | 1.17 (0.73, 1.86) | 0.89 (0.65, 1.22) |
| Lower SR Vision | 1.11 (0.96, 1.28) | 1.08 (0.97, 1.21) | 1.25 (0.73, 2.15) | 1.03 (0.78, 1.37) |
| Lung Disease | 0.88 (0.64, 1.19) | 1.09 (0.83, 1.42) | 1.21 (0.48, 3.06) | 1.45 (0.78, 2.70) |
| Pain | 1.12 (0.93, 1.35) | 1.07 (0.94, 1.22) | 1.07 (0.61, 1.86) | 1.15 (0.84, 1.56) |
| Psychiatric Illness | 1.36 (0.96, 1.93) | 1.16 (0.94, 1.43) | 0.84 (0.25, 2.77) | 1.25 (0.72, 2.15) |
| Short of Breath | 1.13 (0.89, 1.42) | 1.01 (0.85, 1.20) | 1.25 (0.71, 2.17) | 1.17 (0.79, 1.73) |
| Stroke | 1.22 (0.93, 1.61) | 1.13 (0.90, 1.41) | 1.05 (0.49, 2.27) | 0.97 (0.52, 1.81) |
| Wheezing | 1.13 (0.90, 1.41) | 1.03 (0.83, 1.28) | 0.87 (0.43, 1.78) | 1.16 (0.77, 1.73) |
| *Behaviors* | | | | |
| Active Smoker | 1.15 (0.89, 1.50) | 1.06 (0.85, 1.32) | 0.96 (0.54, 1.69) | 1.15 (0.70, 1.88) |
| Ever Smoked | 1.05 (0.89, 1.23) | 1.02 (0.90, 1.14) | 1.14 (0.66, 1.99) | 1.06 (0.81, 1.39) |
| Heavy Alcohol Use | 0.93 (0.72, 1.20) | 1.13 (0.76, 1.69) | 0.96 (0.43, 2.13) | 0.85 (0.30, 2.39) |
| Higher BMI | 0.97 (0.79, 1.19) | 0.98 (0.85, 1.12) | 0.89 (0.53, 1.52) | 0.97 (0.78, 1.20) |
| Low/No Vigorous Physical Activity | 1.06 (0.92, 1.23) | 1.03 (0.91, 1.16) | 0.91 (0.58, 1.40) | 1.01 (0.77, 1.33) |
| *Social Connections* | | | | |
| Ever Divorced | 0.99 (0.80, 1.21) | 1.10 (0.93, 1.30) | 0.96 (0.58, 1.56) | 1.17 (0.87, 1.56) |
| Ever Widowed | 0.94 (0.76, 1.16) | 1.00 (0.89, 1.12) | 0.96 (0.57, 1.60) | 0.95 (0.72, 1.25) |
| Less Religious | 0.91 (0.79, 1.04) | 0.92 (0.81, 1.05) | 0.92 (0.51, 1.69) | 0.85 (0.53, 1.37) |
| Lonely | 1.03 (0.79, 1.33) | 1.14 (0.99, 1.32) | 1.57 (1.04, 2.38) | 1.23 (0.88, 1.72) |
| No Friends Nearby | 0.98 (0.82, 1.16) | 0.94 (0.82, 1.08) | 1.00 (0.62, 1.62) | 1.00 (0.73, 1.37) |
| No Relatives Nearby | 1.09 (0.93, 1.27) | 1.09 (0.96, 1.22) | 0.98 (0.58, 1.65) | 1.07 (0.81, 1.40) |
| Not Married/Partnered | 1.02 (0.81, 1.27) | 0.98 (0.87, 1.10) | 1.01 (0.64, 1.58) | 0.98 (0.72, 1.32) |

| *Genetic* | | | | |
|---|---|---|---|---|
| Higher AD PGS | 1.03 (0.89, 1.18) | 1.02 (0.91, 1.13) | 0.80 (0.38, 1.70) | 1.33 (0.75, 2.36) |
| Higher Coronary Artery Disease PGS | 0.96 (0.84, 1.11) | 1.01 (0.89, 1.14) | 0.72 (0.48, 1.07) | 1.09 (0.87, 1.36) |
| Higher Diabetes PGS | 1.02 (0.89, 1.18) | 1.02 (0.91, 1.14) | 0.70 (0.41, 1.20) | 0.89 (0.62, 1.27) |
| Higher Myocardial Infarction PGS | 0.94 (0.82, 1.08) | 1.03 (0.91, 1.17) | 0.72 (0.47, 1.11) | 0.83 (0.63, 1.09) |
| Higher Parity PGS | 0.94 (0.82, 1.08) | 1.01 (0.90, 1.14) | 0.79 (0.52, 1.20) | 0.89 (0.68, 1.16) |
| Lower Age at First Birth PGS | 1.01 (0.88, 1.17) | 1.00 (0.89, 1.13) | 1.11 (0.74, 1.66) | 1.01 (0.80, 1.28) |
| Lower Age at Menarche PGS | — | 1.01 (0.90, 1.13) | | 0.96 (0.73, 1.26) |
| Lower Age at Menopause PGS | — | 1.03 (0.92, 1.15) | — | 0.93 (0.72, 1.19) |
| Lower Education PGS | 0.98 (0.86, 1.13) | 1.02 (0.91, 1.14) | 0.81 (0.53, 1.25) | 0.89 (0.69, 1.15) |
| Lower General Cognition PGS | 0.95 (0.82, 1.10) | 1.05 (0.93, 1.18) | 0.80 (0.54, 1.20) | 1.02 (0.79, 1.30) |
| Lower Height PGS | 0.97 (0.84, 1.11) | 1.02 (0.91, 1.14) | 0.71 (0.43, 1.15) | 1.05 (0.77, 1.42) |
| Lower Longevity PGS | 0.98 (0.84, 1.14) | 1.00 (0.89, 1.12) | 1.11 (0.83, 1.49) | 1.00 (0.78, 1.28) |

**S1 File. Variable Construction.**

Sociodemographic characteristics.

Respondents were classified as *Childless* (1) if they reported never having children compared to reporting at least one child (-1).

Respondents were classified as *Foreign Born* if they reported being born outside of the United States (1) compared to being born in the US (-1).

Among respondents who reported having at least one child, *Higher Age at Last Birth* was a standardized measure (with mean zero, standard deviation one) of the respondent's age when they last gave birth. Among respondents who were classified as childless, H*igher Age at Last Birth* was set to zero.

*Higher Parity* was defined as the number of children the respondent reported ever having. Values were reverse coded and standardized such that higher parity reflected higher risk.

Among respondents who reported having at least one child, *Lower Age at First Birth* was a standardized measure (with mean zero, standard deviation one) of the respondent's age when they first gave birth. Values were reverse coded such that younger ages reflected higher risk. Among respondents who were classified as childless, *Lower Age at First Birth* was set to zero.

Respondents were classified as *Southern Born* if they reported being born in the South Atlantic, Eastern South Central, or Western South Central Census Divisions of the United States (1) compared to being born outside of these divisions (-1).

Respondents who served in the military were classified as a *Veteran* (1) compared to not (-1).

Early-Life Characteristics.

Respondent's years of completed education were reverse coded and standardized with mean zero and standard deviation one to create the variable *Lower Education*. This same process was used to create *Lower Father's Education* and *Lower Mother's Education*.

*Lower Father Occupational Status* was scored by categorizing the respondent's father's occupation as (1) Executives and managers, (2) Professional specialty, (3) Sales and administration, (4) Protection services and armed forces, (5) Cleaning, building, food preparation, and personal services, and (6) Production, construction, and operation occupations. Scores were then standardized with mean zero and standard deviation one. Scores were then standardized with mean zero and standard deviation one.

Respondents self-reported their overall health from birth to age 16 as excellent, very good, good, fair, and poor. Responses were scored such that higher values corresponded to worse health and then standardized with mean zero and standard deviation one to generate the variable *Lower SR Childhood Health.*

Respondents self-reported their family's financial well-being from birth to age 16 as being pretty well off financially, about average, or poor. Responses were scored such that higher values corresponded to lower socioeconomic status (SES) and then standardized with mean zero and standard deviation one to generate the variable *Lower SR Childhood SES.*

Economic Characteristics

*Food Insecurity* was scored as binary (-1/1), with 1 indicating the respondent had reported not having enough money to buy the food they needed over the two year prior to their interview in 2000, and -1 indicating the respondent had enough money to buy the food they needed.

Income was measured in nominal dollars during the respondent's interview in 2000 and is the sum of all income in a household, including: respondent's and spouse's wage/salary income, bonuses/overtime pay/commissions/tips, 2nd job or military reserve earnings, professional practice or trade income; household business or farm income, self-employment earnings, business income, gross rent, dividend and interest income, trust funds or royalties, and other asset income; respondent's and spouse's income from all pensions and annuities; respondent's and spouse's total Social Security income (including that which is and is not received due to disability); respondent's and spouse's income from unemployment and worker's compensation; respondent's and spouse's income from veterans' benefits, welfare, and food stamps; and alimony, other income, and lump sums from insurance, pension, and inheritance at the household level. This value was then log-transformed, reverse coded, and standardized with mean zero and standard deviation one to generate the variable *Lower Income*.

Respondents reported the safety of their neighborhood. Response options were excellent, very good, good, fair, and poor. Responses were coded such that higher values corresponded to *Lower Neighborhood Safety* and then standardized with mean zero and standard deviation one.

Wealth was measured in nominal dollars during the respondent's interview in 2000 as the sum of all wealth components (except secondary home) less all debt. This value was then log-transformed, reverse coded, and standardized with mean zero and standard deviation one to generate the variable *Lower Wealth.*

*Medicaid* was coded as binary (-1/1), with 1 indicating the respondent reported being covered by Medicaid during their interview in 2000 and -1 indicating otherwise.

*Medicare* was coded as binary (-1/1), with 1 indicating the respondent reported being covered by Medicare during their interview in 2000 and -1 indicating otherwise.

*No Insurance* was coded as binary (-1/1), with 1 indicating the respondent reported being uninsured during their interview in 2000 and -1 indicating otherwise.

*Food Stamps* was scored as binary (-1/1), with 1 indicating the respondent or a family member living with them received government food stamps over the two year prior to their interview in 2000, and -1 indicating otherwise.

*Retired* was coded as binary (-1/1), with 1 indicating the respondent reported being retired during their interview in 2000 and -1 indicating otherwise.

*Unemployed* was coded as binary (-1/1), with 1 indicating the respondent reported was not working for pay while actively looking for a job in the last four weeks prior to their interview in 2000 and -1 indicating otherwise.

Behaviors

*Active Smoker* was scored as binary (-1/1), with 1 indicating the respondent reported being a current smoker and -1 otherwise.

*Ever Smoked* was scored as binary (-1/1), with 1 indicating the respondent reported ever smoking and -1 indicating otherwise.

*Heavy Alcohol Use* was scored as binary (-1/1), with 1 indicating the respondent reported drinking three or more drinks per day on days they drank and -1 otherwise.

Body mass index (BMI) was calculated by dividing the respondent's weight by their height-squared and then standardizing with mean zero and standard deviation one. Higher scores reflected *Higher BMI***.**

*Low/No Vigorous Activity* was scored as binary (-1/1), with -1 indicating the respondent reported completing vigorous activity (including, for example, sports, heavy housework, or a job that involves physical labor) three or more times per week over the 12 month period prior to their interview in 2000 and 1 indicating otherwise.

Health Characteristics

All binary health characteristics were coded as 1 if the respondent self-reported the condition or was otherwise coded as -1.

Respondents were asked whether a doctor ever told them that they had *Arthritis* (arthritis or rheumatism), *Cancer* (cancer or a malignant tumor, excluding minor skin cancers), *Diabetes* (diabetes or high blood sugar), *Heart Problems* (heart attack, coronary heart disease, angina, congestive heart failure, or other heart problems), *Hypertension* (high blood pressure or hypertension), *Lung Disease* (chronic lung disease such as chronic bronchitis or emphysema), *Psychiatric Illness* (any emotional, nervous, or psychiatric problems), *Stroke*.

Respondents were also asked to report whether they had persistent *Back Pain, Dizziness, Fatigue, Headaches*, were *Short of Breath, Wheezing*, and whether they were often troubled with *Pain* over the 12 month period prior to the interview in 2000.

Respondents self-reported their overall health (*SR Health*), hearing (*SR Hearing*), and vision (*SR Vision*). Response options were excellent, very good, good, fair, and poor. Responses were coded such that higher values corresponded to worse health/hearing/vision and then standardized with mean zero and standard deviation one.

Social Ties

*Ever Divorced* was coded as binary (-1/1), with 1 indicating the respondent reported being divorced or separated and -1 indicating the respondent never reported being divorced no separated.

*Ever Widowed* was coded as binary (-1/1), with 1 indicating the respondent reported being widowed and -1 indicating the respondent never reported being widowed.

Respondents reported whether religion was very important, somewhat important, or not too important in their lives. Responses were coded such that higher values corresponded to being *Less Religious* and then standardized with mean zero and standard deviation one.

Respondents reported whether they felt *Lonely* much of the week prior to their interview in 2000. Responses were scored (-1/1), with 1 indicating a response of "yes" and -1 indicating a response of "no."

Respondents reported whether they had any good friends or relatives living in their neighborhood. Responses were used to create binary indicators for *No Friends Nearby* (1 if "no"; -1 if "yes") and *No Relatives Nearby* (1 if "no"; -1 if "yes").

*Not Married/Partnered* was coded as binary (-1/1), with 1 indicating the respondent was not married and not partnered at the time of their interview in 2000 and -1 indicating otherwise.

Genetic

All polygenic scores (PGSs) were residualized by regressing the PGS on the first 10 single nucleotide polymorphisms principal components, computing the residuals from the predictions, and standardizing the residual values to a normal distribution with mean zero and standard deviation 1. Some PGSs were reverse coded as described below. Detailed information on sample selection, consent procedures, and assay processes are provided by the Health and Retirement study (HRS) investigators [1].

*Higher AD PGS* represents the Alzheimer's disease PGS (with apolipoprotein [ApoE] status variants, rs7412 and rs429358) [2].

*Higher Coronary Artery Disease PGS* represents the PGS for coronary artery disease which was created by HRS investigators using results from a 2011 study conducted by the Coronary Artery Disease Genome wide Replication and Meta-analysis Consortium [3].

*Higher Diabetes PGS* represents the type 2 diabetes PGS which was created by HRS investigators using GWAS meta-analysis results from a 2012 study conducted by the Diabetes Genetics Replication and Meta-analysis Consortium [4].

*Higher Myocardial Infarction PGS* represents the myocardial infarction PGS which was created by HRS investigators using 2015 results from a subgroup analysis of coronary artery disease conducted by the Coronary Artery Disease Genome wide Replication and Meta-analysis Consortium [5].

*Higher Parity PGS* represents a PGS for the number of children ever born created by HRS investigators from a 2016 study conducted by the Sociogenome Consortium [6]. Values were coded such that higher PGSs corresponded to higher genetic propensity for higher parity levels.

*Lower Age at First Birth PGS* represents a PGS for age at first birth created by HRS investigators from a 2016 study conducted by the Sociogenome Consortium [6]. Values were reverse coded such that higher PGSs corresponded to lower genetic propensity for younger age at first birth.

*Lower Age at Menarche PGS* represents a PGS for age at menarche created by HRS investigators from a 2014 study conducted by the Reproductive Genetics (ReproGen) Consortium [7]. Values were reverse coded such that higher PGSs corresponded to lower genetic propensity for younger age at menarche.

*Lower Age at Menopause PGS* represents a PGS for age at menopause created by HRS investigators from a 2014 study conducted by the Reproductive Genetics Consortium [8]. Values were reverse coded such that higher PGSs corresponded to lower genetic propensity for younger age at menopause.

*Lower Education PGS* represents a PGS for educational attainment which was created by HRS investigators using results from a 2018 study by the Social Science Genetic Association Consortium [9]. Values were reverse coded such that higher PGSs corresponded to lower genetic propensity for higher educational attainment

*Lower General Cognition PGS* represents a PGS for general cognition created by HRS investigators from the Cohorts for Heart and Aging Research in Genomic Epidemiology Consortium, 2015 [10]. Values were reverse coded such that higher PGSs corresponded to lower genetic propensity for higher cognitive functioning

*Lower Height PGS* represents a PGS for height created by HRS investigators from a 2014 study conducted by the Genetic Investigation of Anthropometric Traits Consortium [11]. Values were reverse coded such that higher PGSs corresponded to lower genetic propensity for taller height

*Lower Longevity PGS* represents a PGS for longevity created by HRS investigators from the Cohorts for Heart and Aging Research in Genomic Epidemiology Consortium, 2015 [12]. Values were reverse coded such that higher PGSs corresponded to lower genetic propensity for higher longevity

**S2 File. Statistical Methods.**

**Fine-Gray Estimator**

The Kaplan-Meier (KM) estimator is a non-parametric statistical technique that was introduced as a way to study time-to-event data with applications to mortality in the presence of attrition (e.g., study drop-out, loss to follow-up) [13]. In the setting for which it was developed (i.e., studies of mortality with incomplete follow-up is incomplete), the required assumption of uninformative censoring—or that subjects who are lost to follow-up (i.e., censored) have the same survival prospects as those who continue to be followed–is likely to be met within the KM framework. It is assumed that these censored subjects for whom we do not observe an event (i.e., death) remain at risk of the event.

However, when using the KM estimator to examine a health outcome, mortality becomes another way in which subjects may be censored. In this application, subjects will be censored of they die prior to developing the health outcome of interest yet they will be considered "at risk" in the same way a subject who was lost to follow-up would be. In this case, the KM estimator does not account for the competing risk of mortality and will yield biased estimates.

The Fine-Gray model was introduced to account for competing risks in these settings [14]. Whereas in the KM framework a subject who died prior to experiencing the health outcome of interest would be censored, the Fine-Gray classification scheme uses weights to represent the conditional probability of an event of interest for subjects who experienced the competing event, carrying these subjects forward in the analysis rather than censoring them.

**Random Forest Algorithm**

To complement our analyses using the Fine-Gray model, we used a data-driven classification scheme referred to as random forest competing risks survival analysis [15, 16]. Random forest is a non-parametric, ensemble machine-learning algorithm which iteratively bifurcates a dataset based on predictor variables over a number of permutations set by the user and then ranks the importance of each predictor based on its ability to "split" the data [17]. The basis for this algorithm is fitting multiple decision trees on the data and pooling them together which overcomes the limitations of singular classification and regression tree (CART) classification schemees. To do this, the algorithm repeatedly draws bootstrap samples from the analytic sample and a random selection of predictors to grow a predetermined number of decision trees (i.e., a forest) set by the user across which results are pooled. A training data set consisting of $n$ of N cases (approximately two-thirds of the original sample) is generated for each of $k$ decision trees and the remaining cases (one third of the original sample) are used as test data to estimate the out of bag (OOB) classification error. A random sample $m$ of M predictors is selected at each node and the one predictor that best discriminates discrepancies in the outcome is chosen for that particular split. As a result, the root node of each decision tree represents the strongest predictor and the splits that follow are based on the successively strongest predictors. A final classification is made using a majority of votes across all trees.

Fig S1

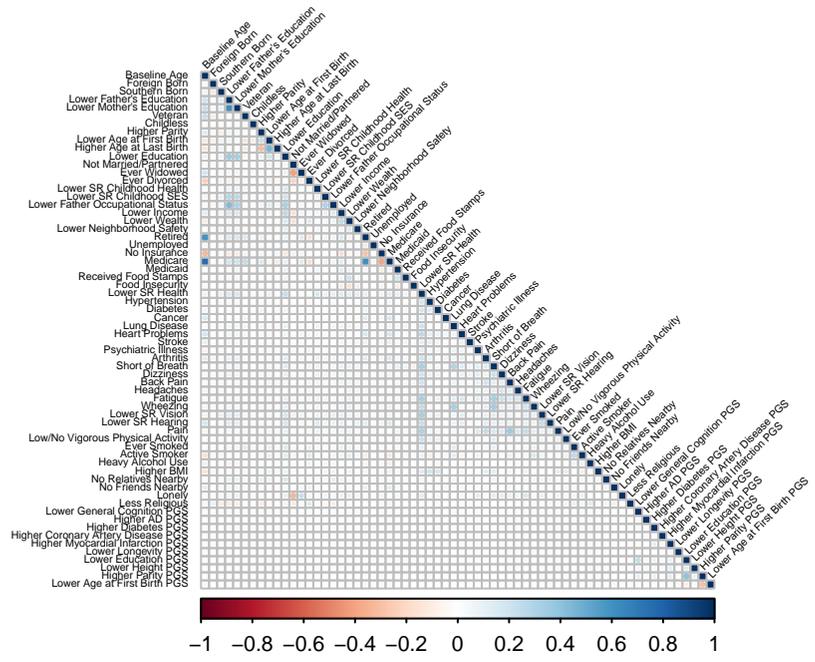

**NH White Men**

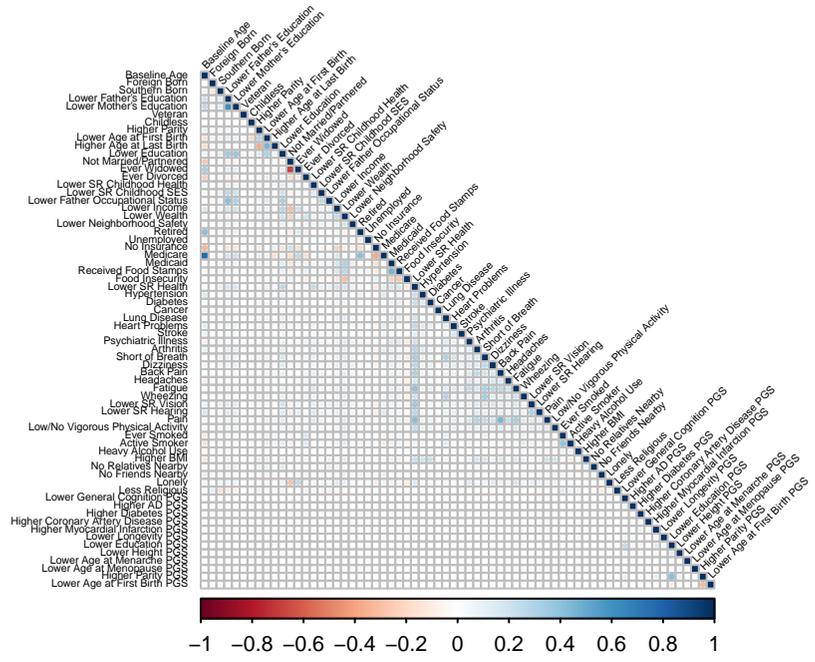

**NH White Women**

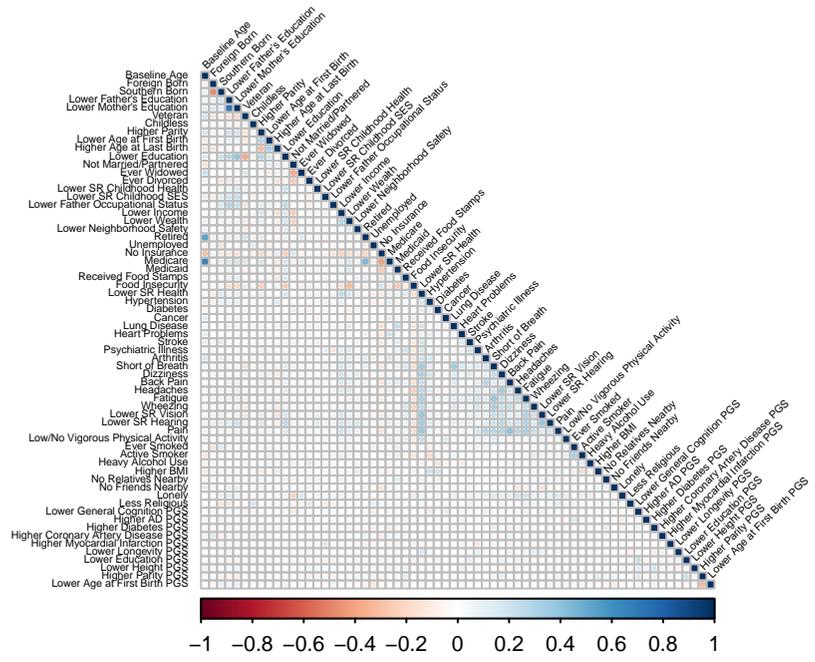

**NH Black Men**

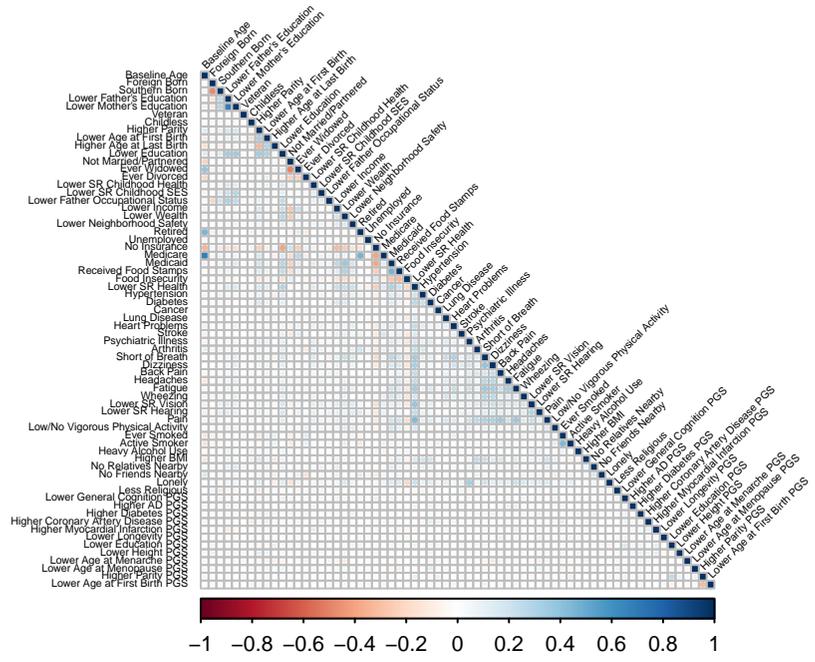

**NH Black Women**

Fig S2

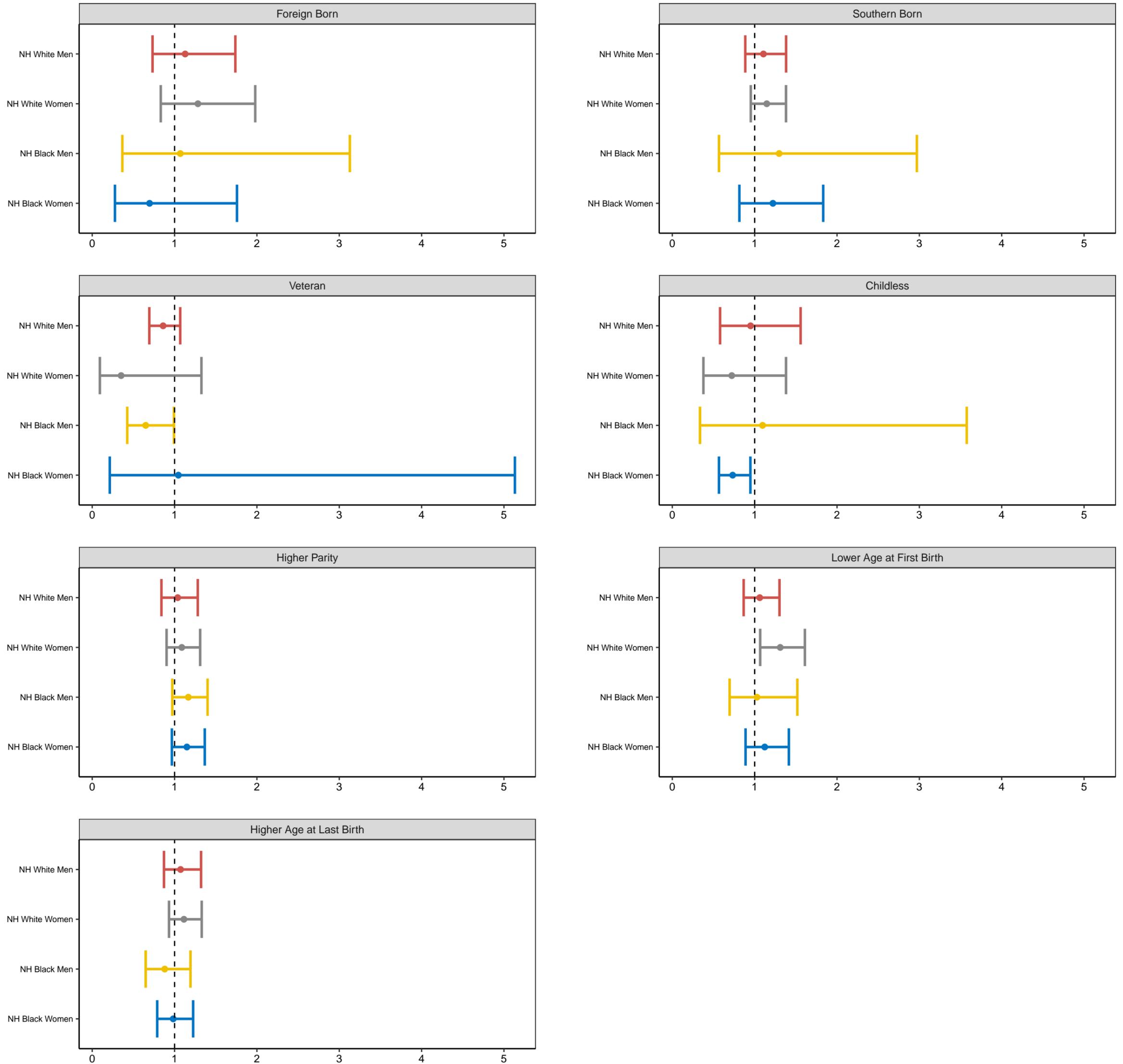

Sociodemographic

Hazard Ratio

## Early-Life

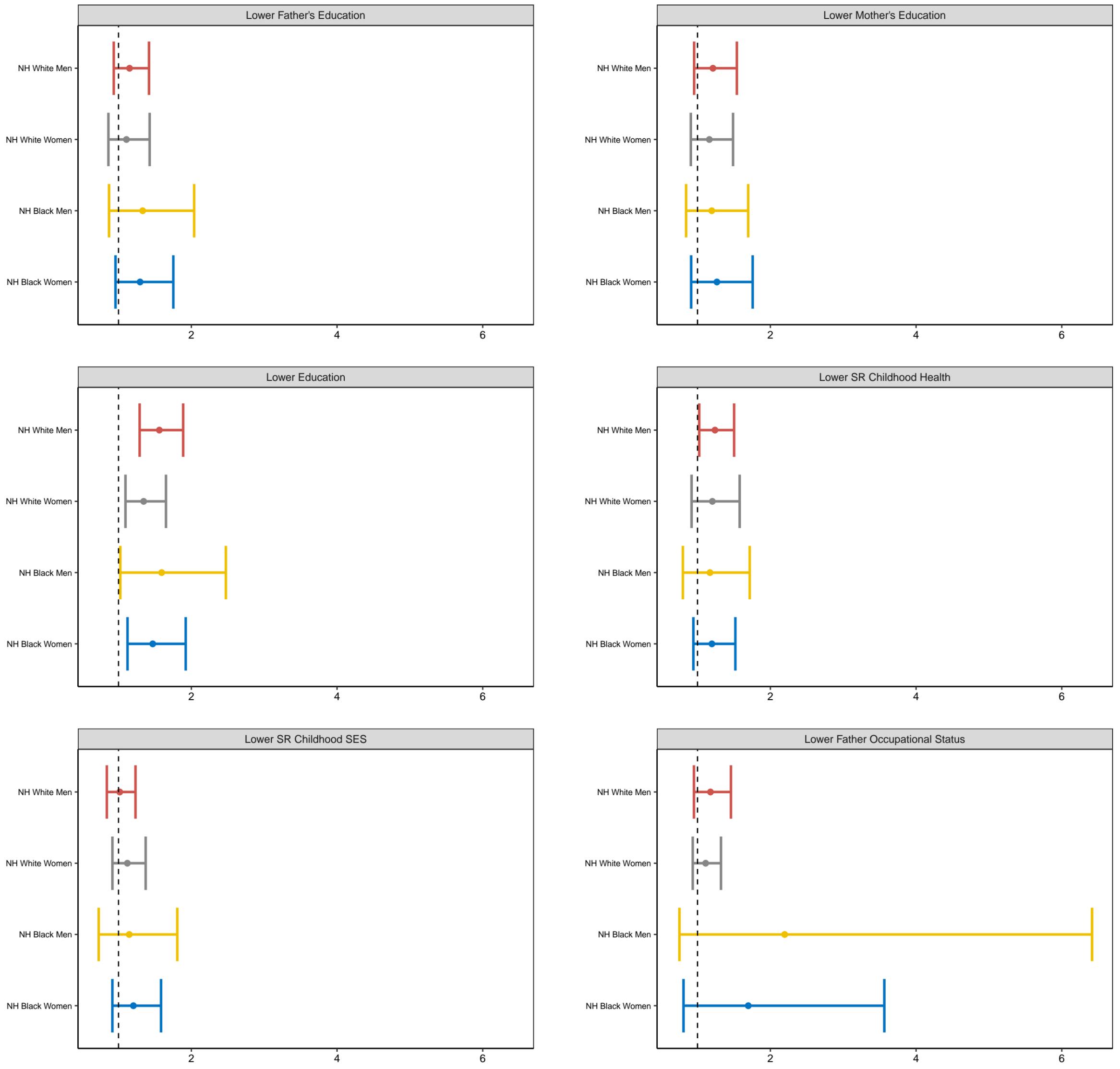

# Economic

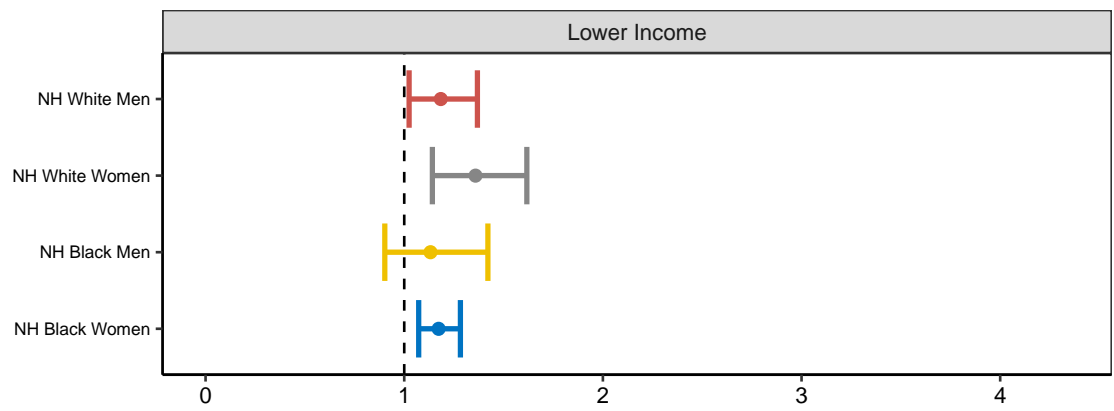
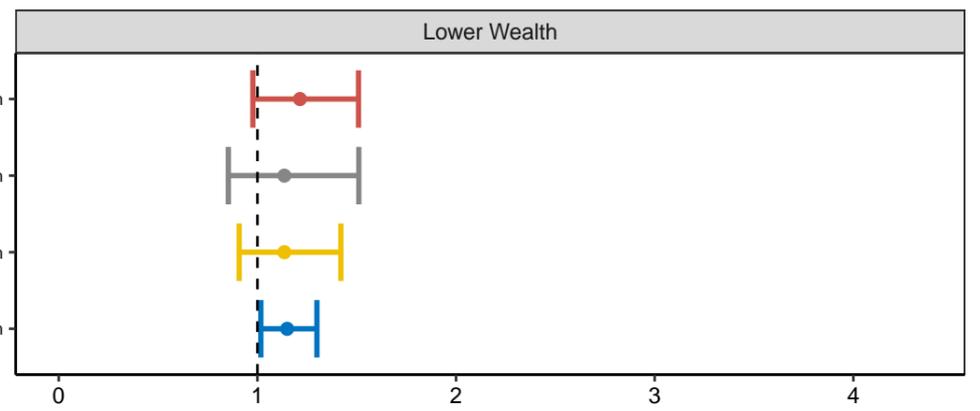
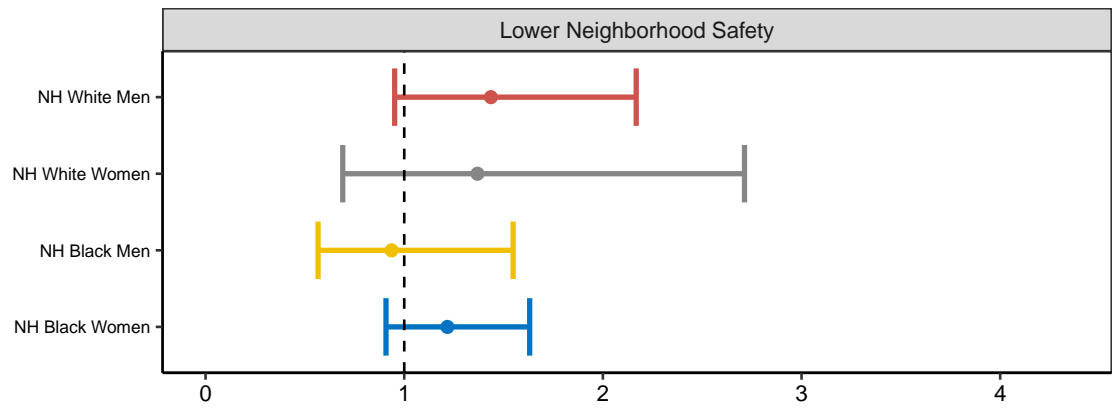
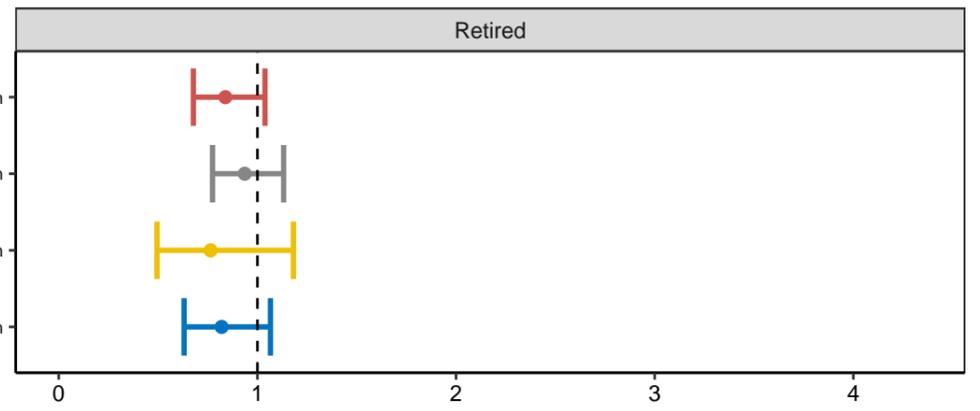
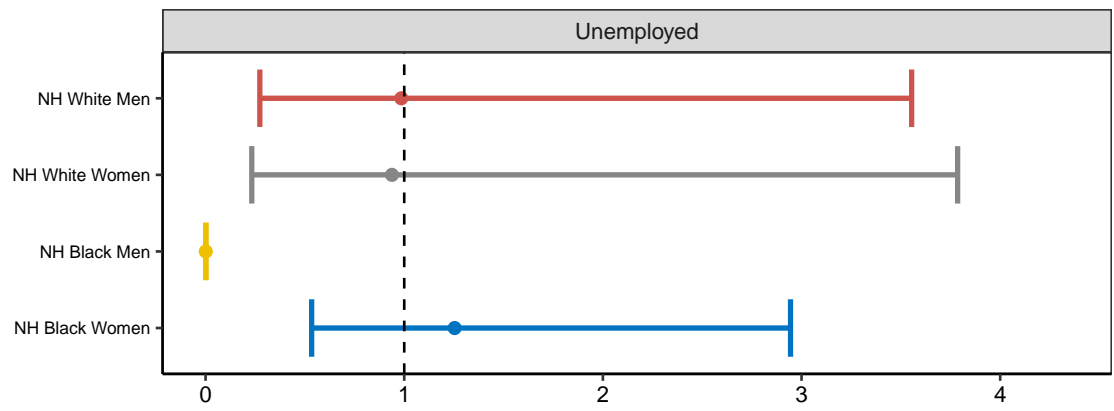
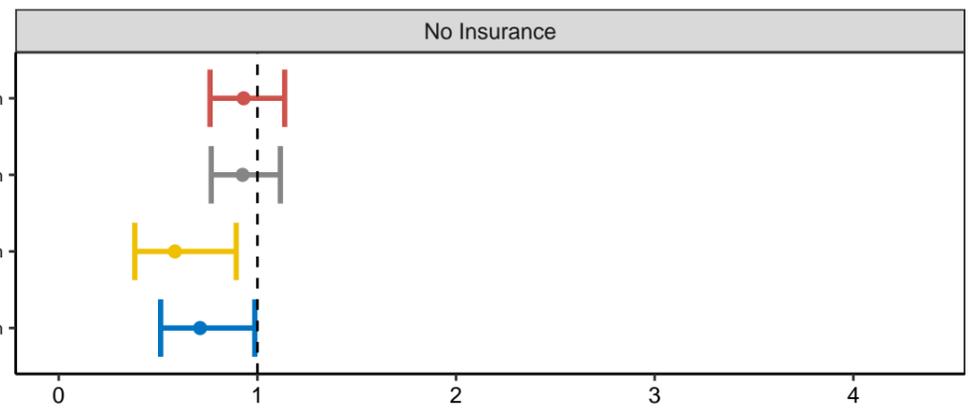
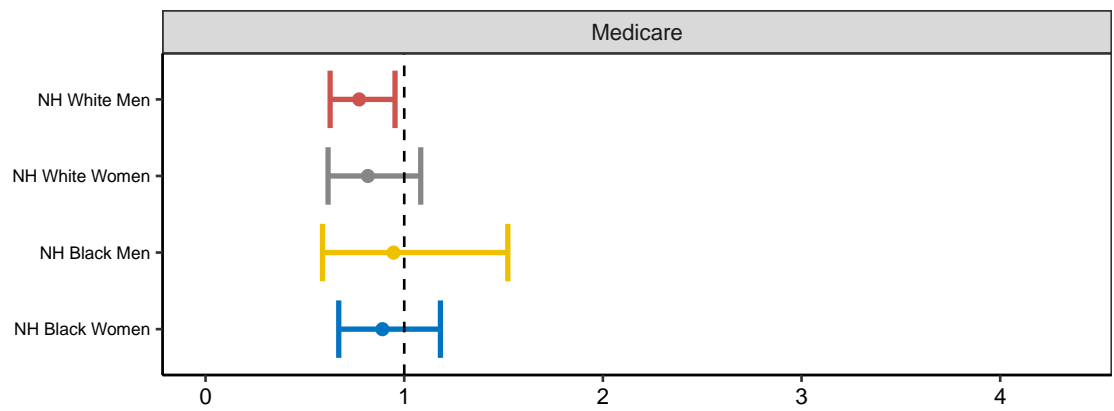
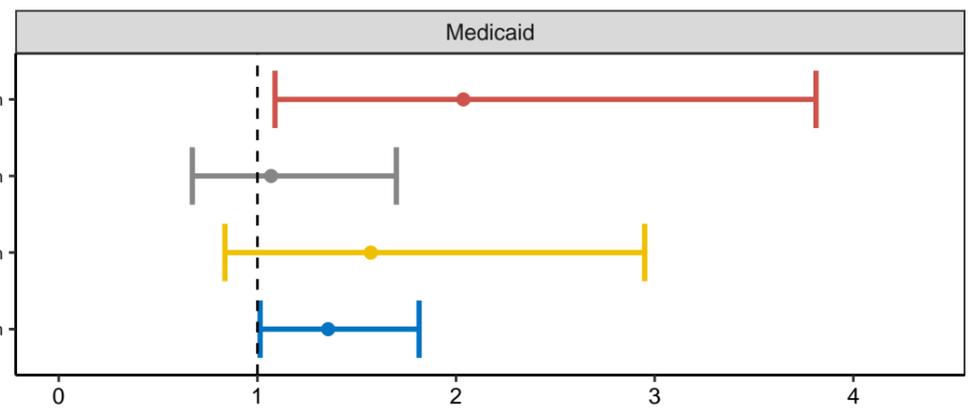
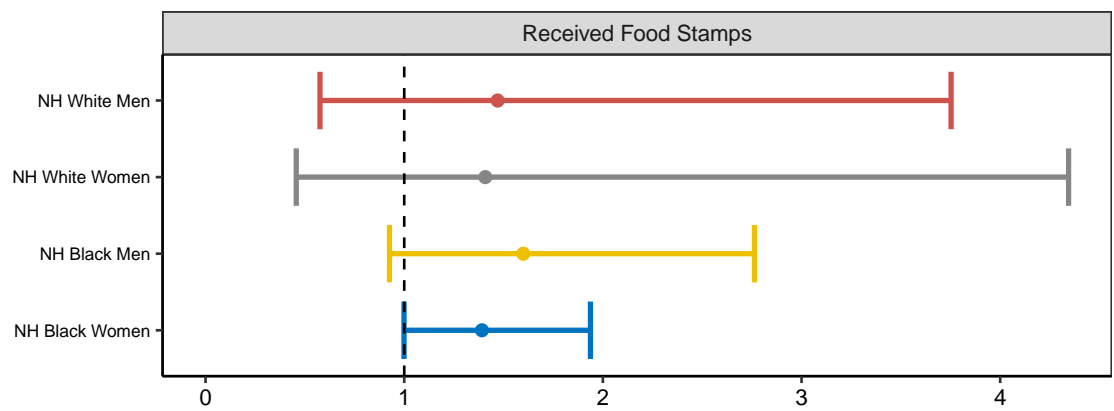
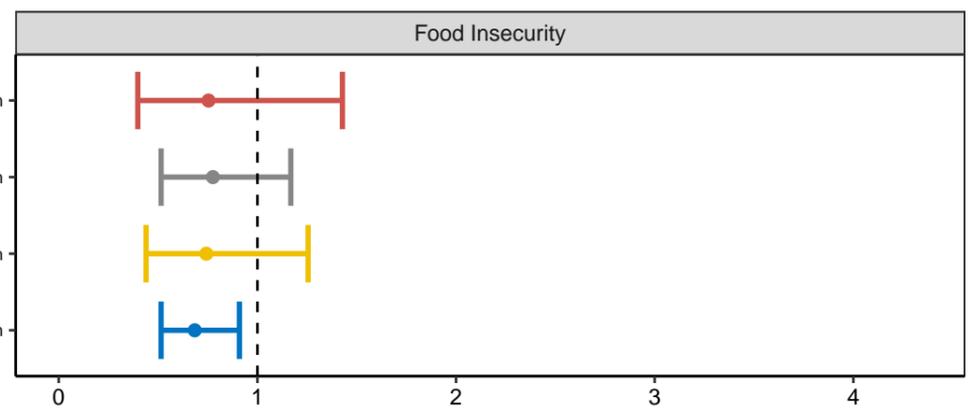

Hazard Ratio

# Health

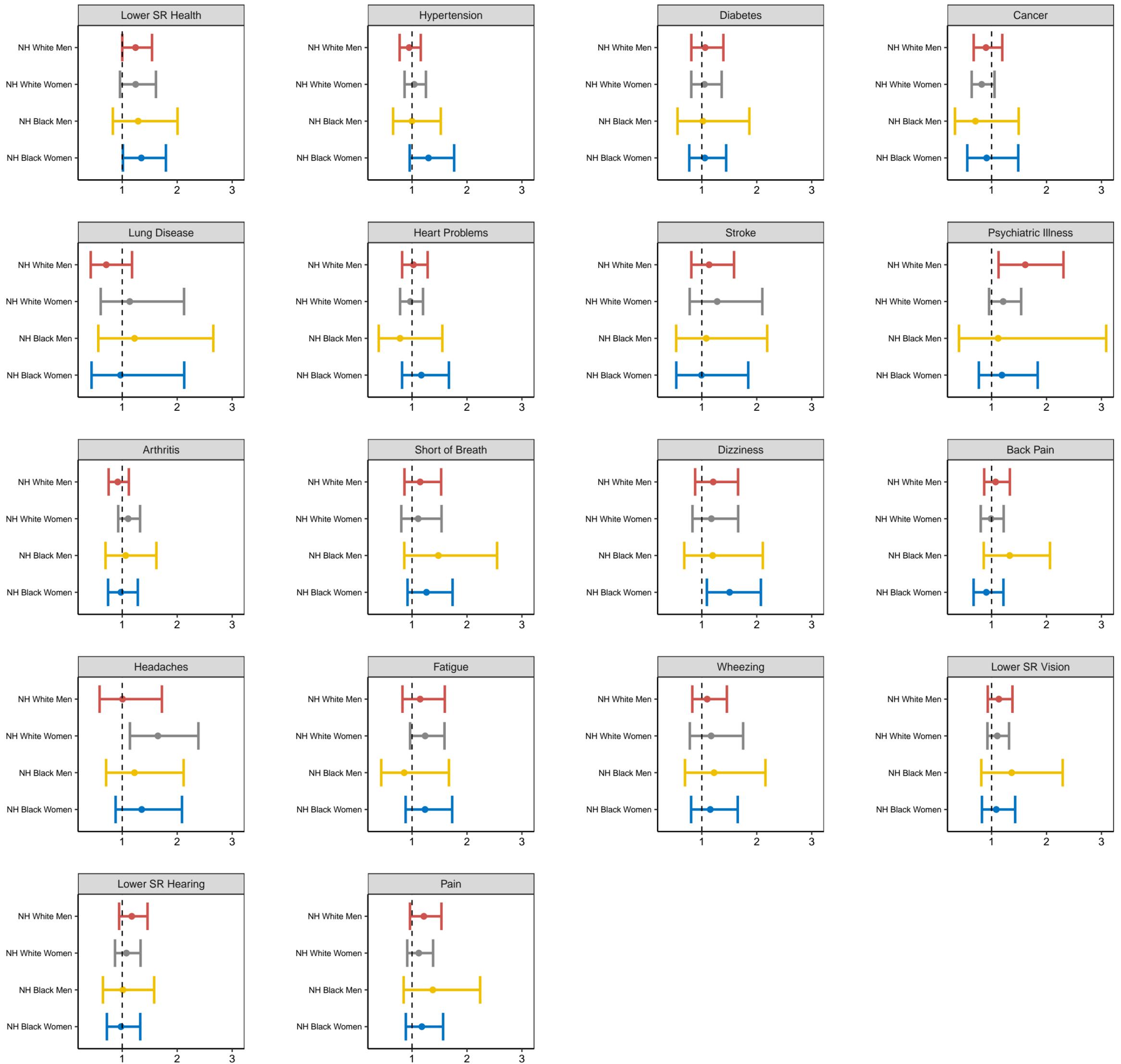

## Behaviors

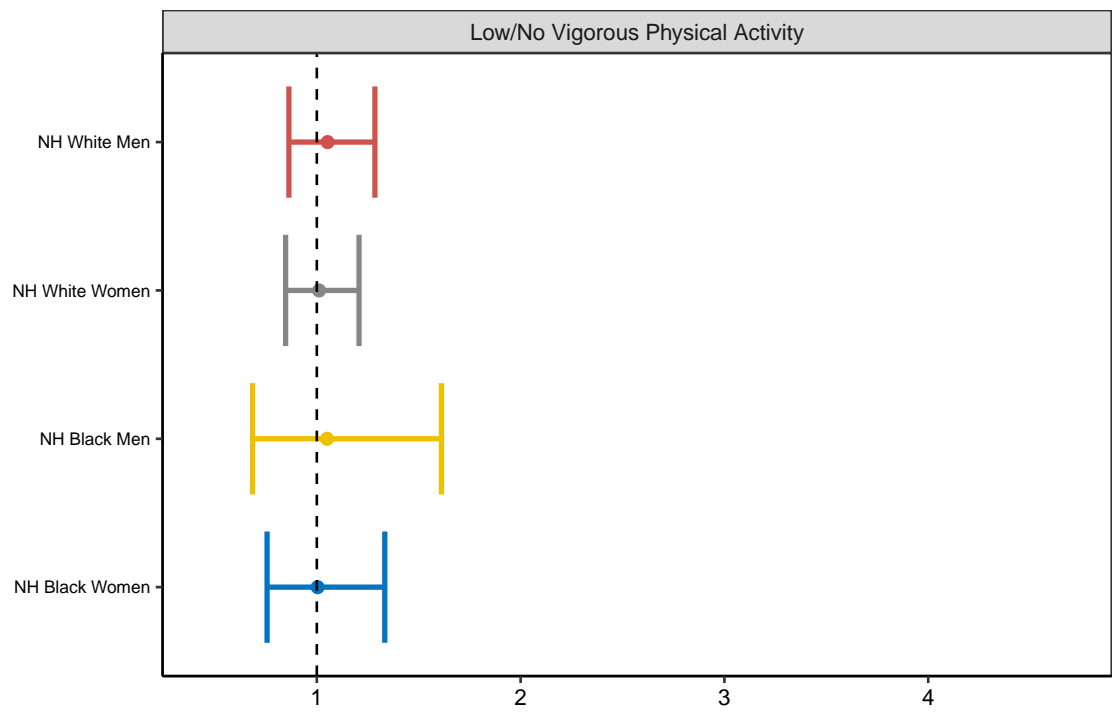
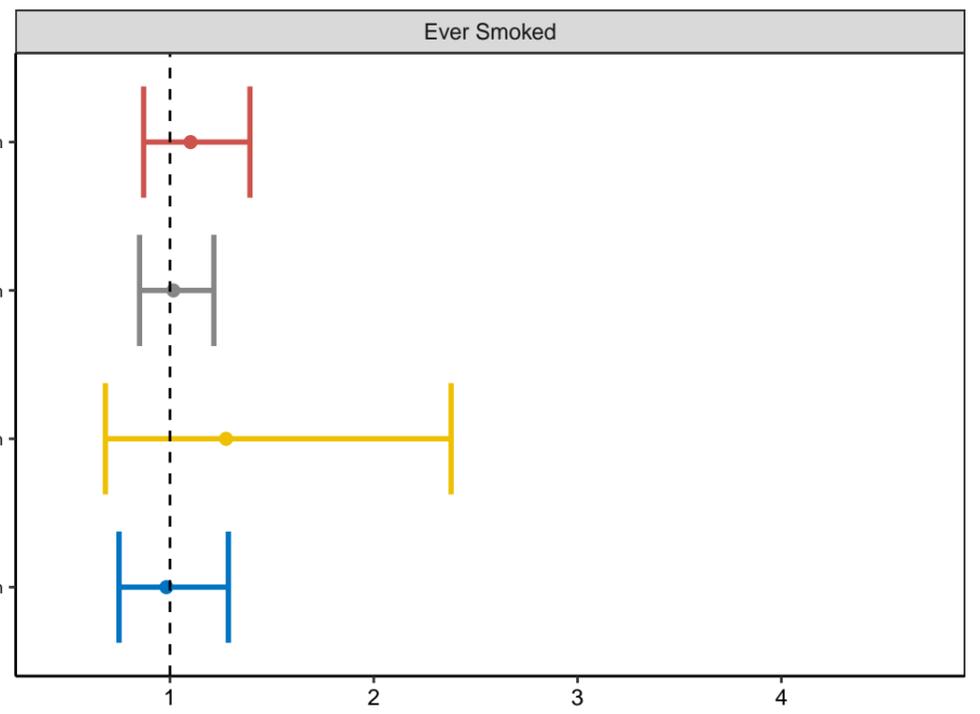
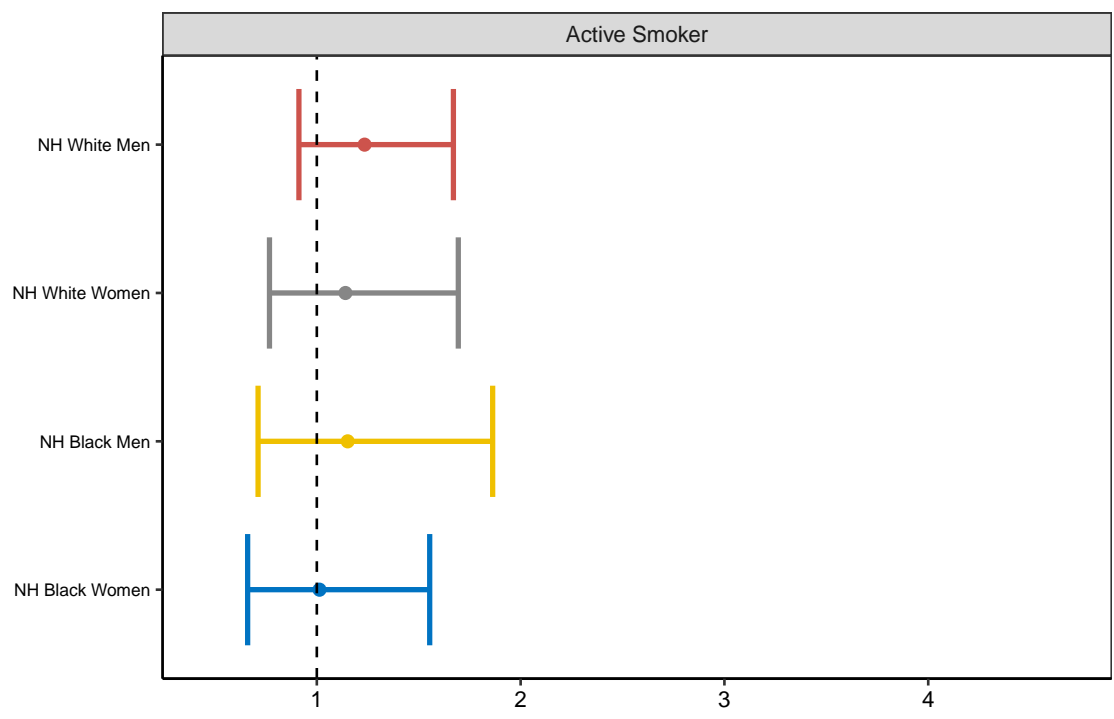
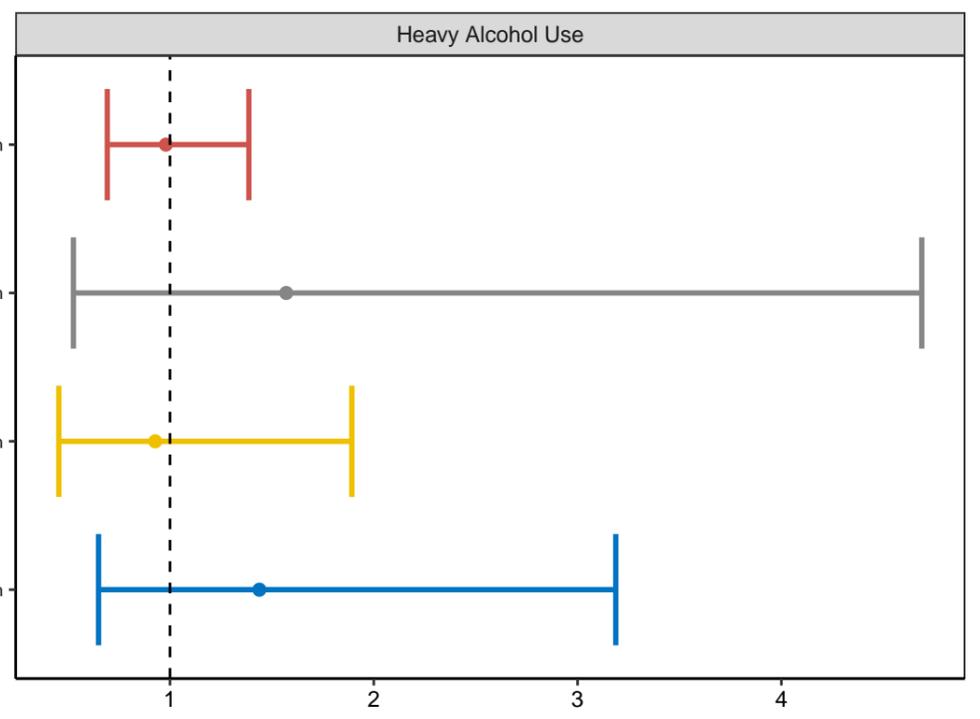
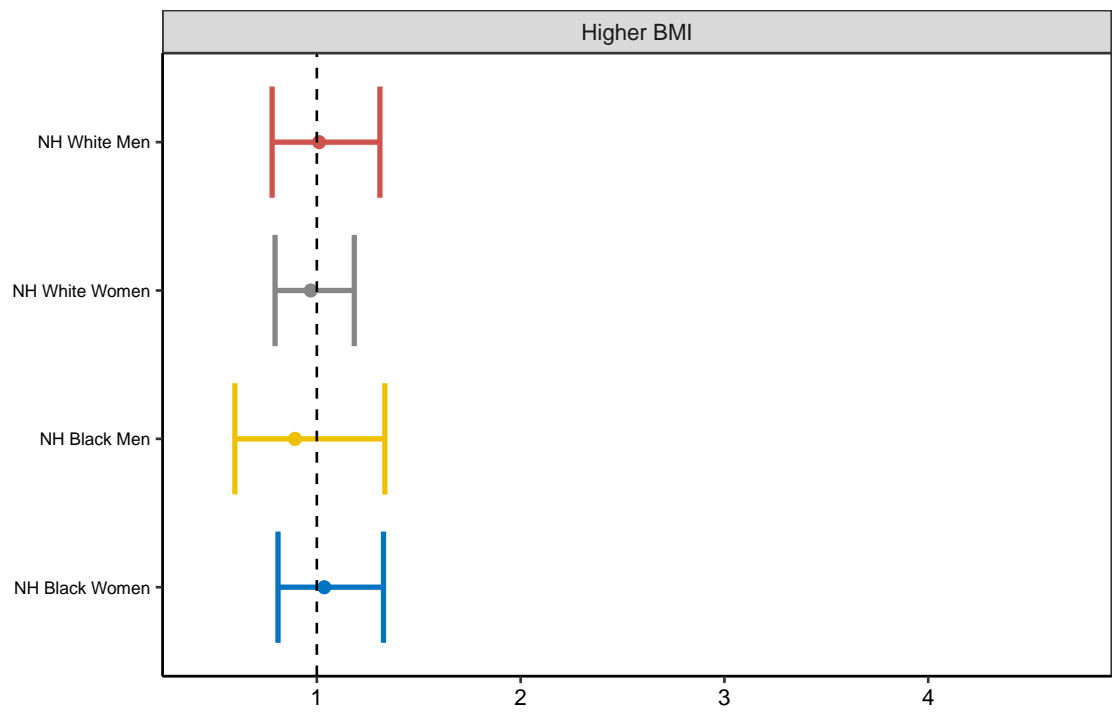

Hazard Ratio

## Social Ties

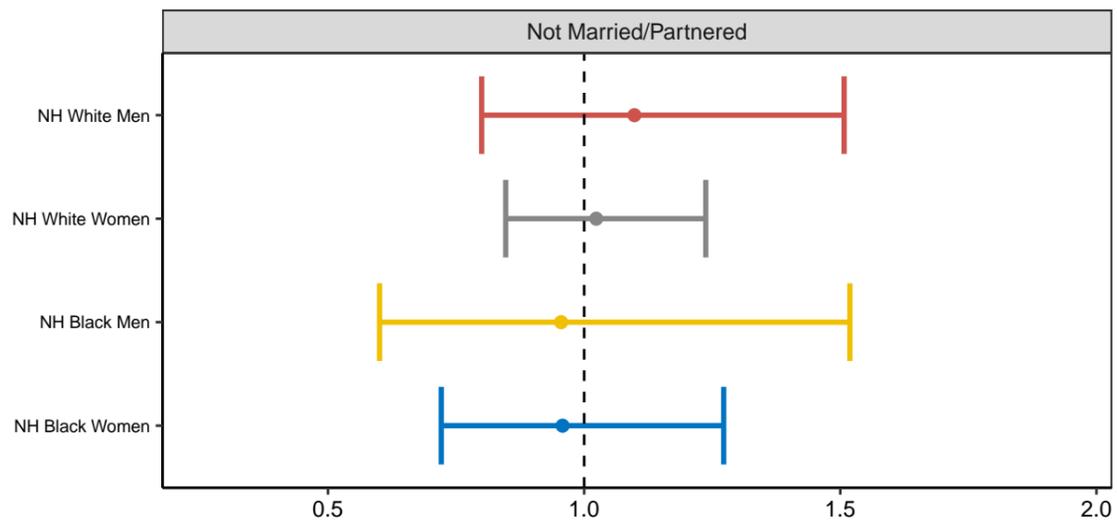
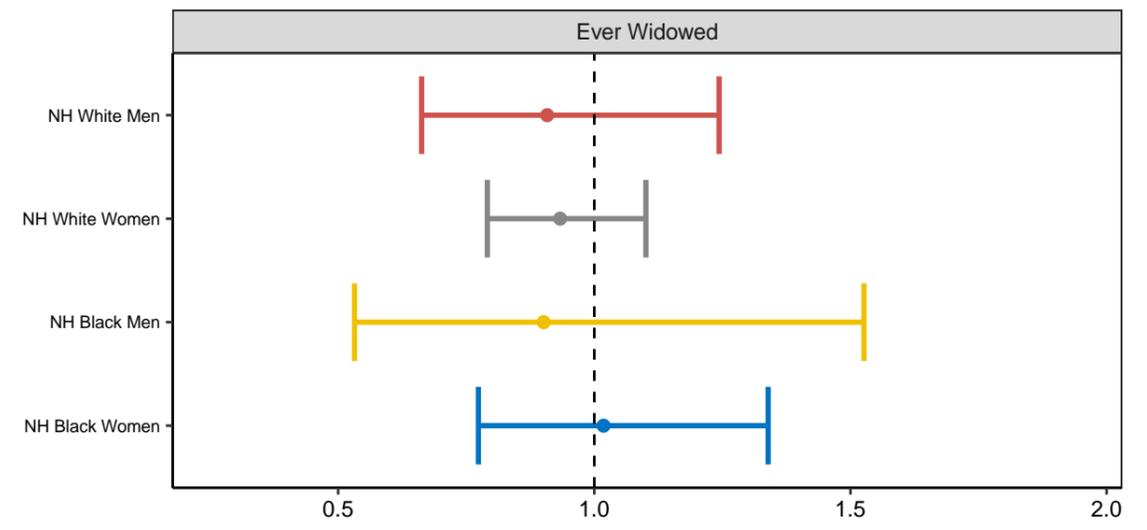
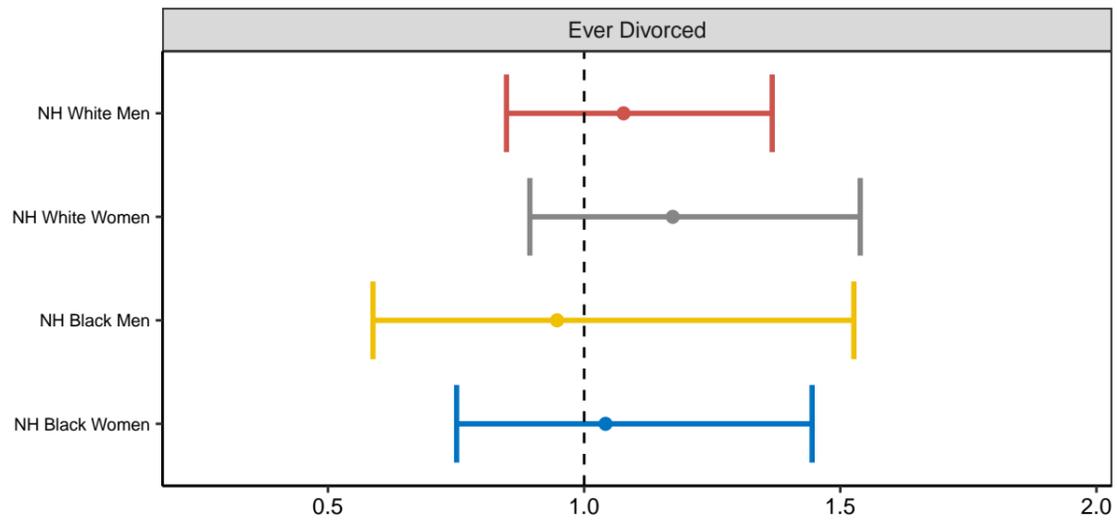
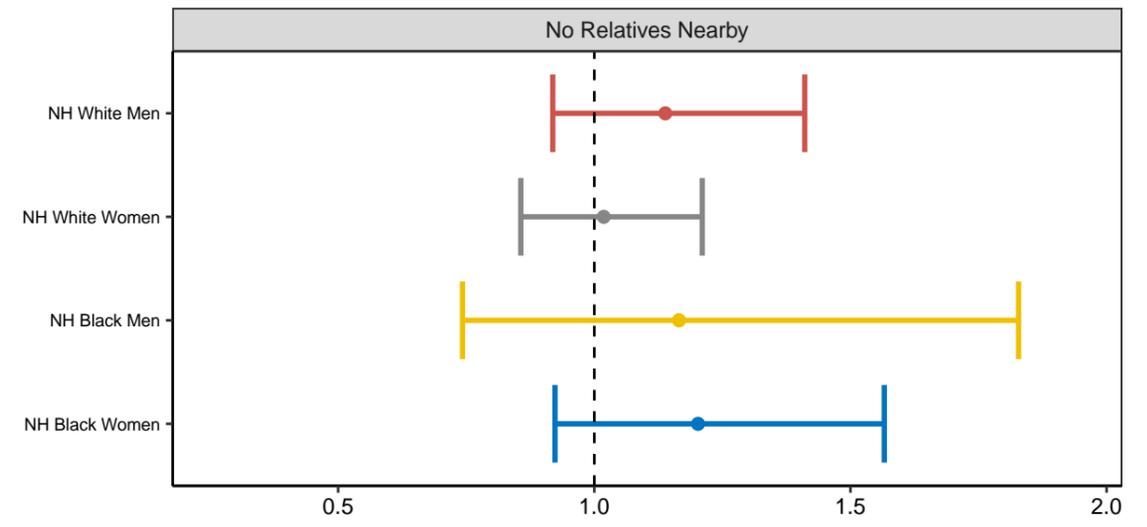
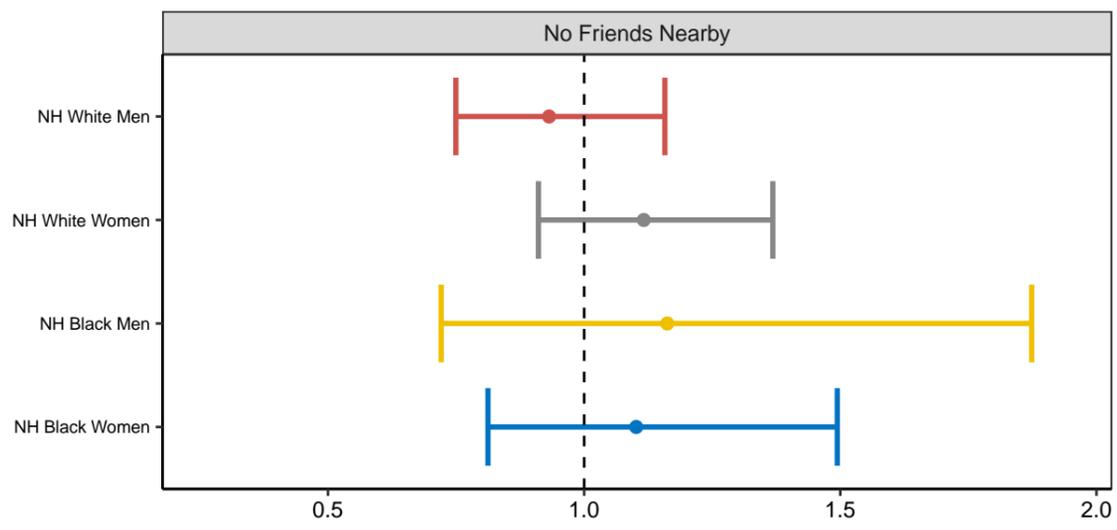
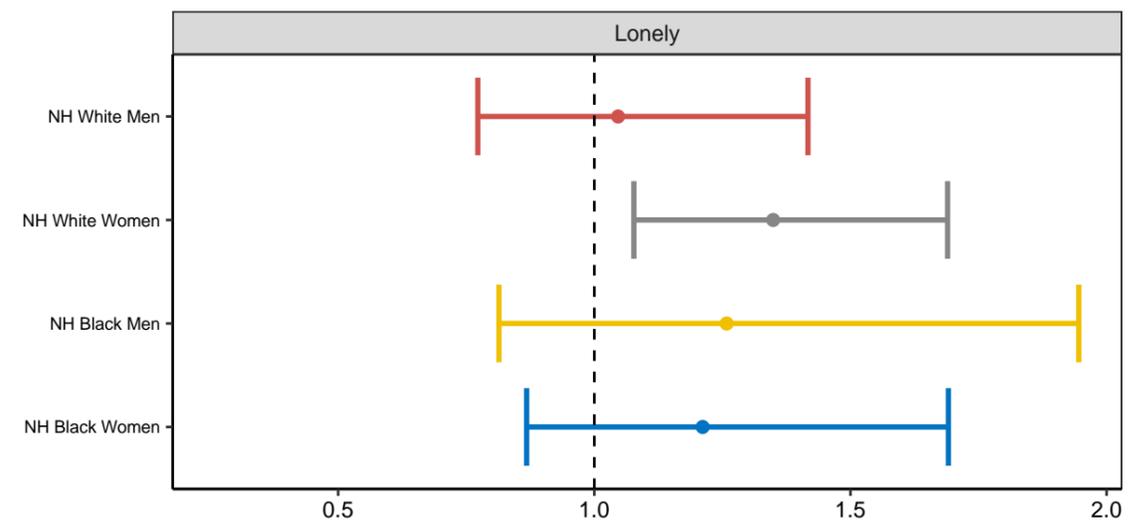
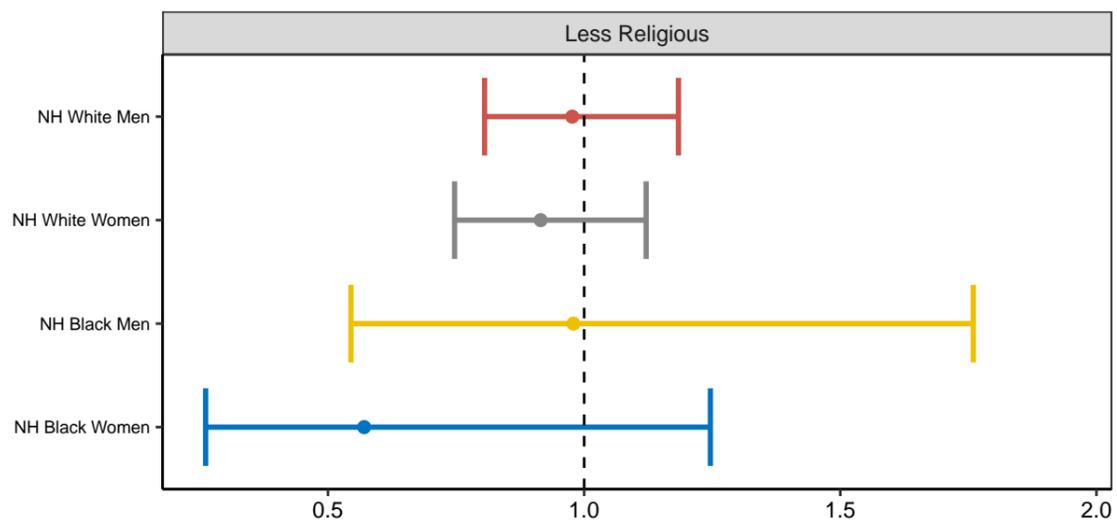

Hazard Ratio

## Genetic

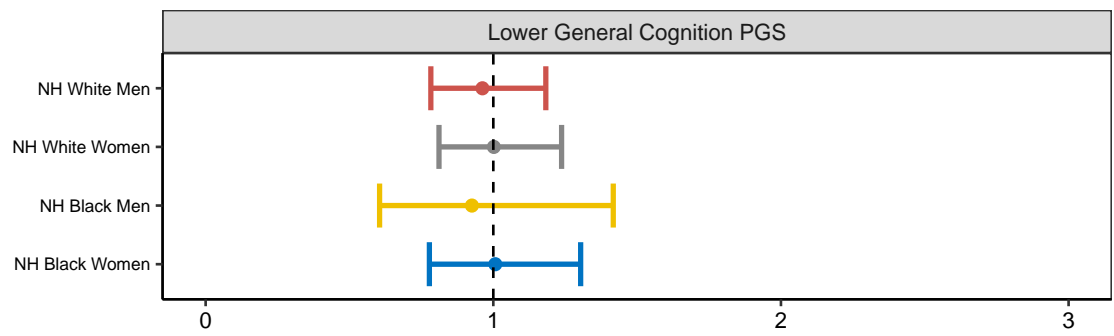

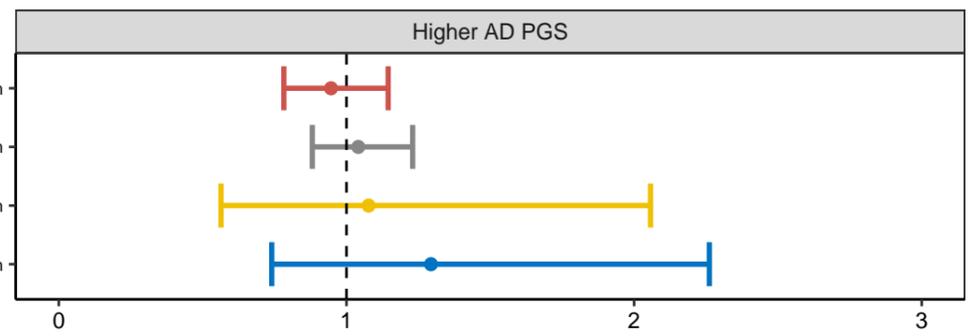

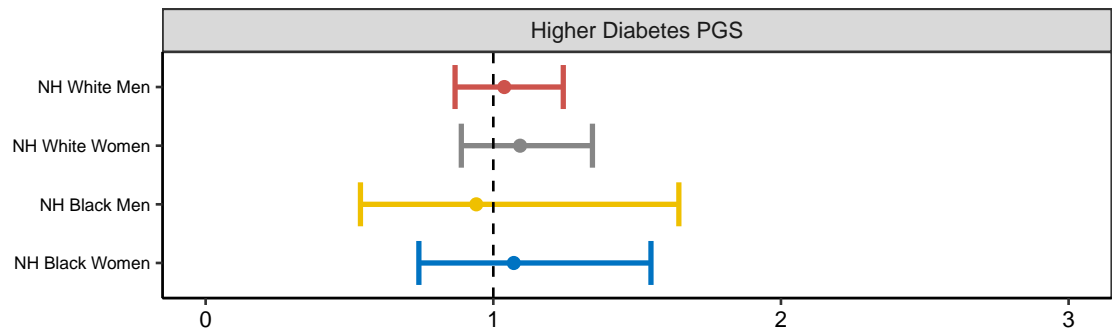

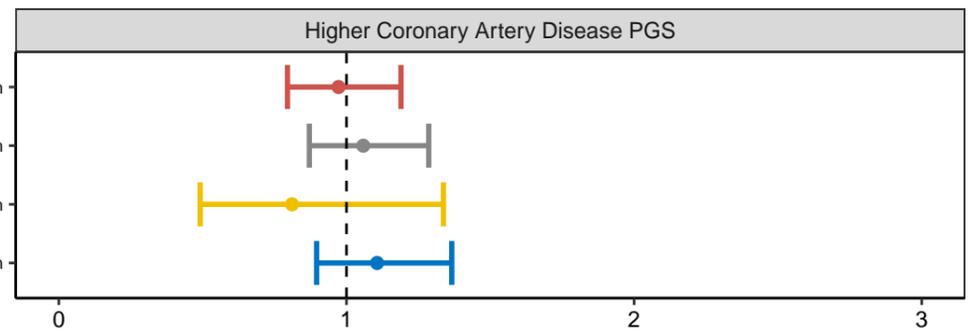

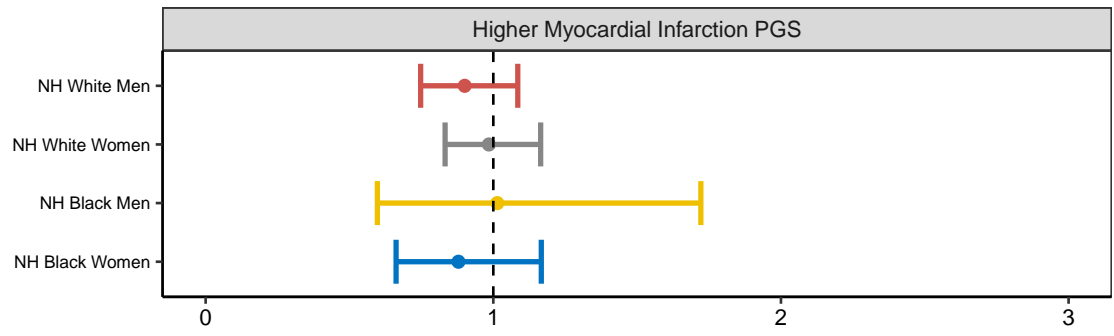

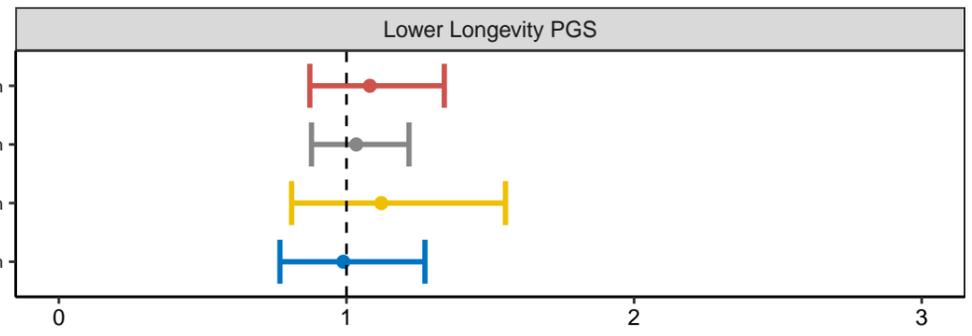

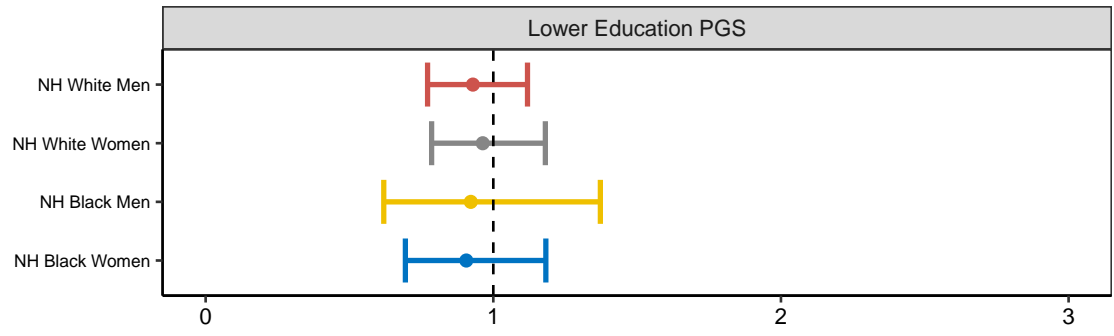

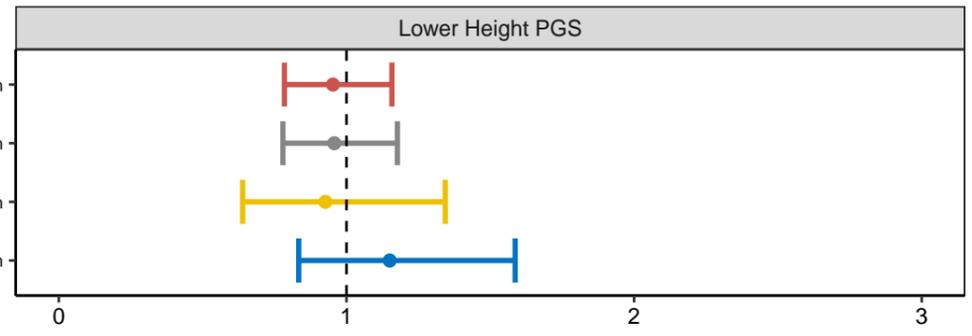

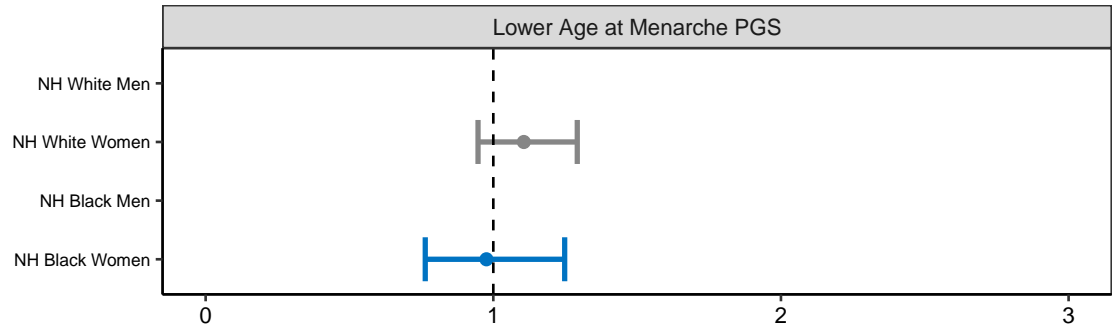

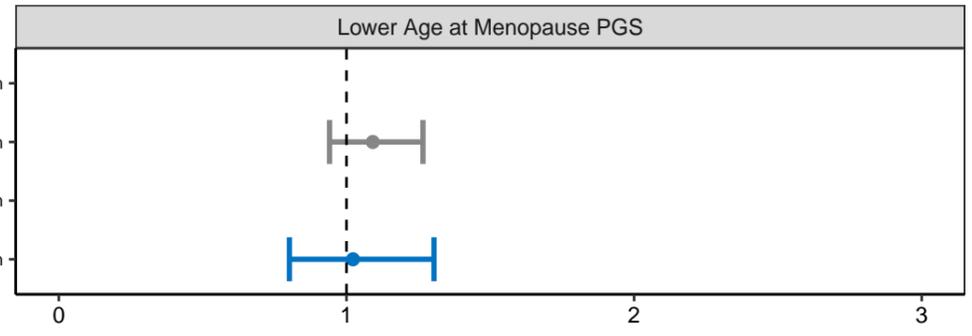

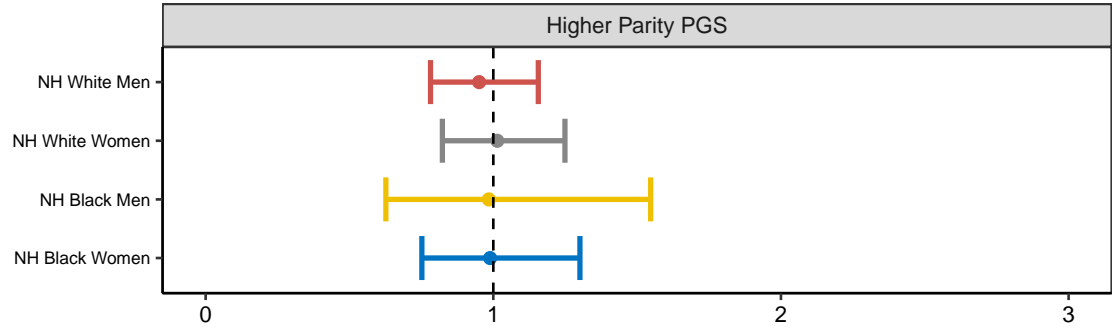

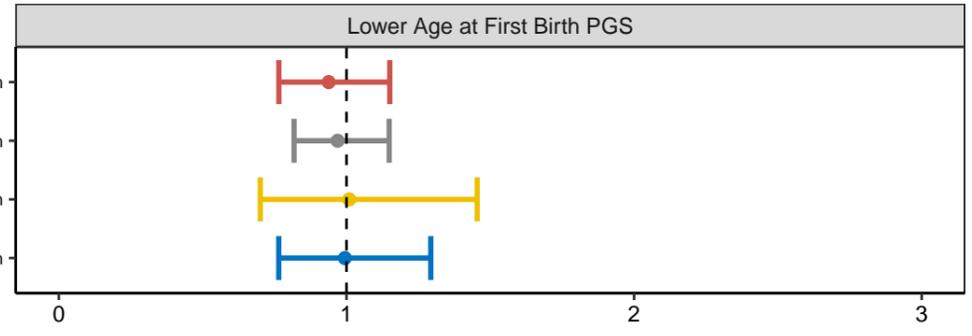

Hazard Ratio

Fig S3

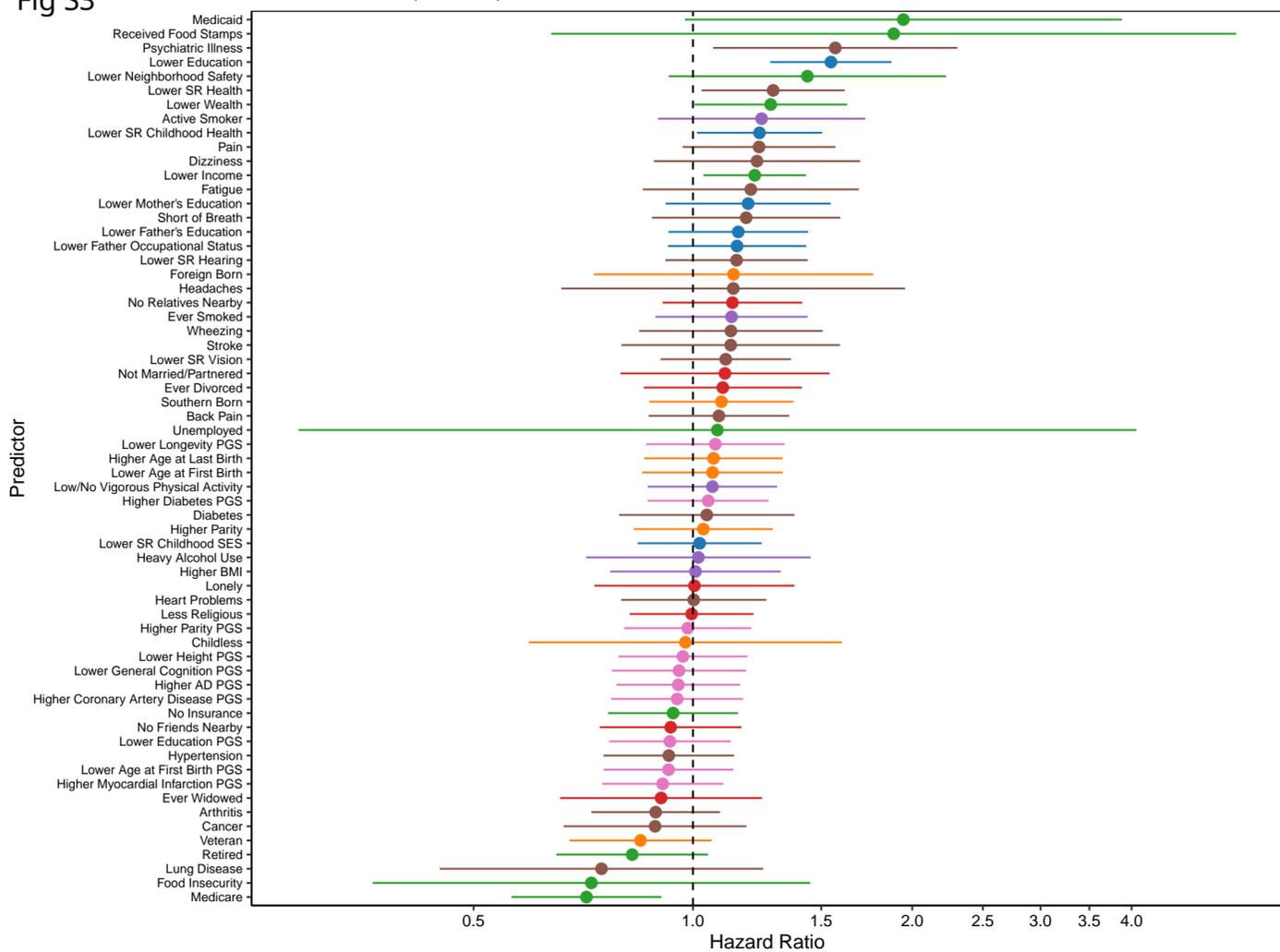

NH White Men (n=2561)

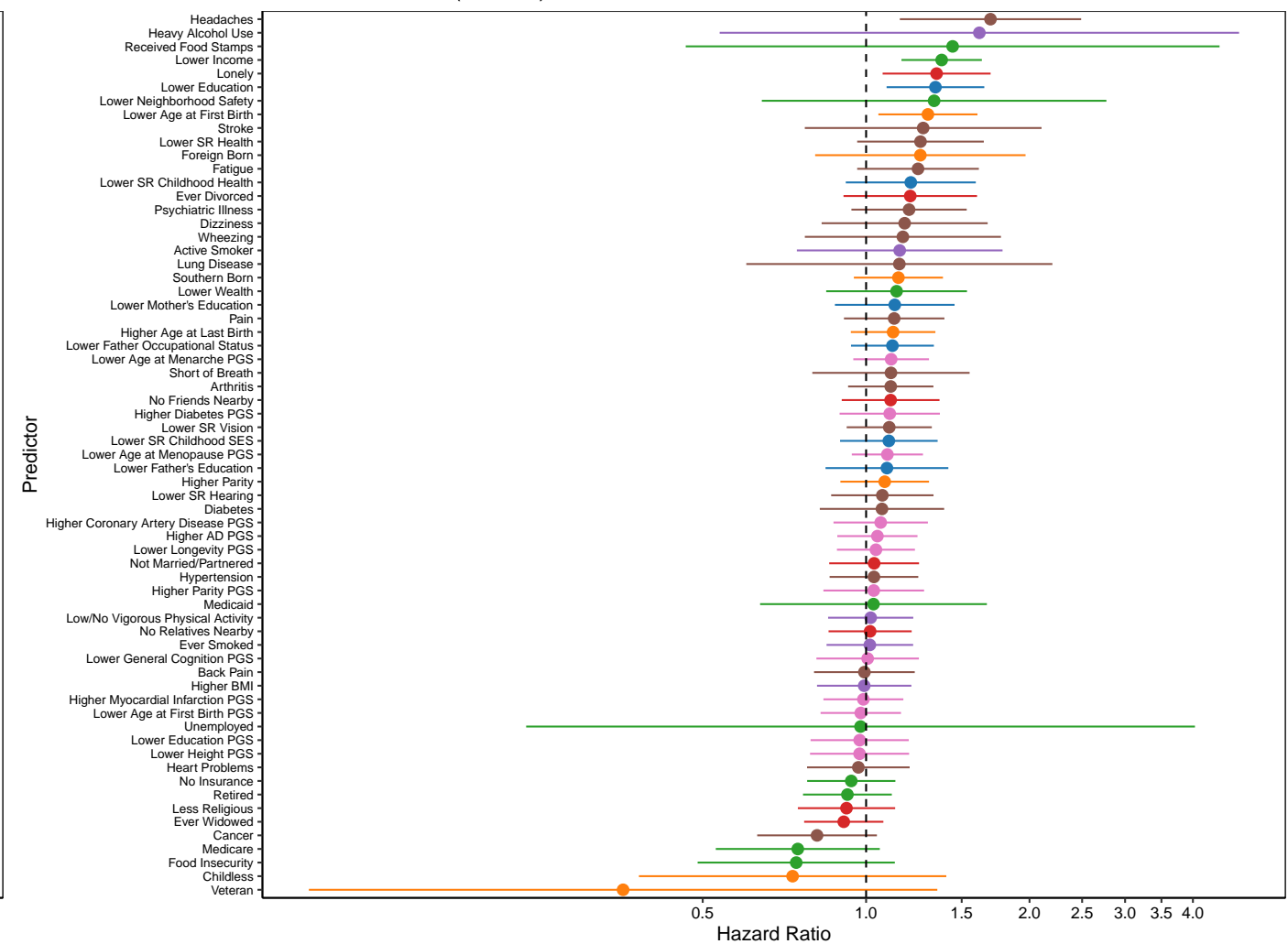

NH White Women (n=3377)

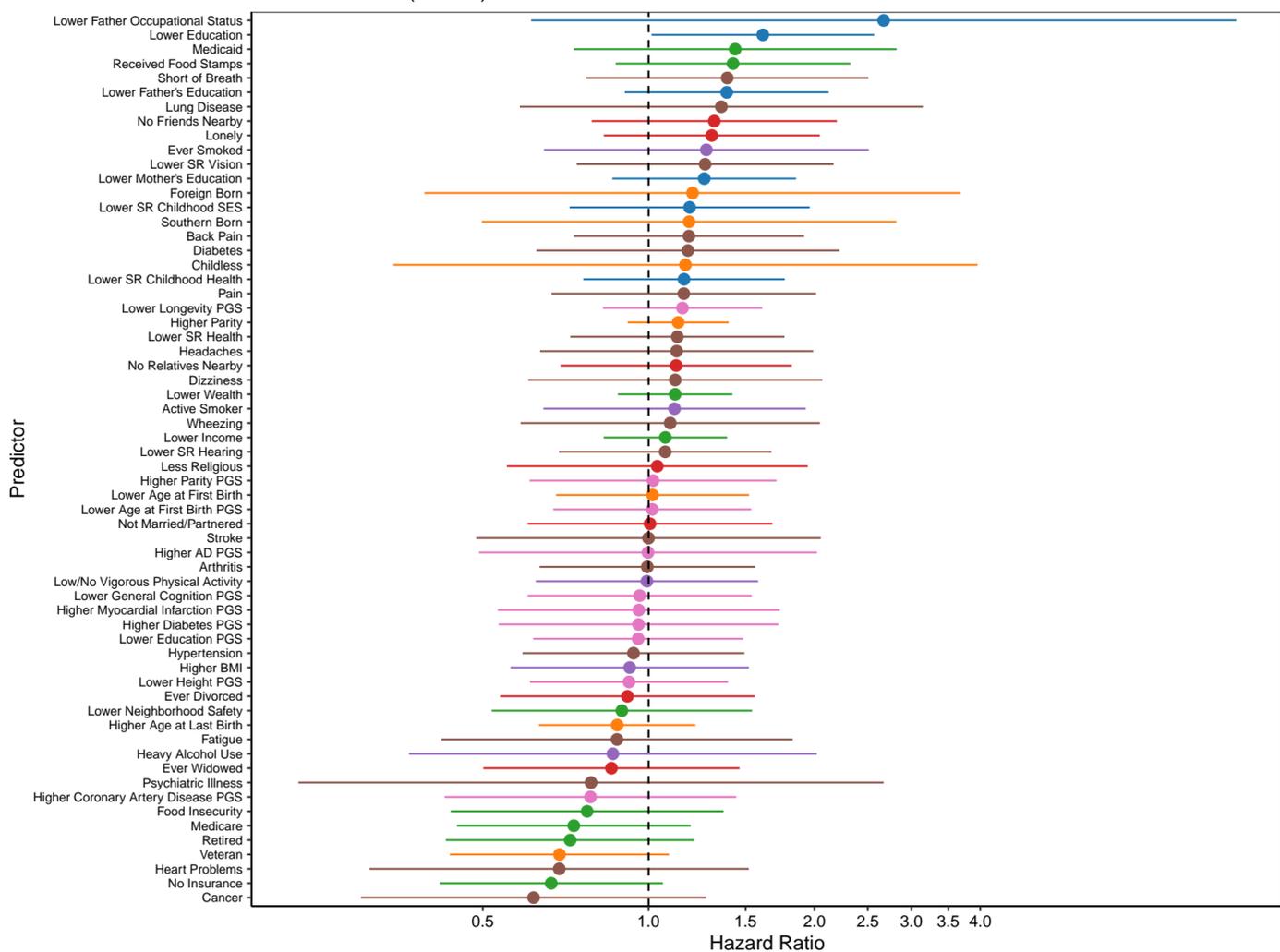

NH Black Men (n=283)

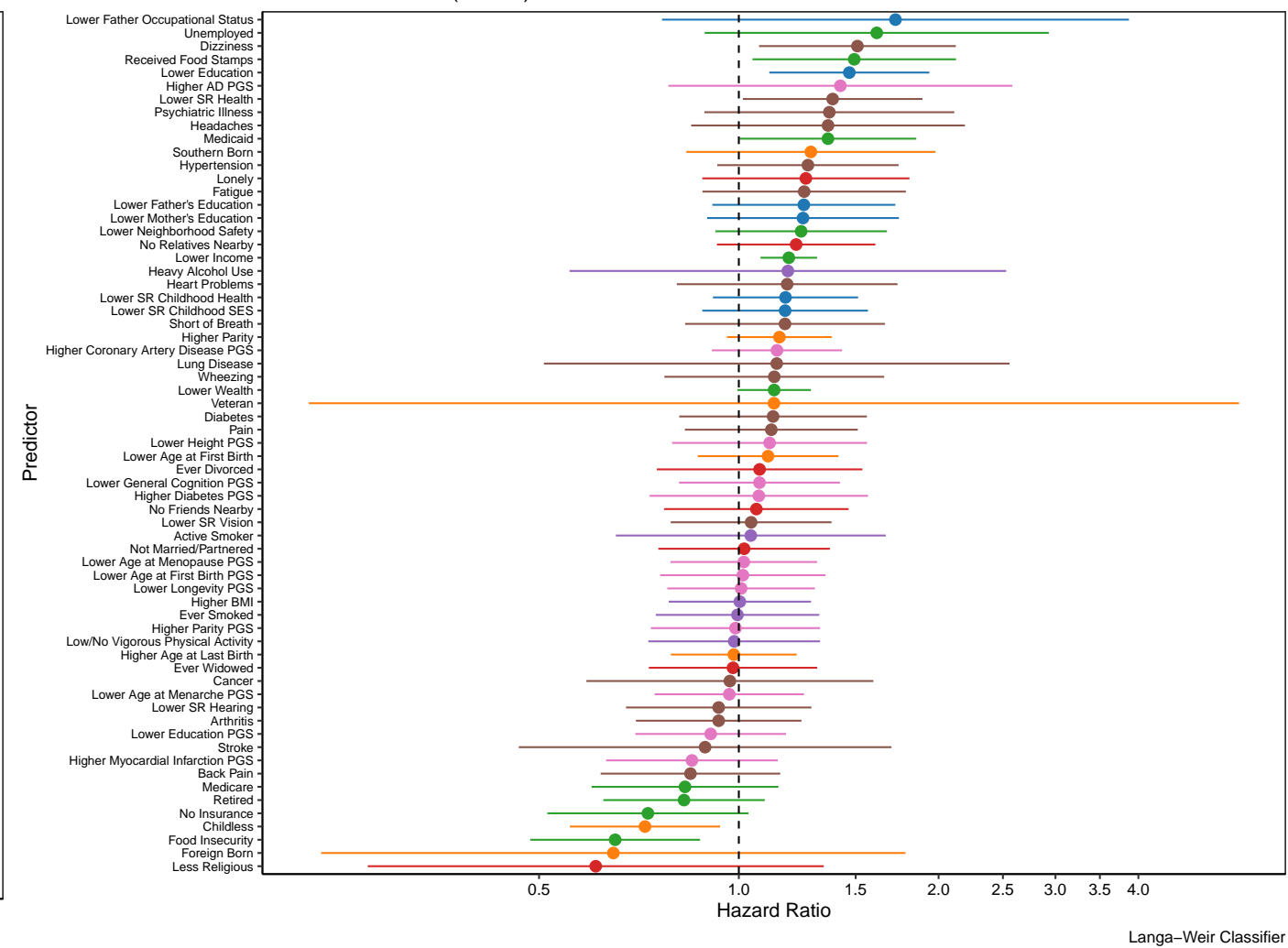

NH Black Women (n=525)

Early–Life    Economic    Behaviors    Genetic

Sociodemographic    Social Ties    Health

Langa–Weir Classifier

Fig S4

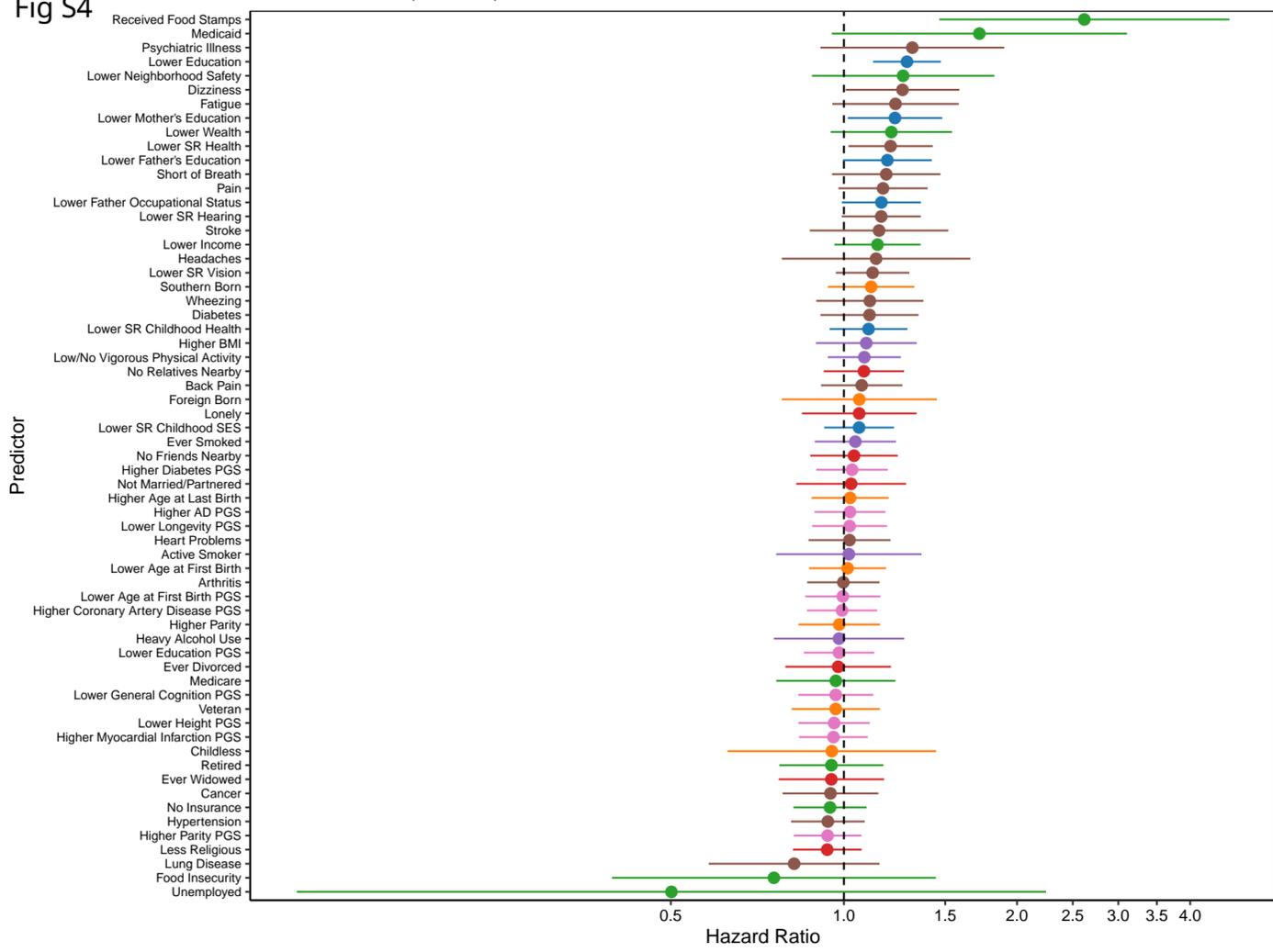

NH White Men (n=2561)

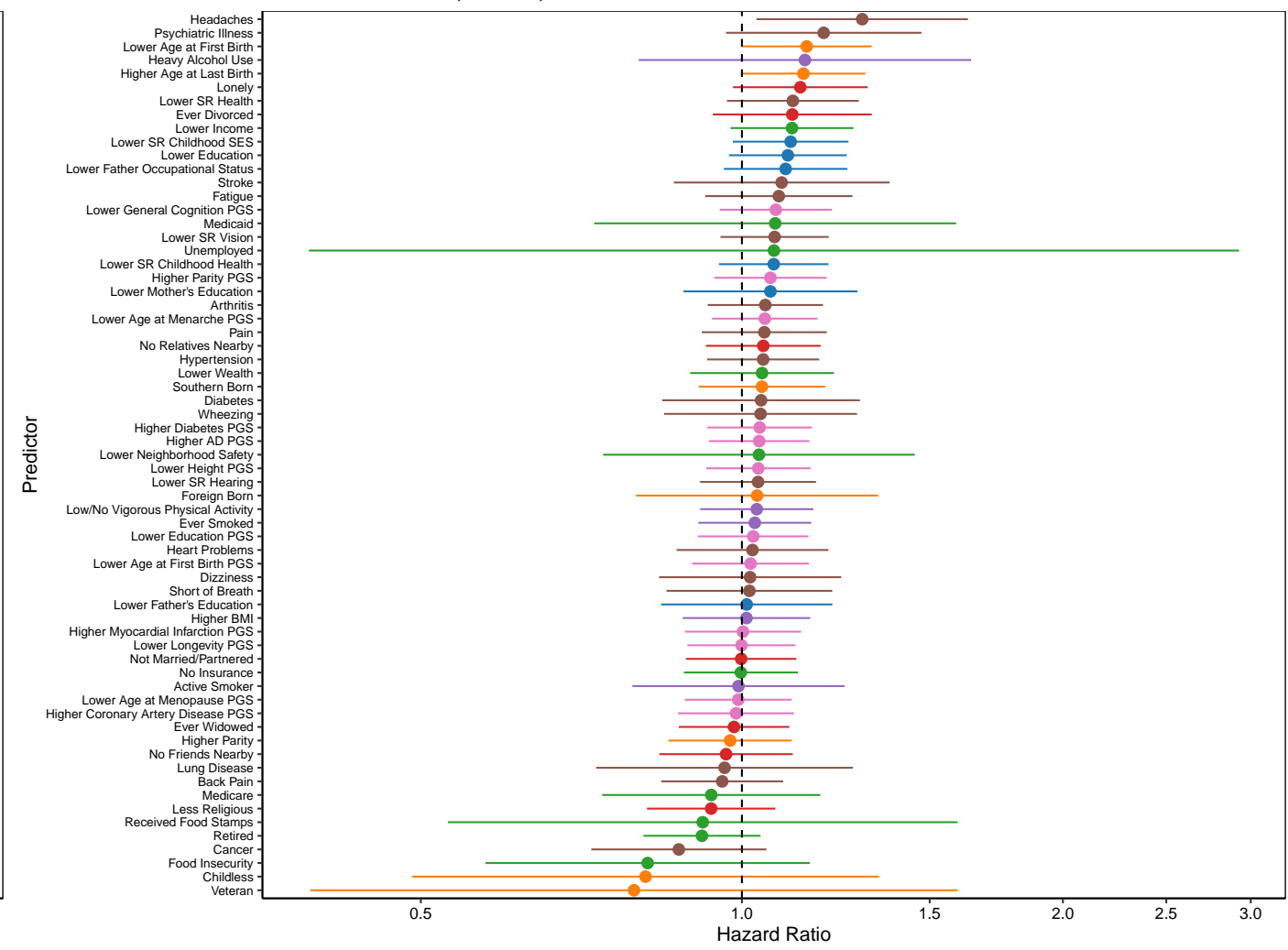

NH White Women (n=3377)

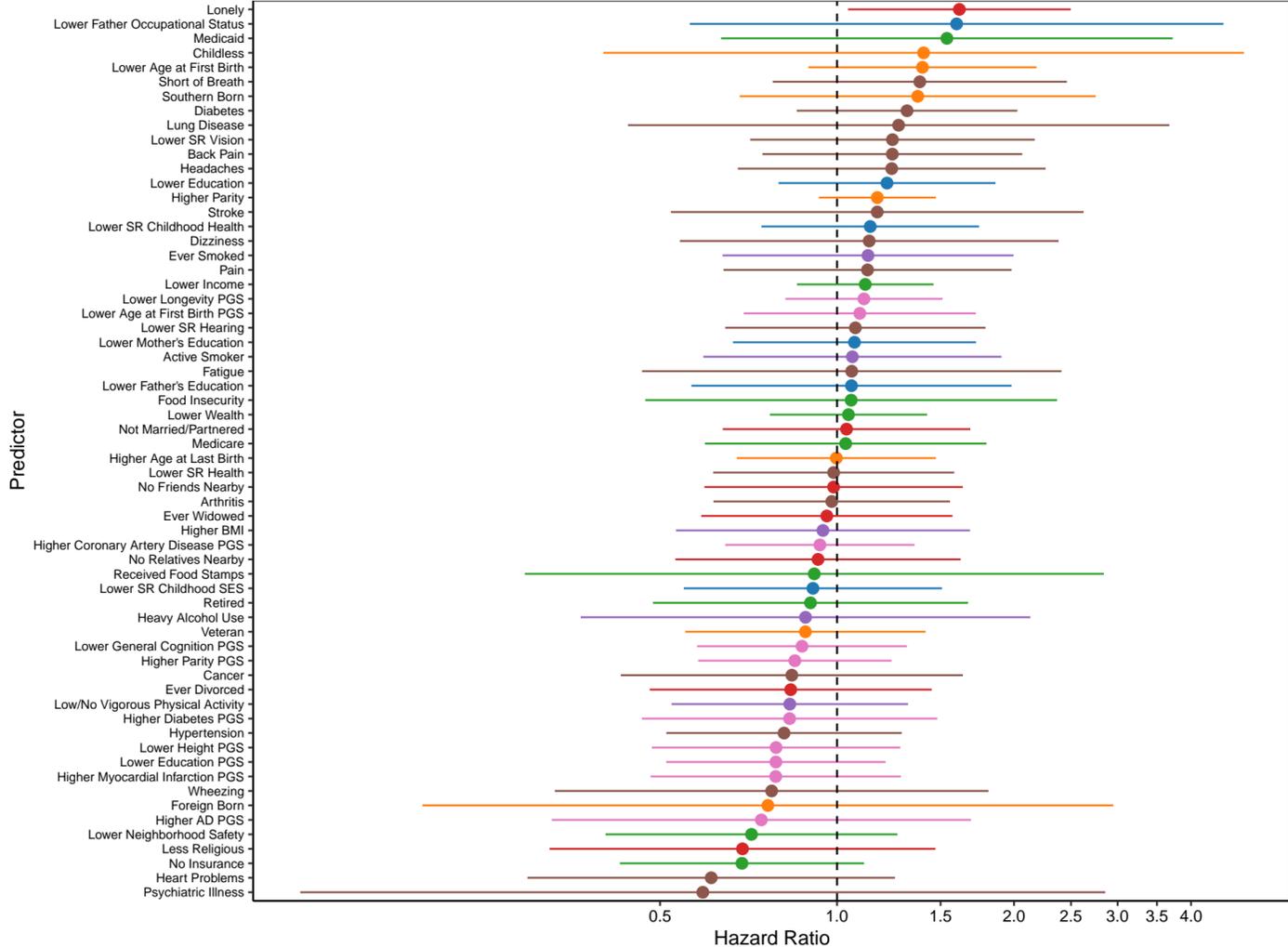

NH Black Men (n=283)

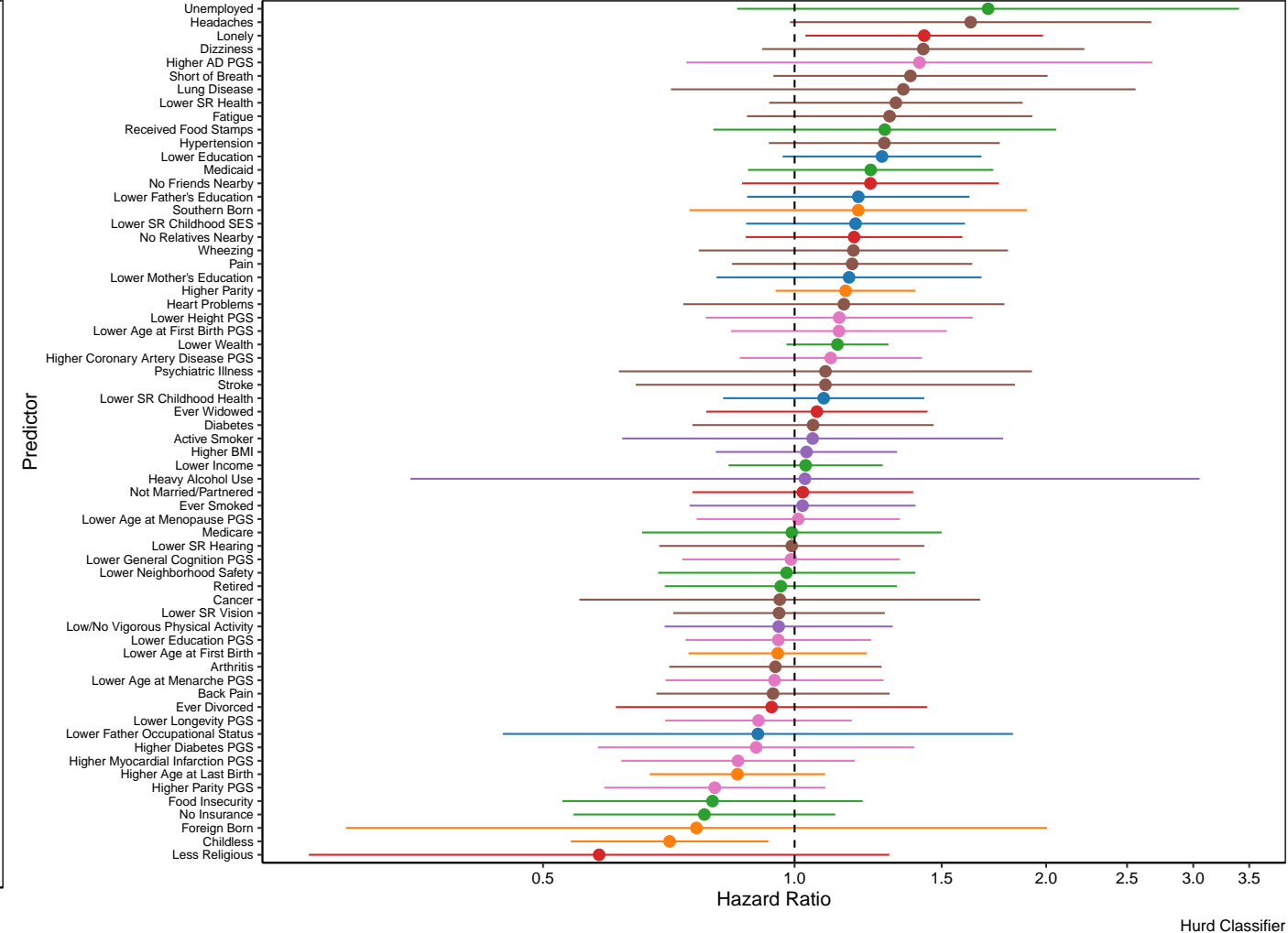

NH Black Women (n=525)

Hurd Classifier: ● Early-Life ● Economic ● Behaviors ● Genetic ● Sociodemographic ● Social Ties ● Health

Fig S5

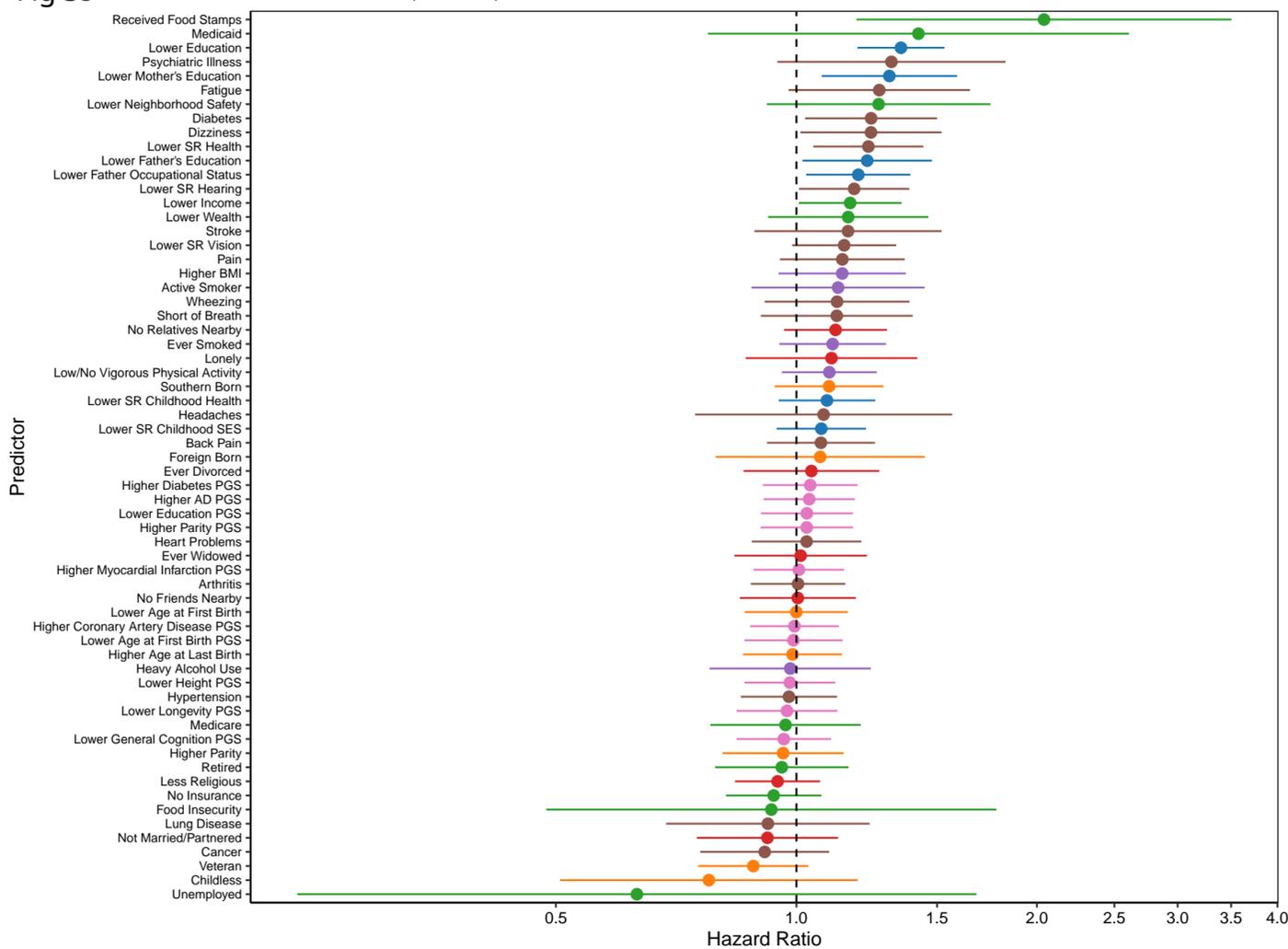

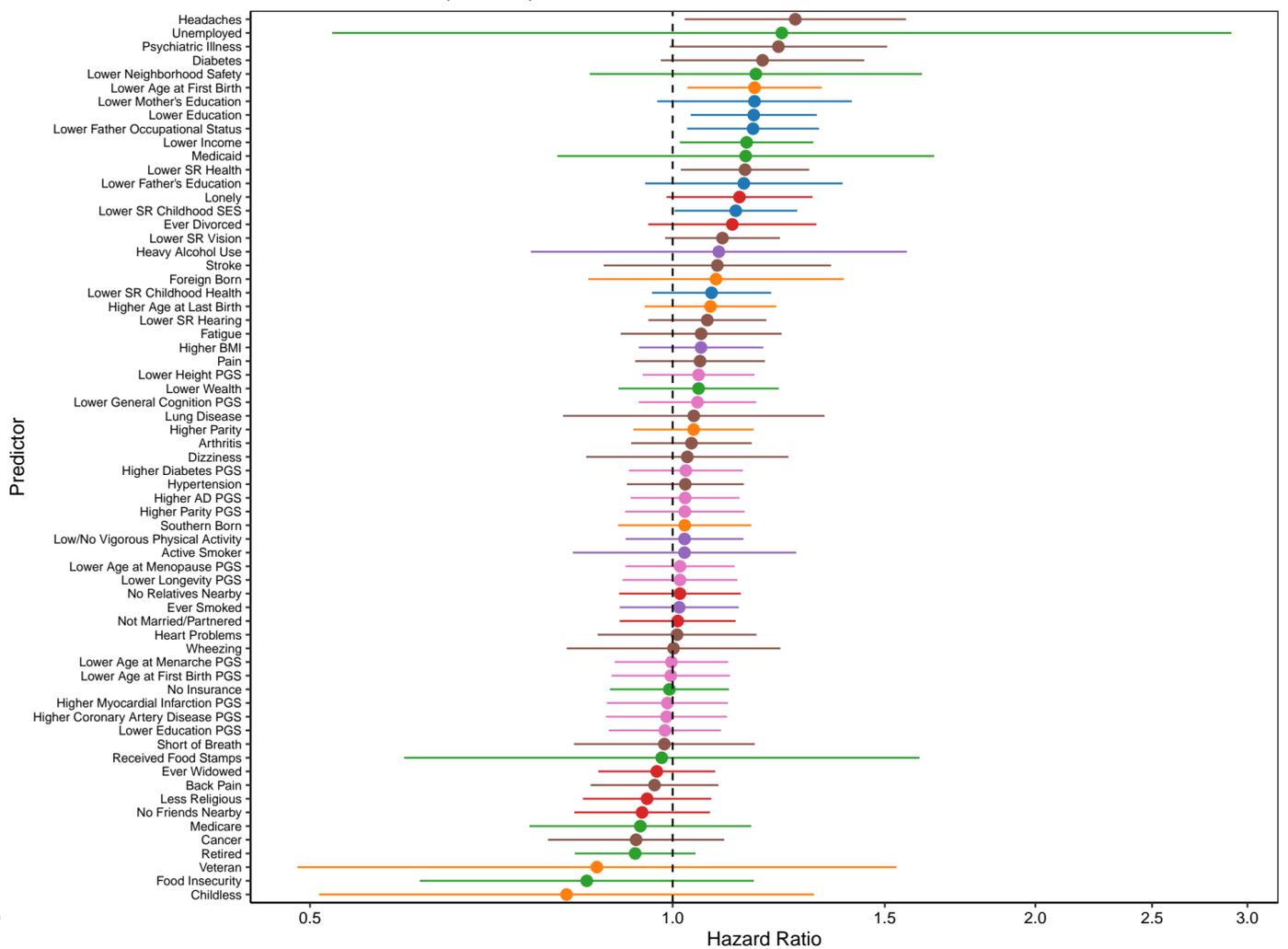

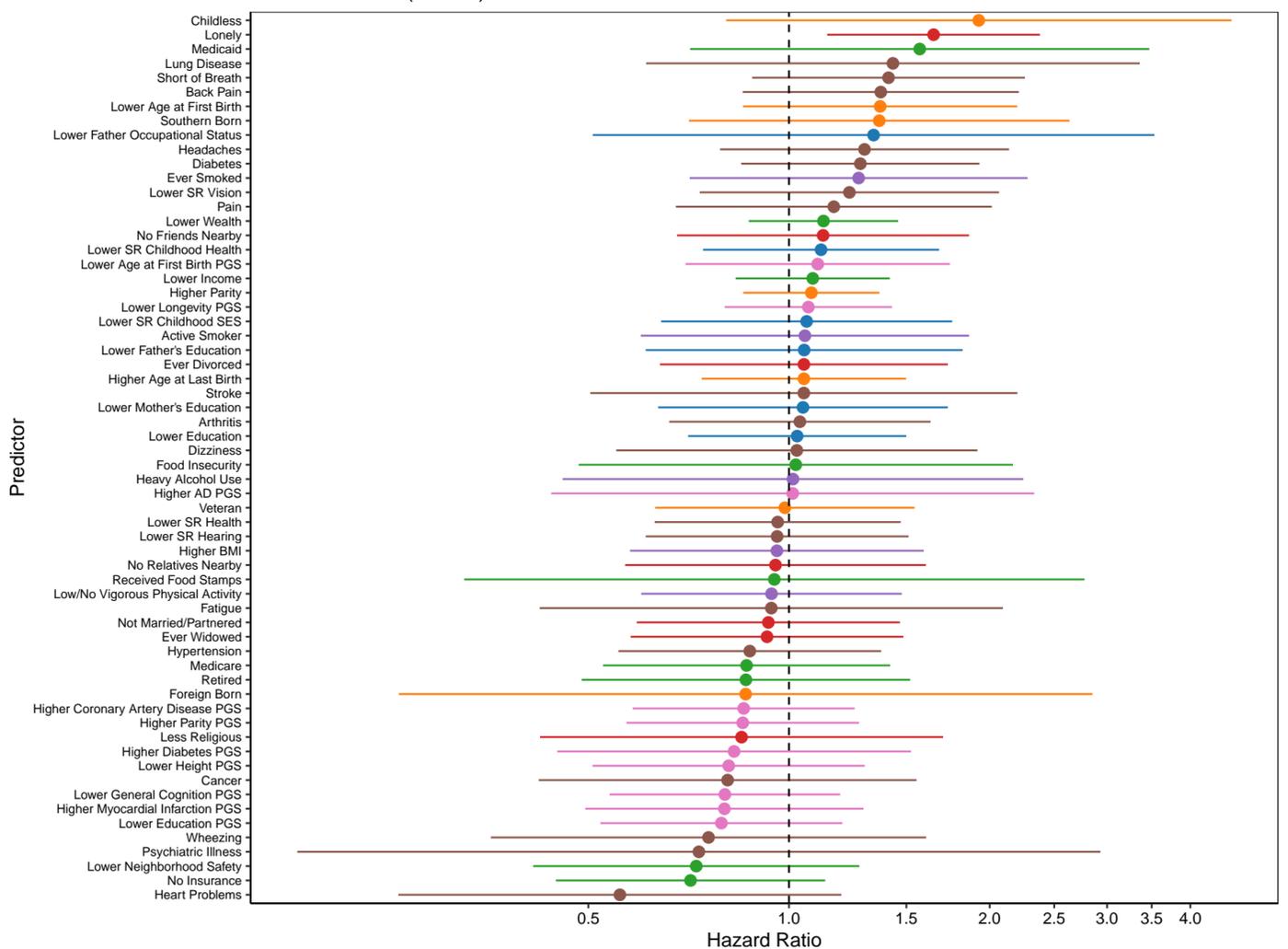

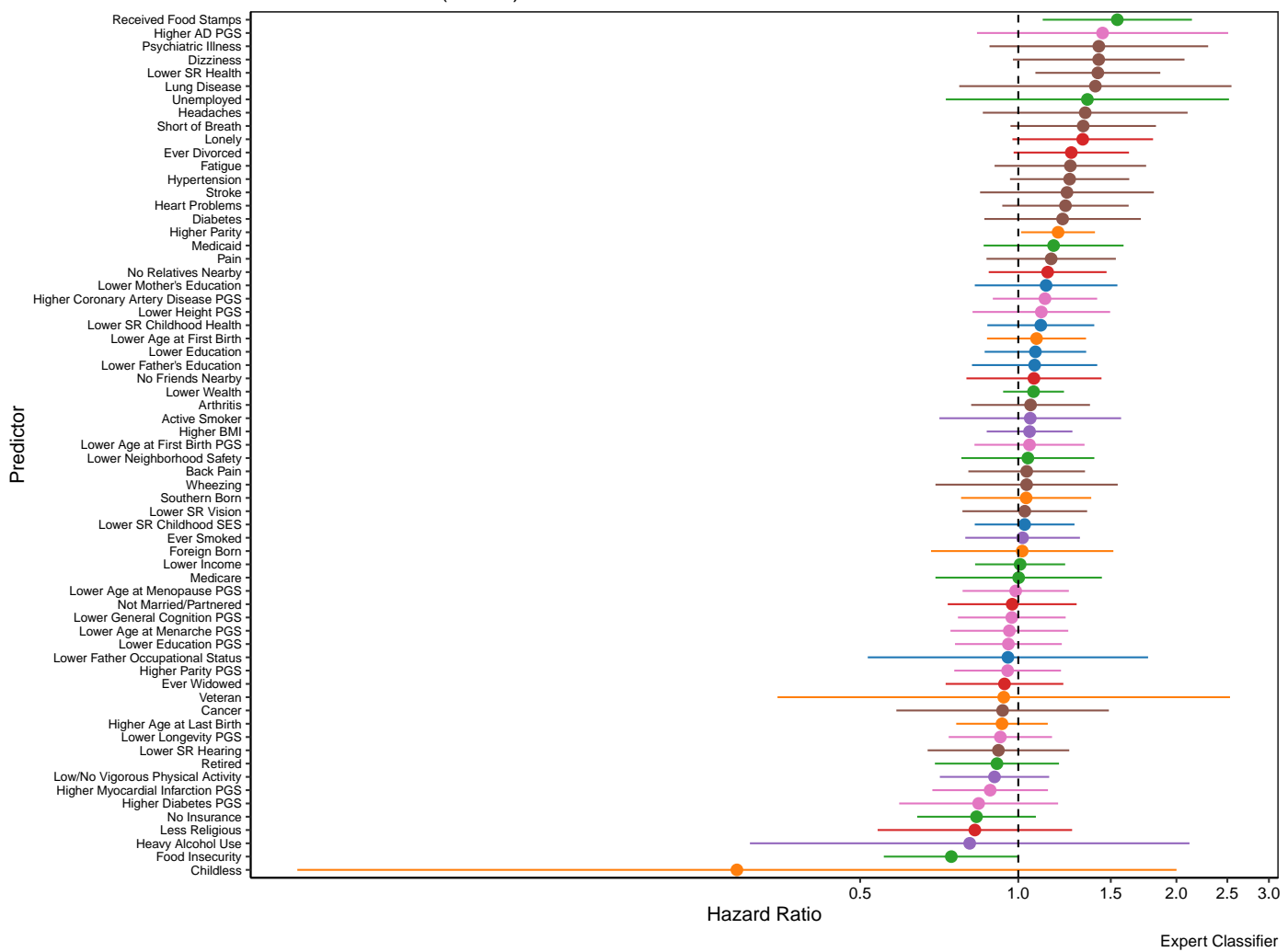

Expert Classifier

Fig S6

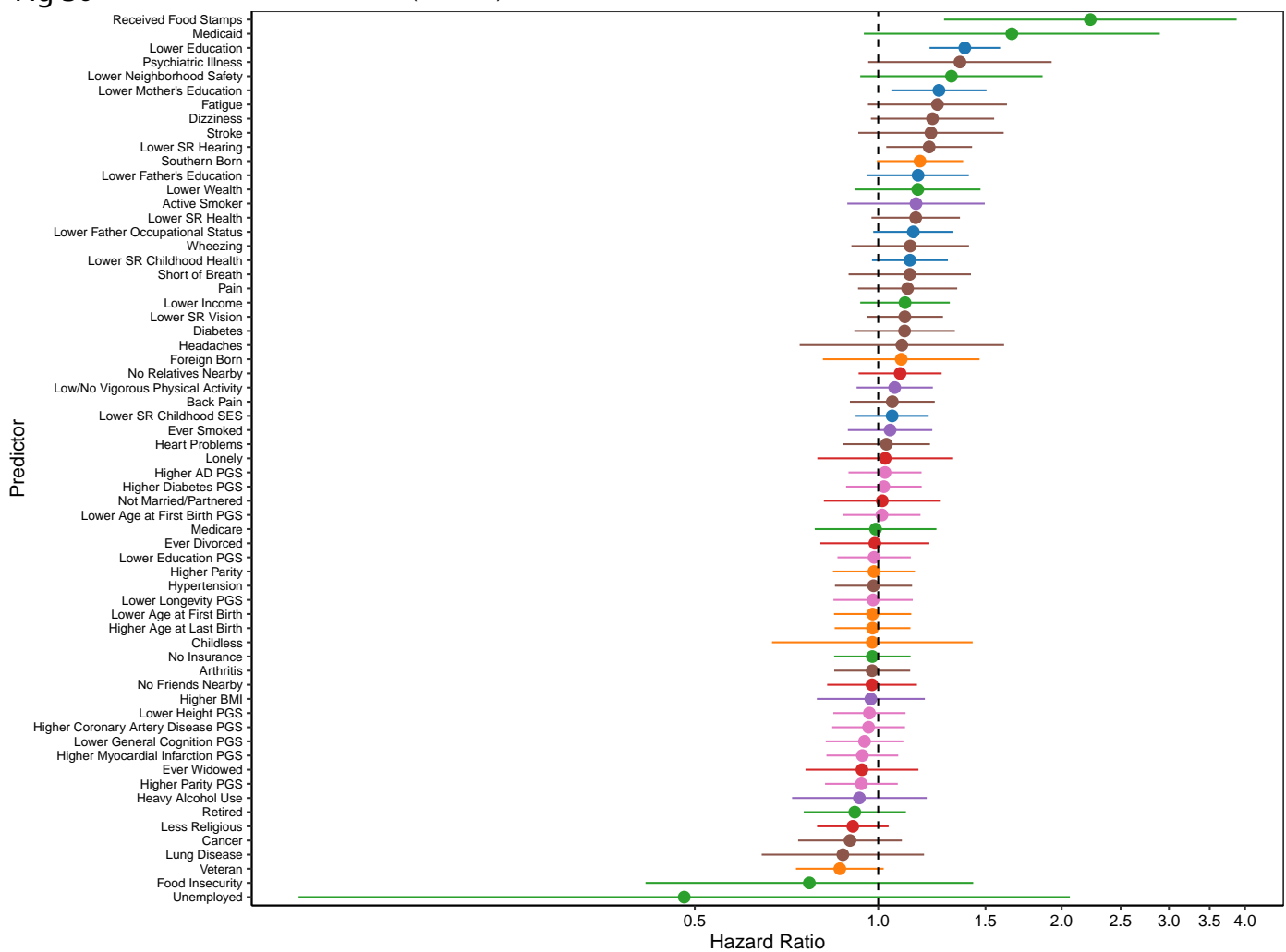

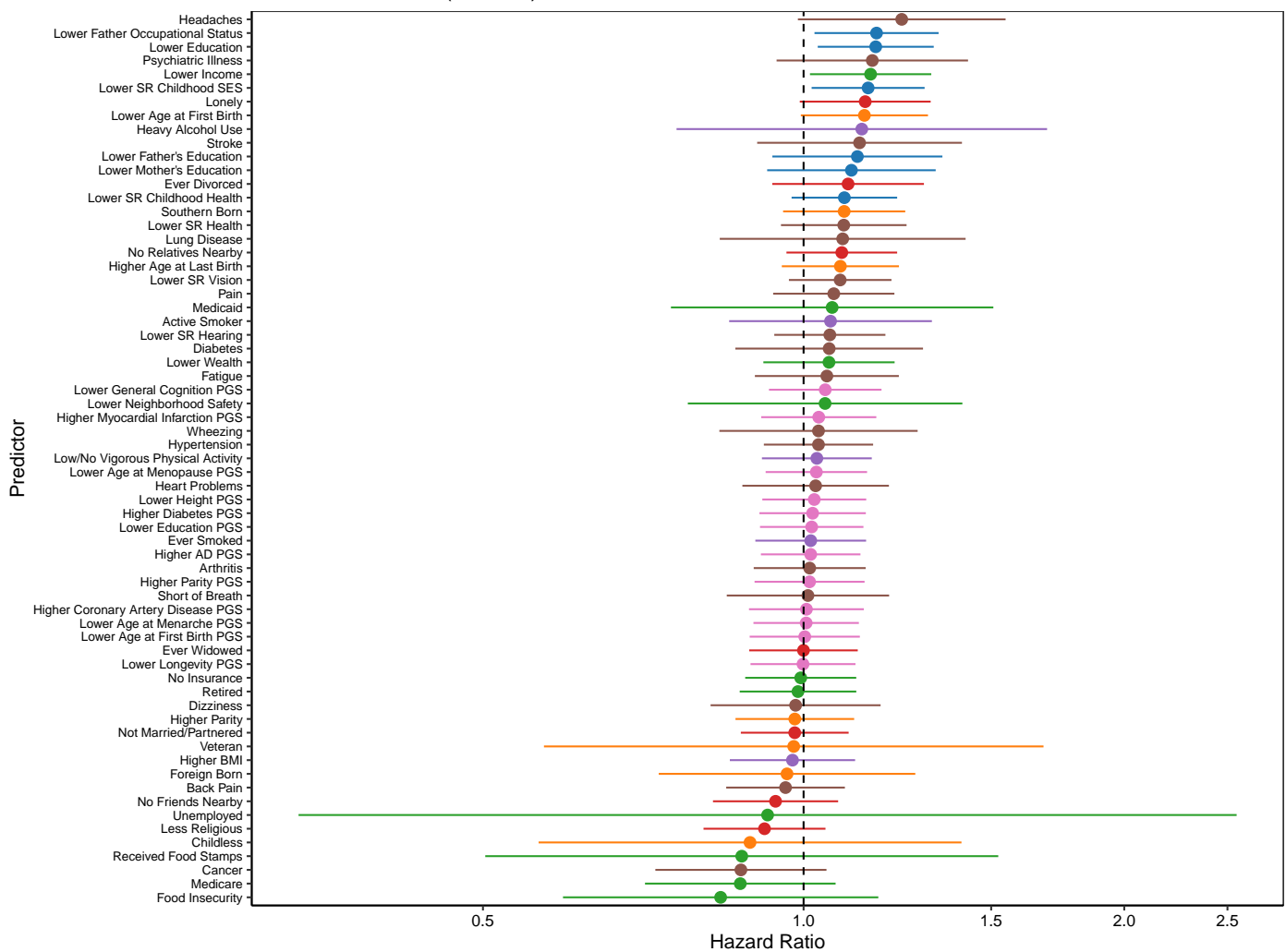

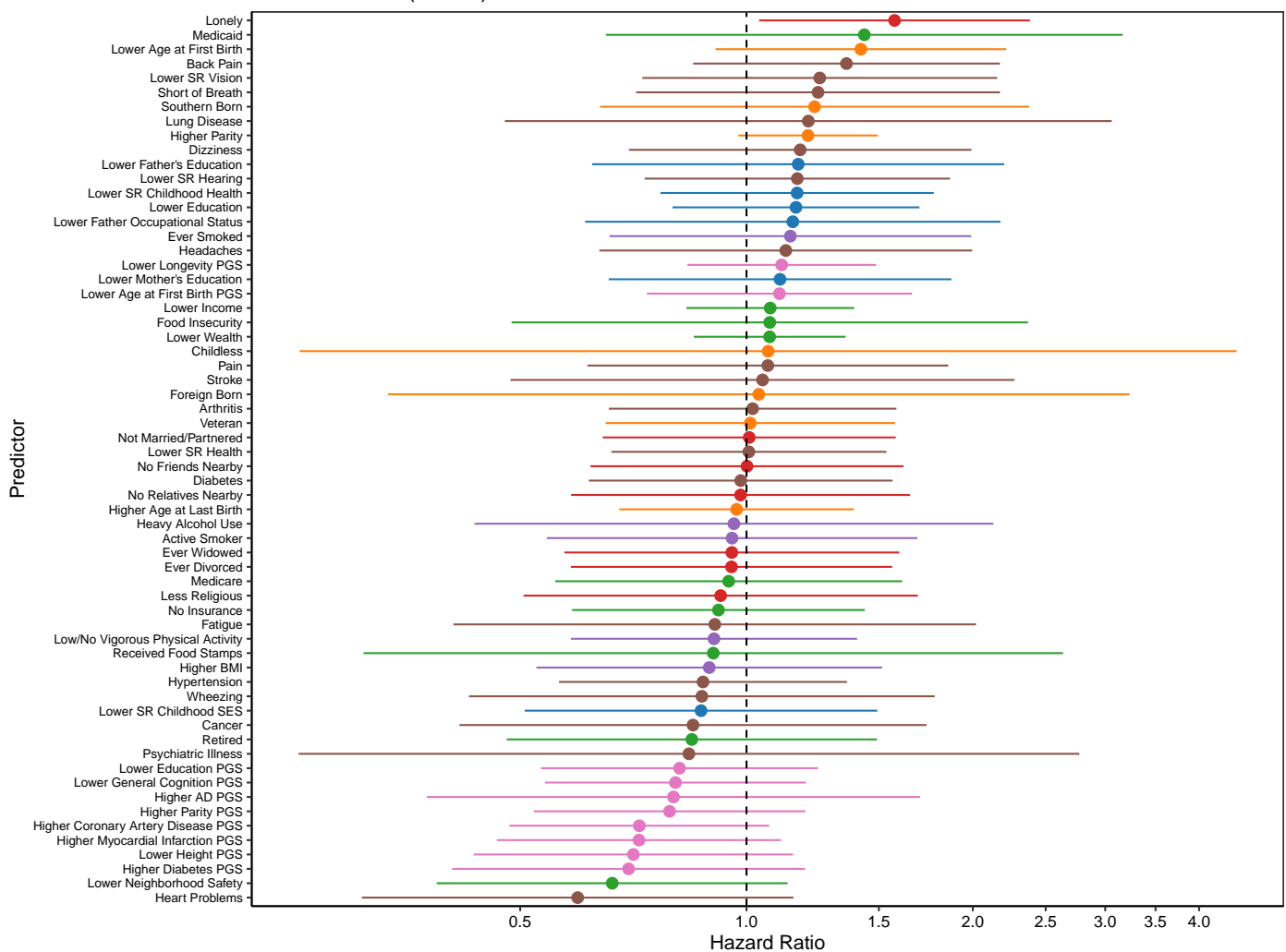

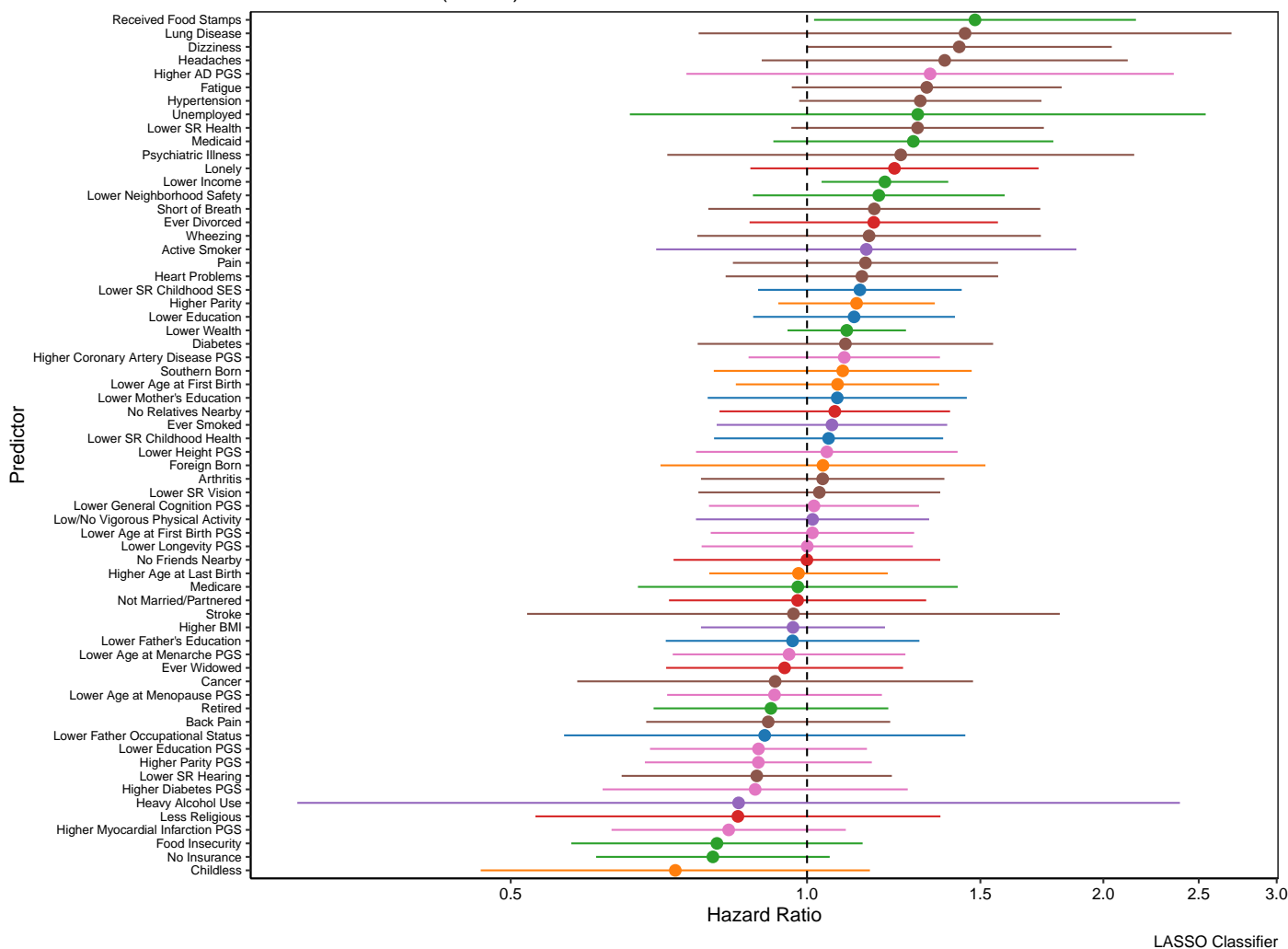

LASSO Classifier

Early–Life    Economic    Behaviors    Genetic
Sociodemographic    Social Ties    Health

Fig S7

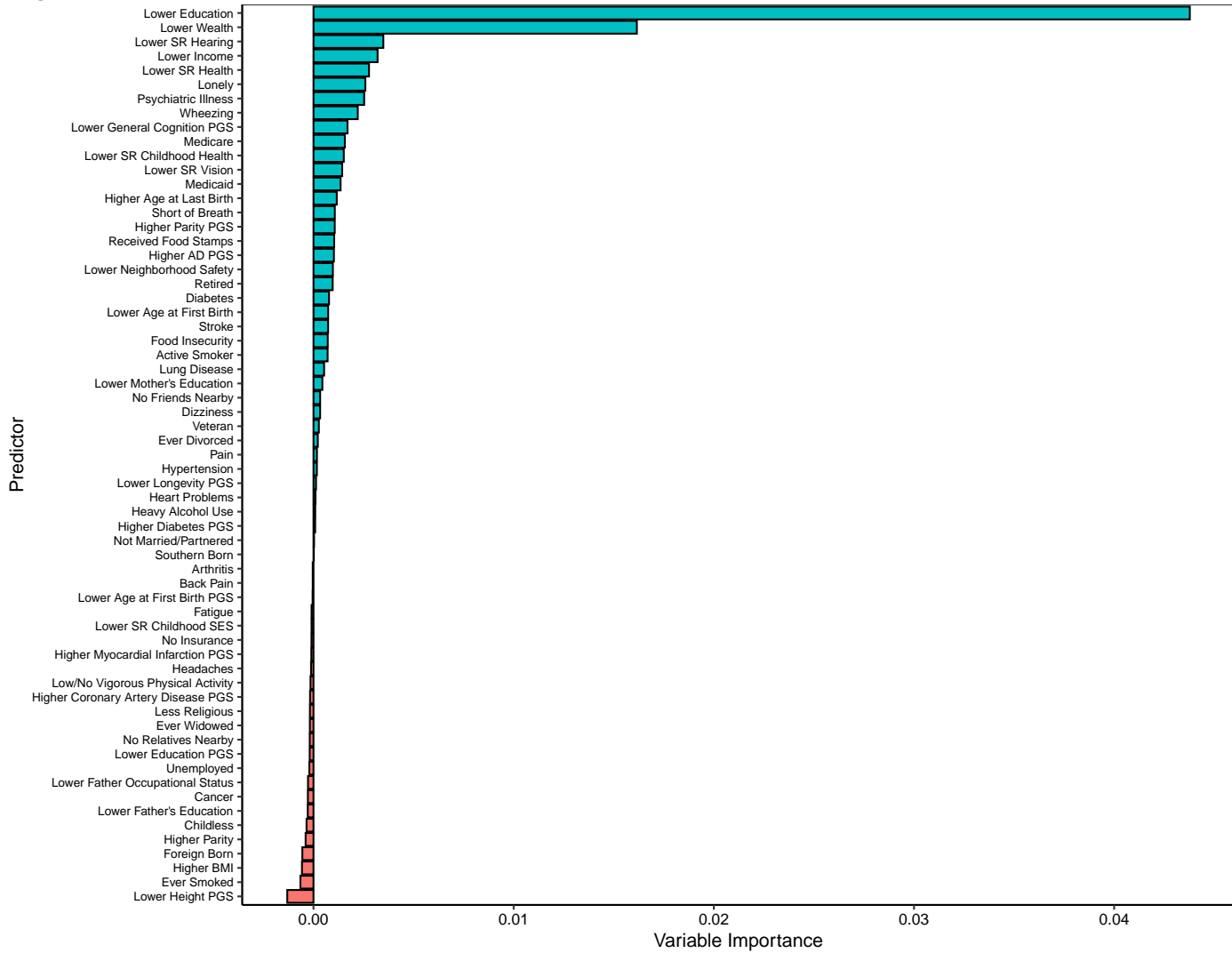

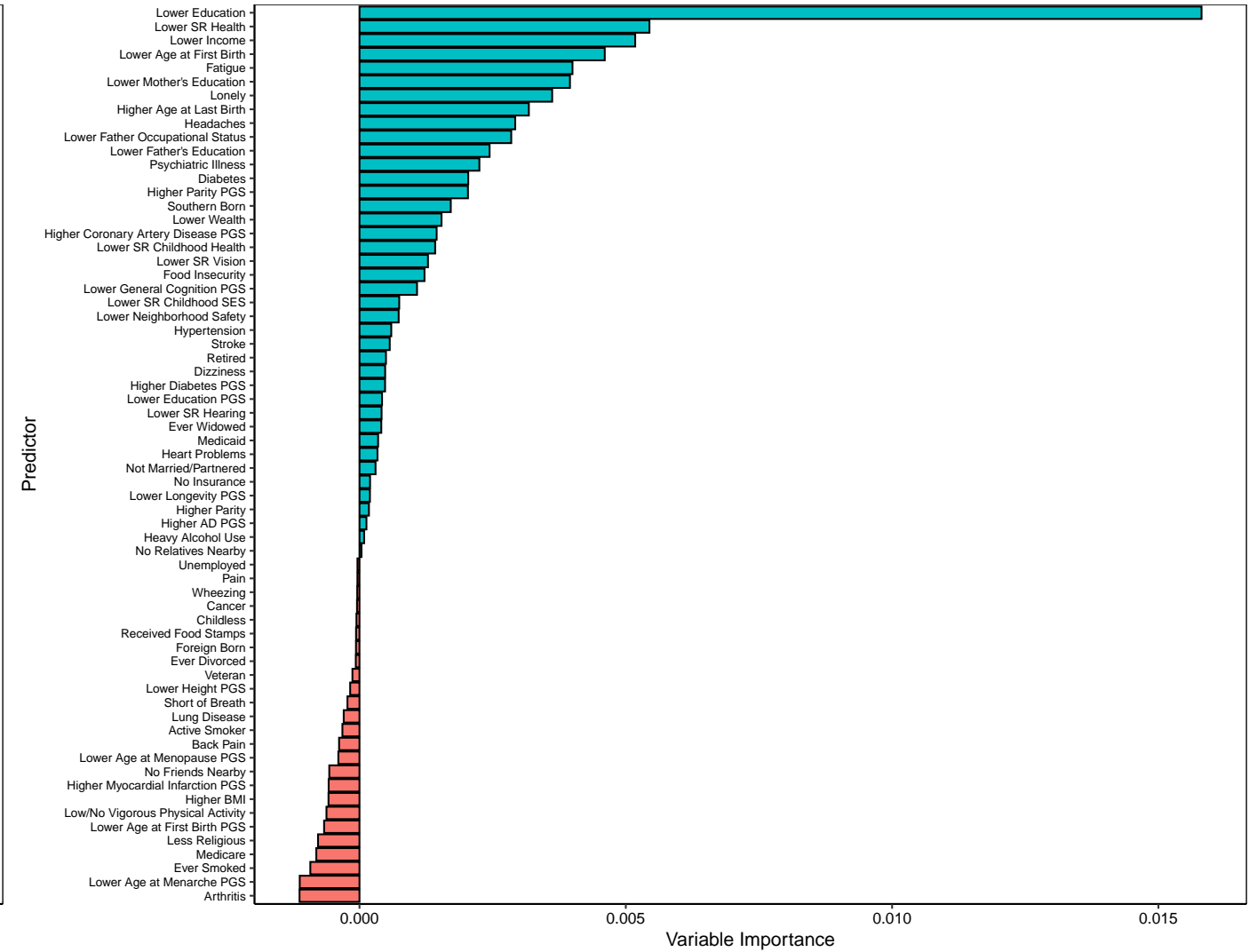

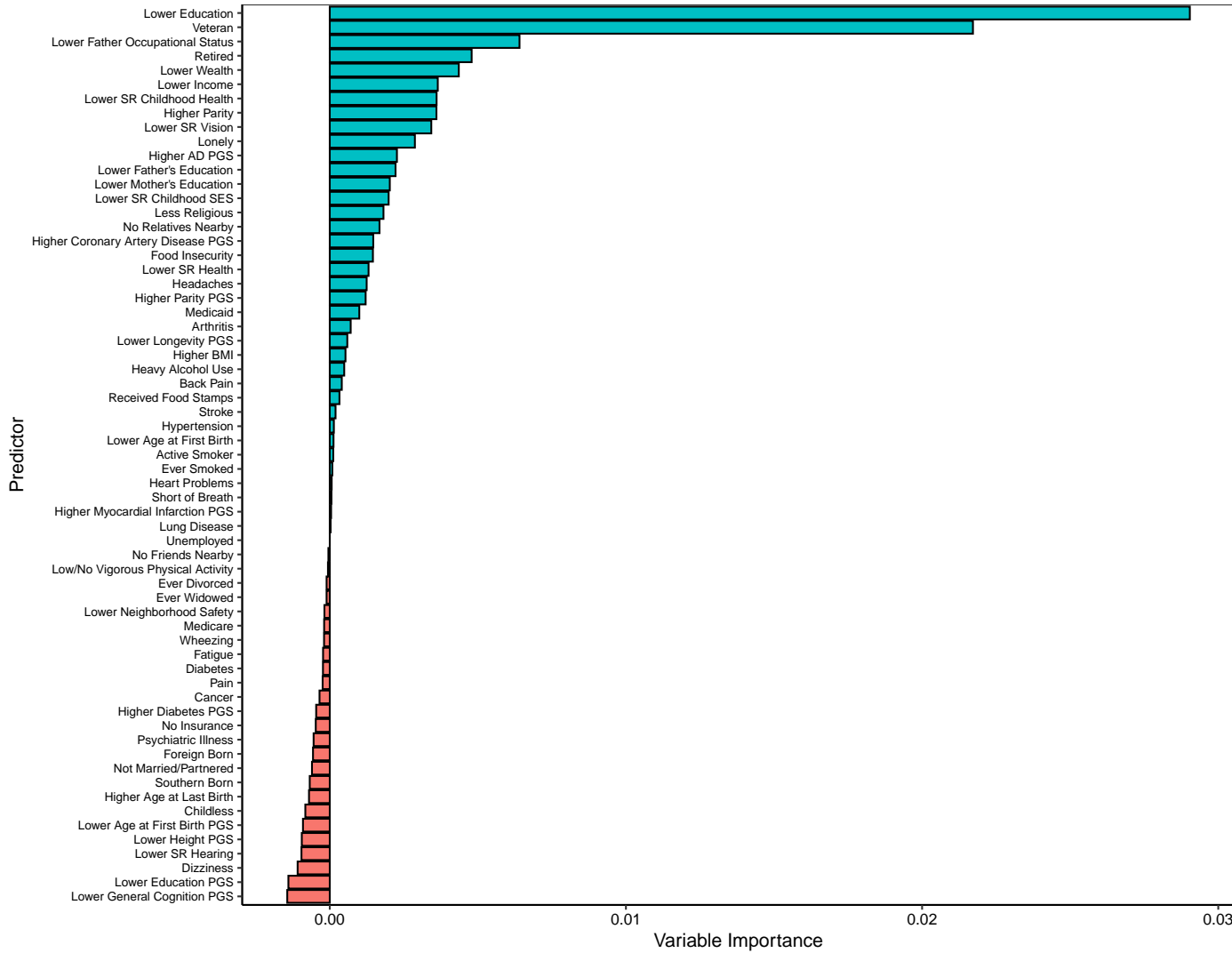

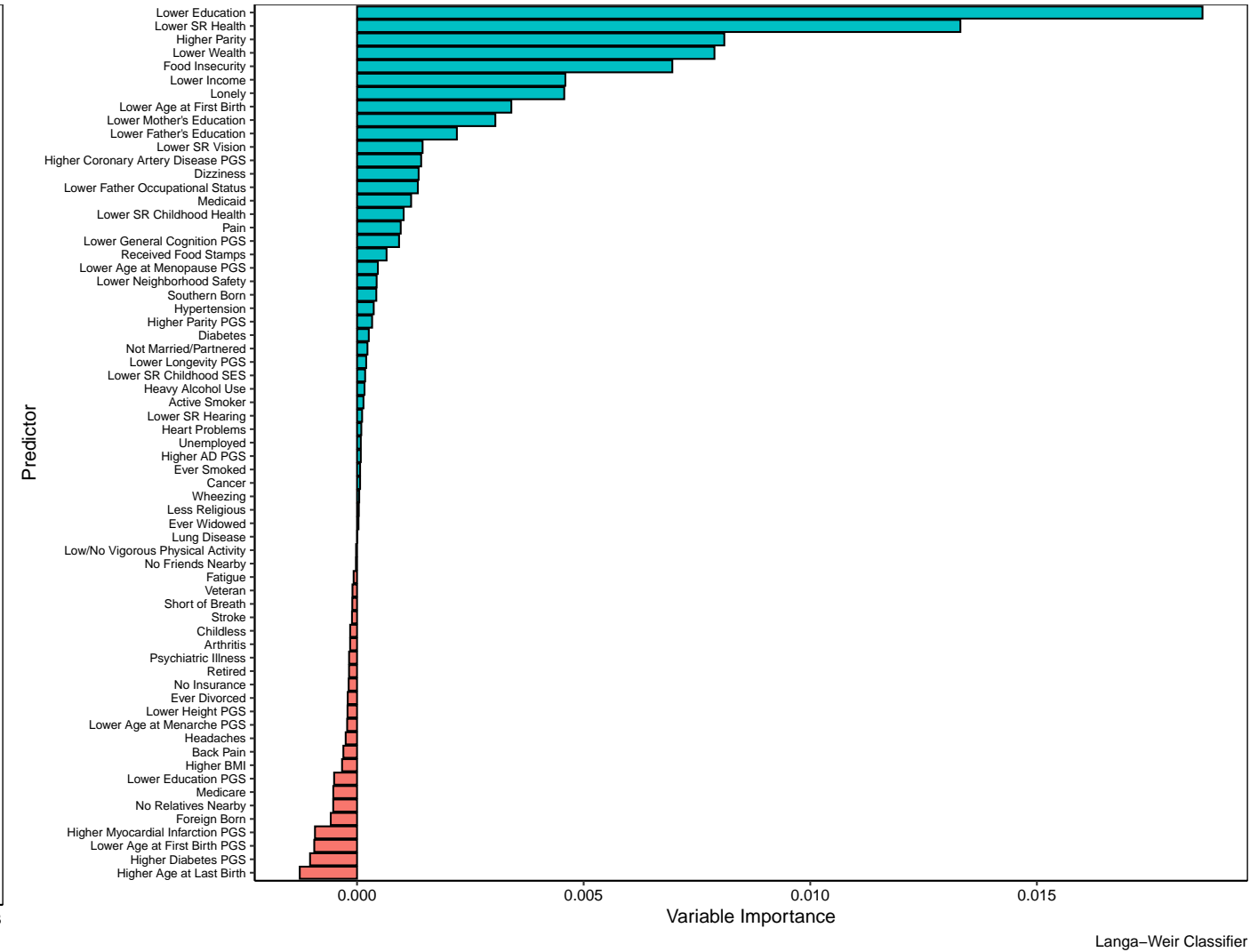

Langa–Weir Classifier

Fig S8

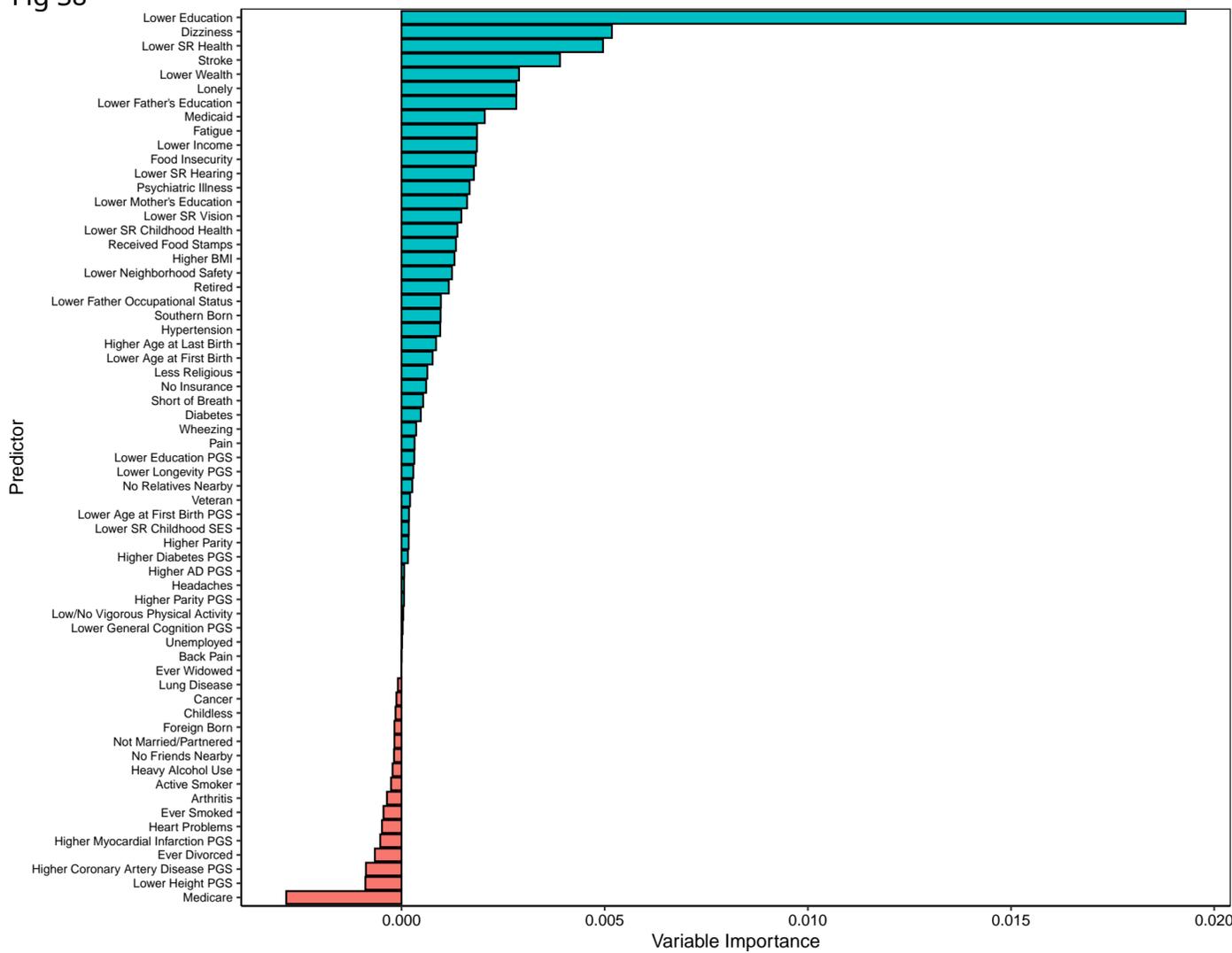

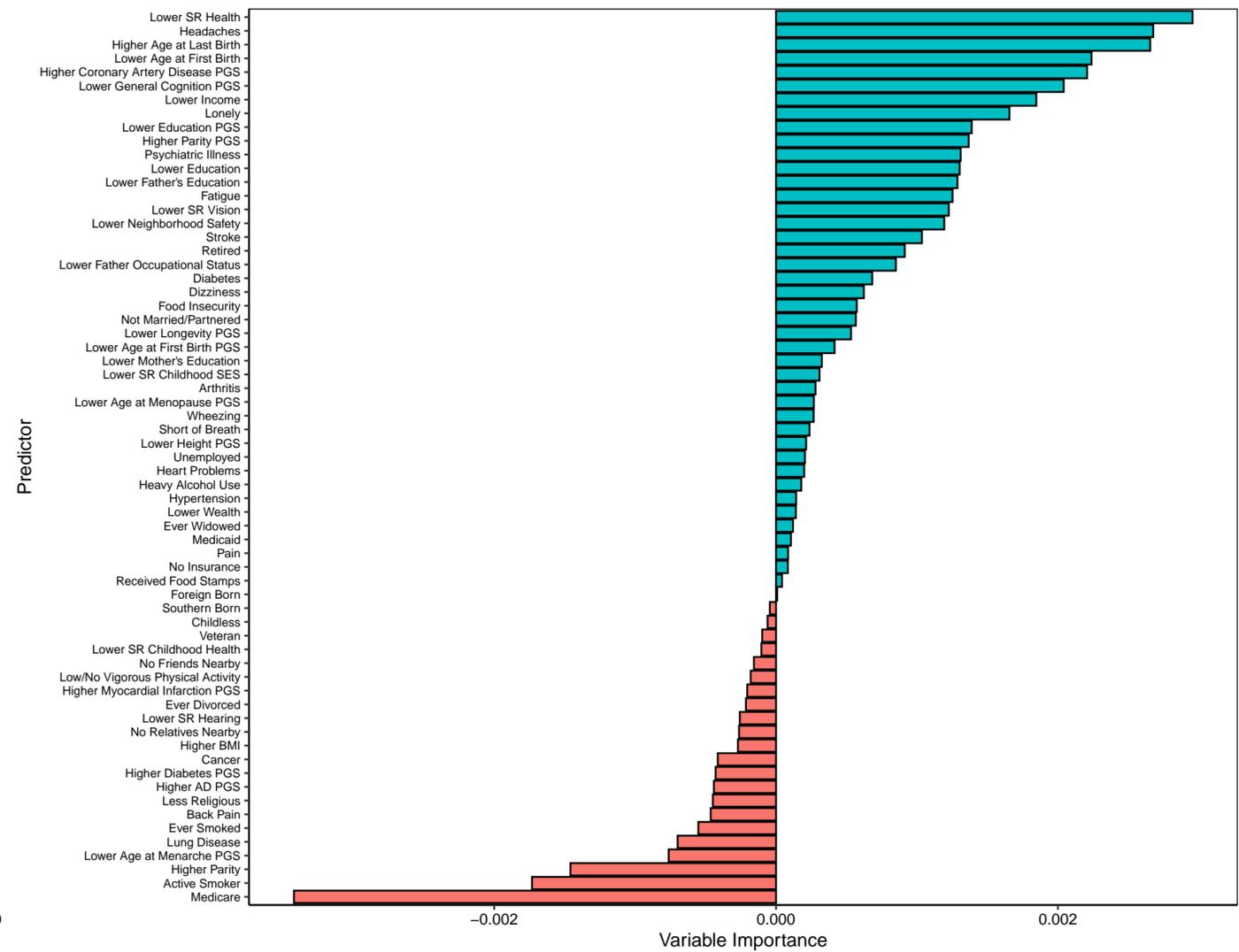

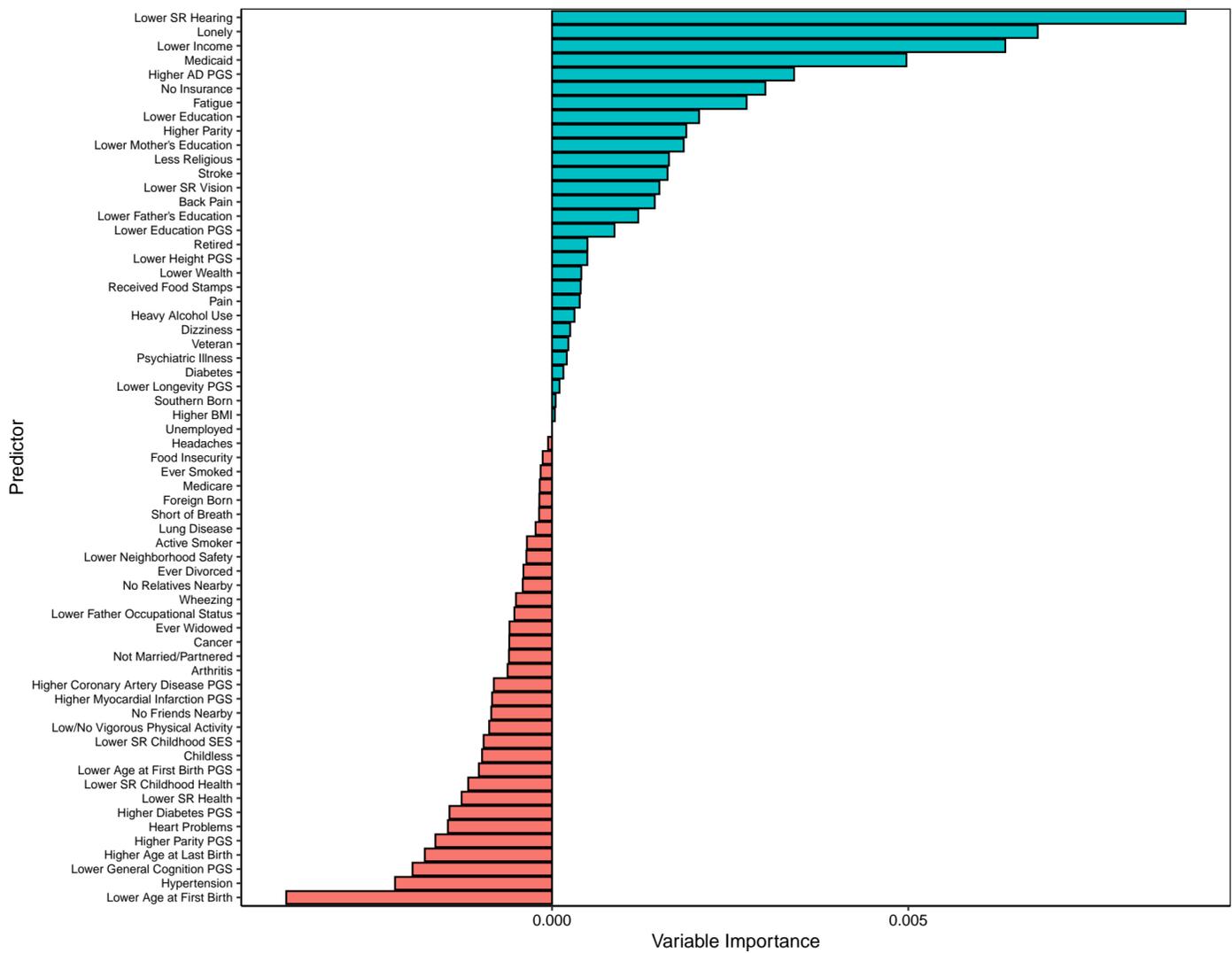

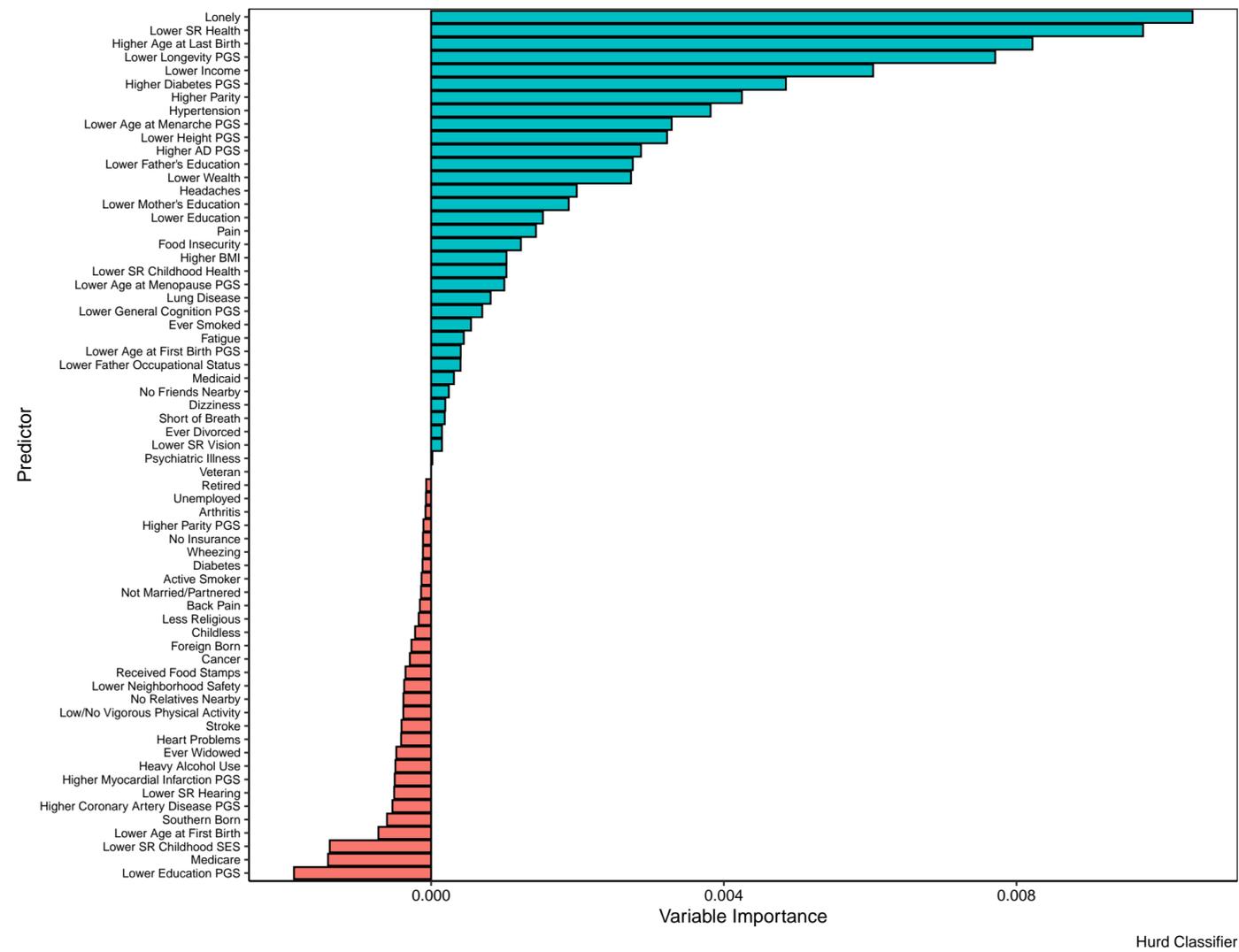

Fig S9

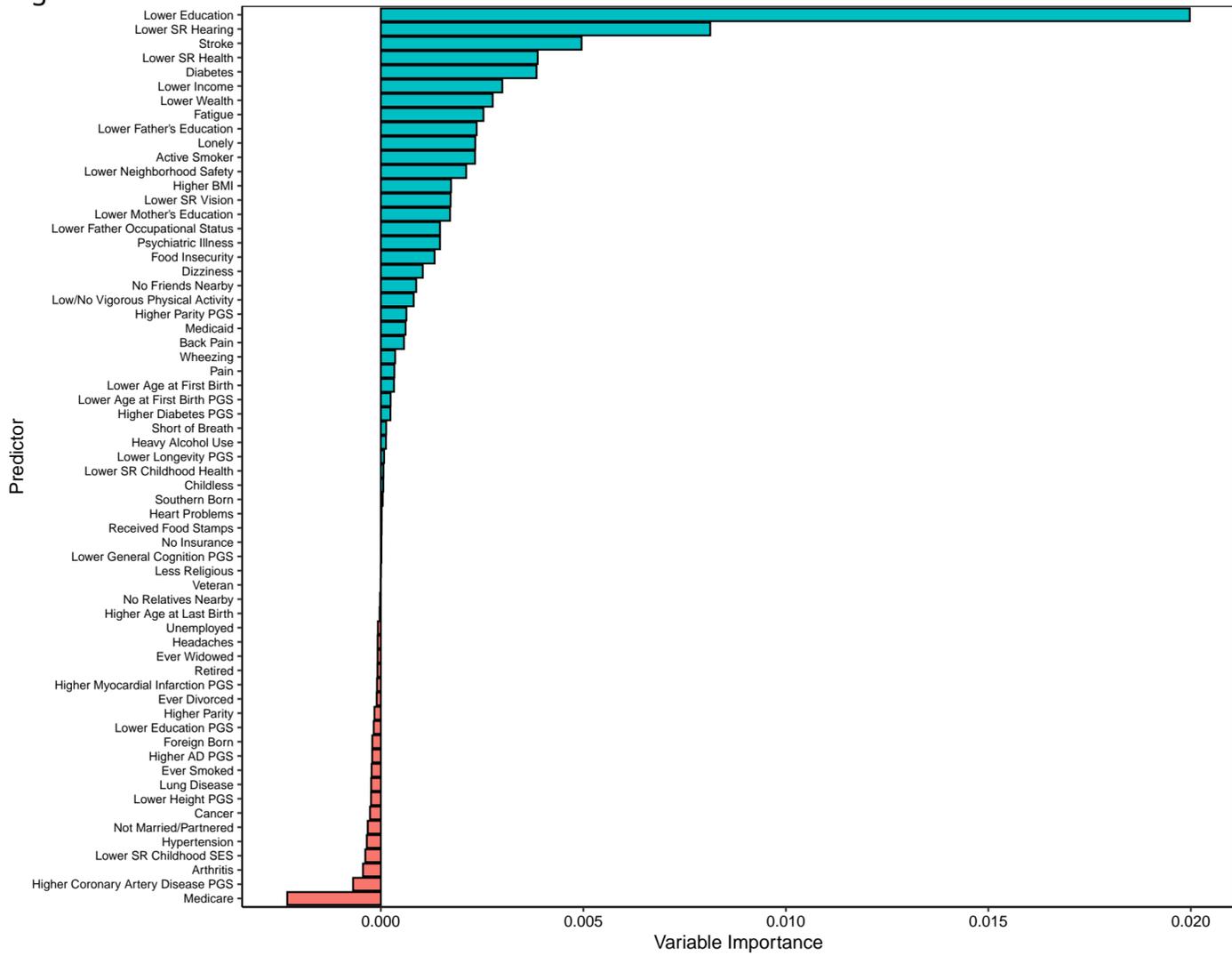

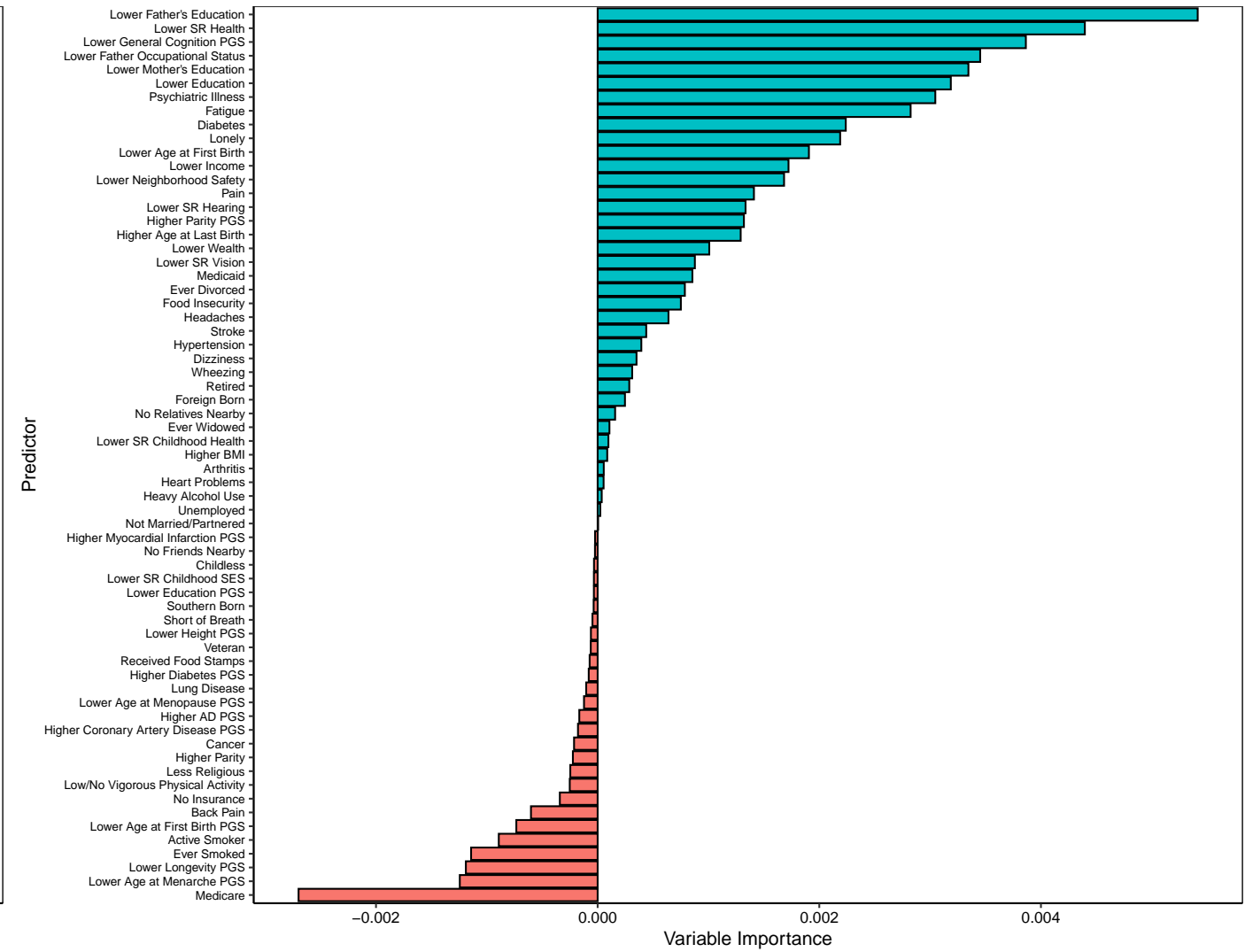

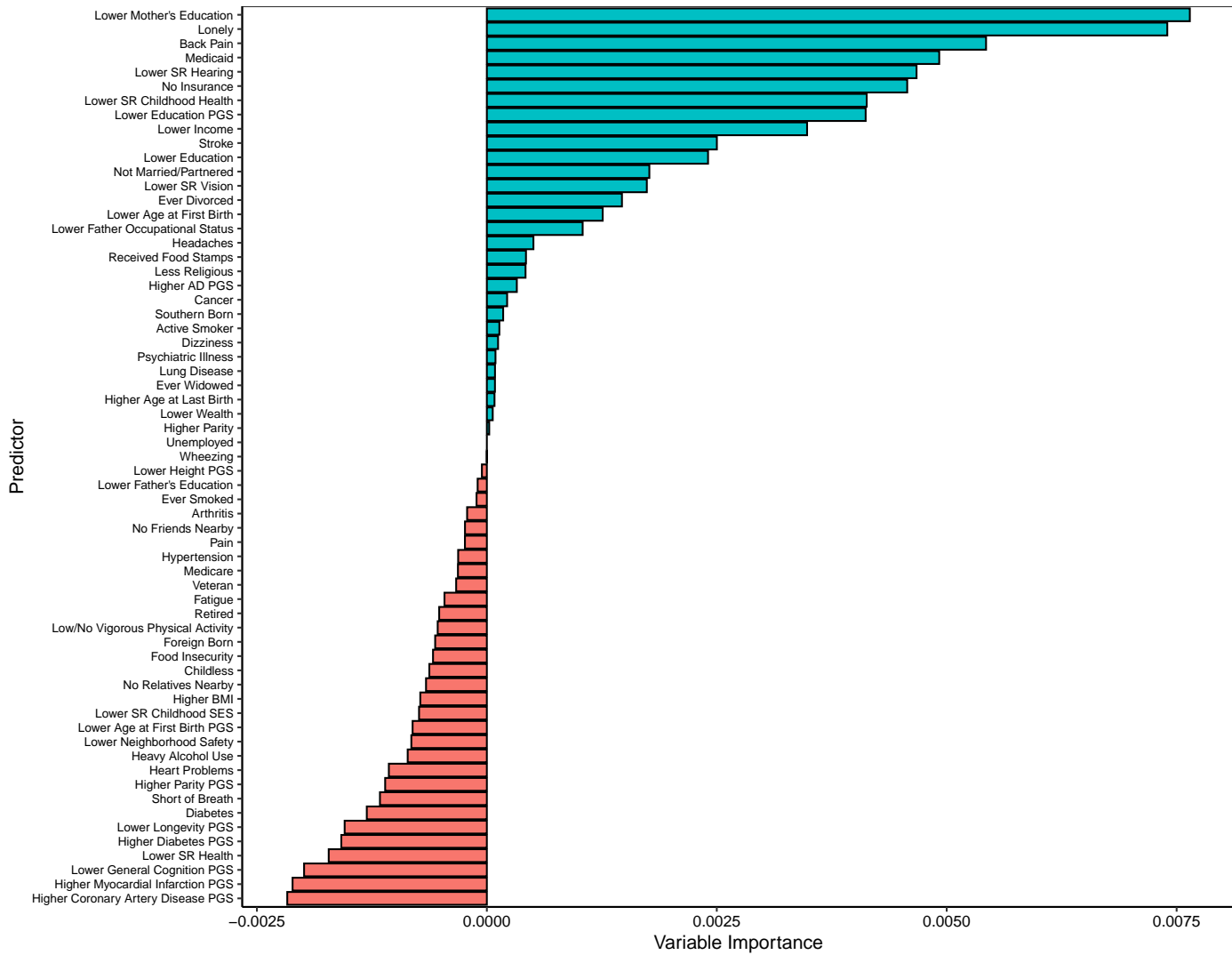

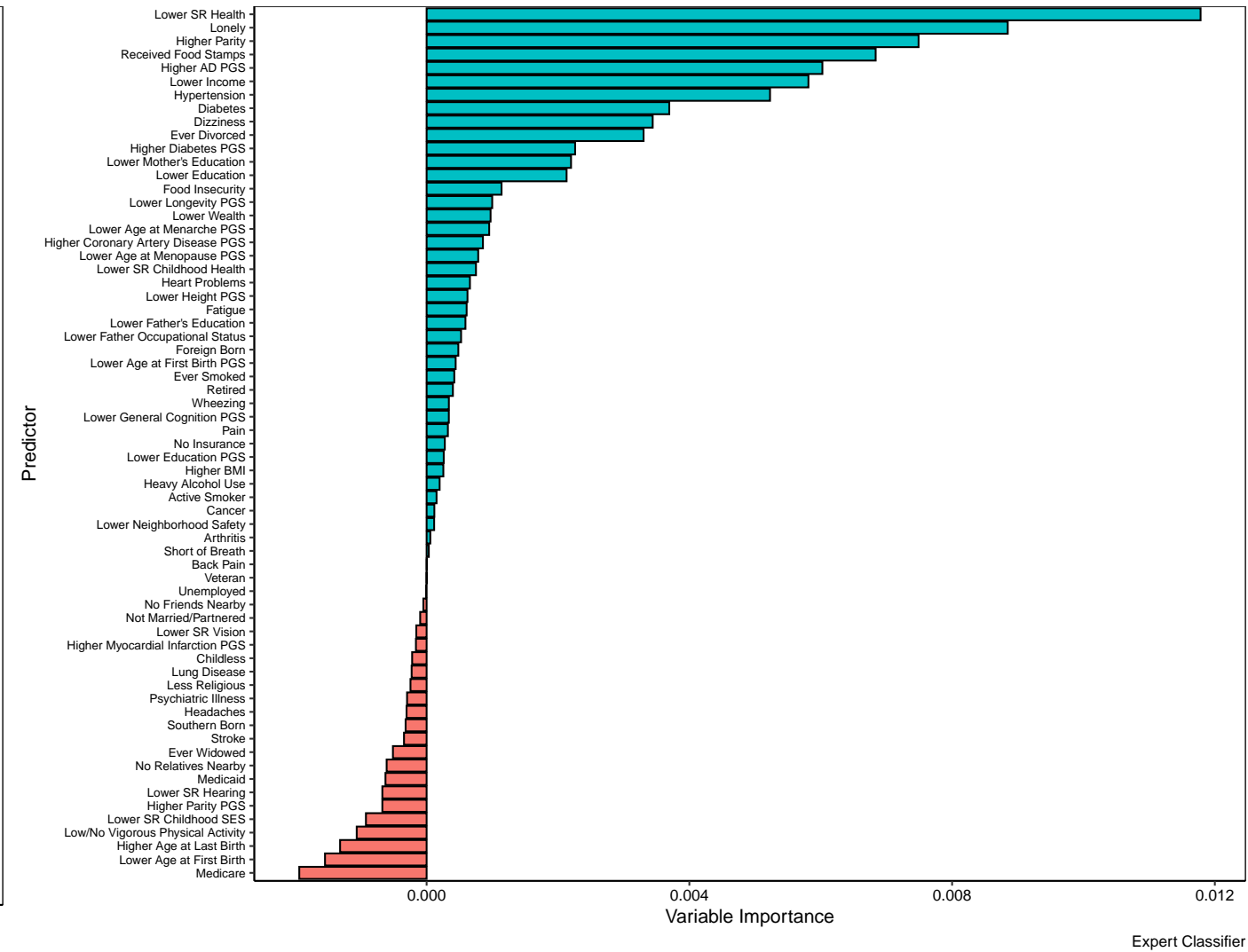

Fig S10

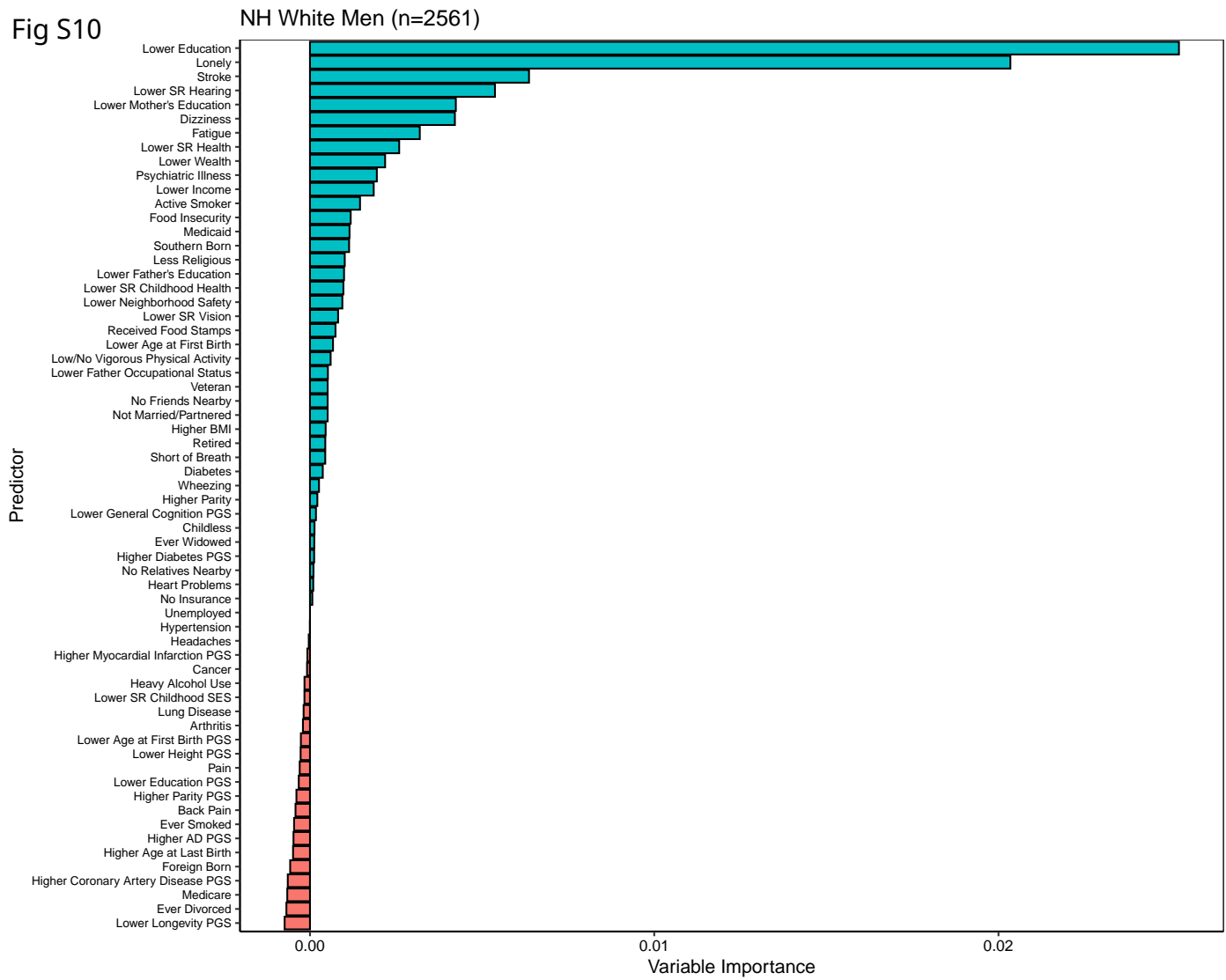

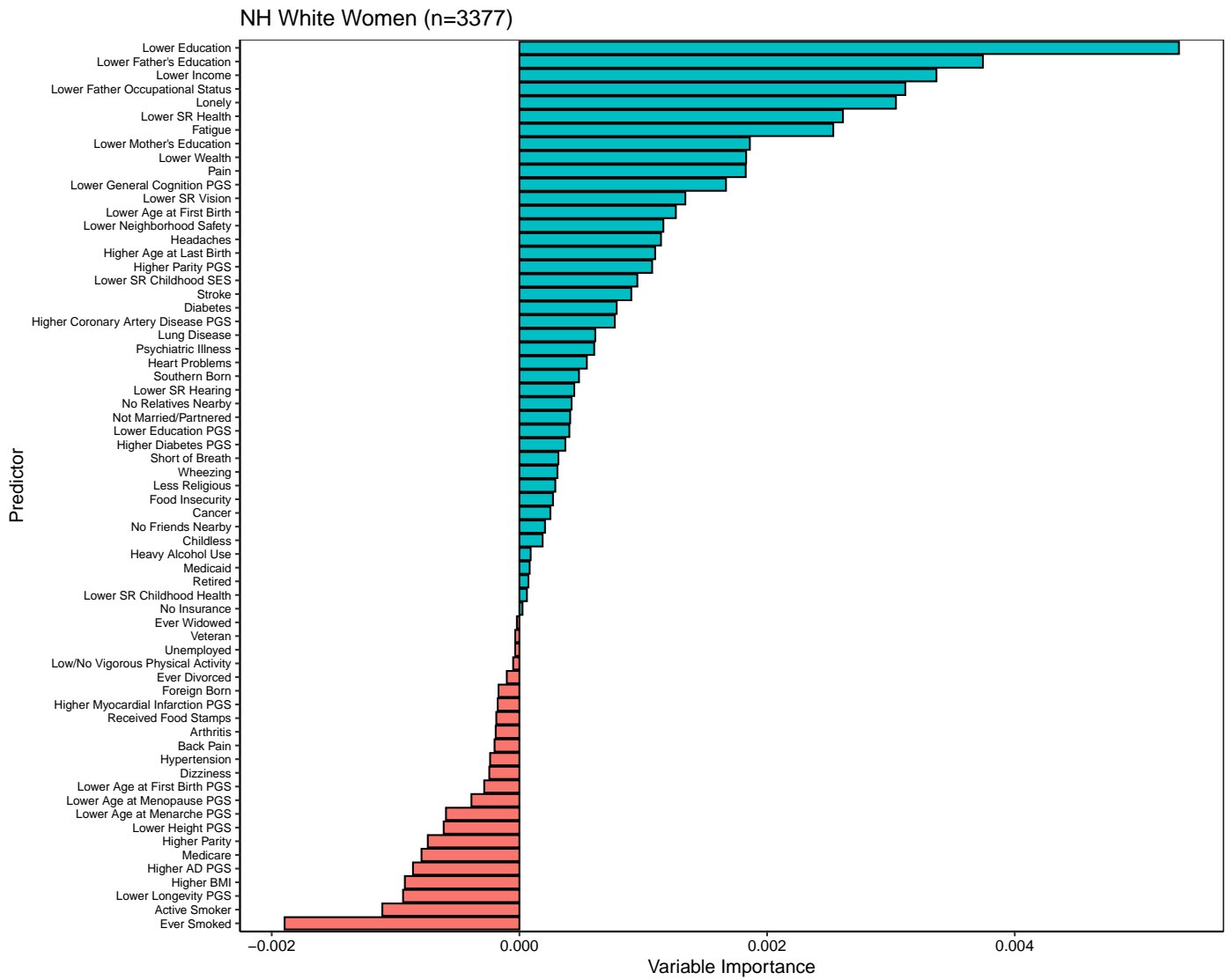

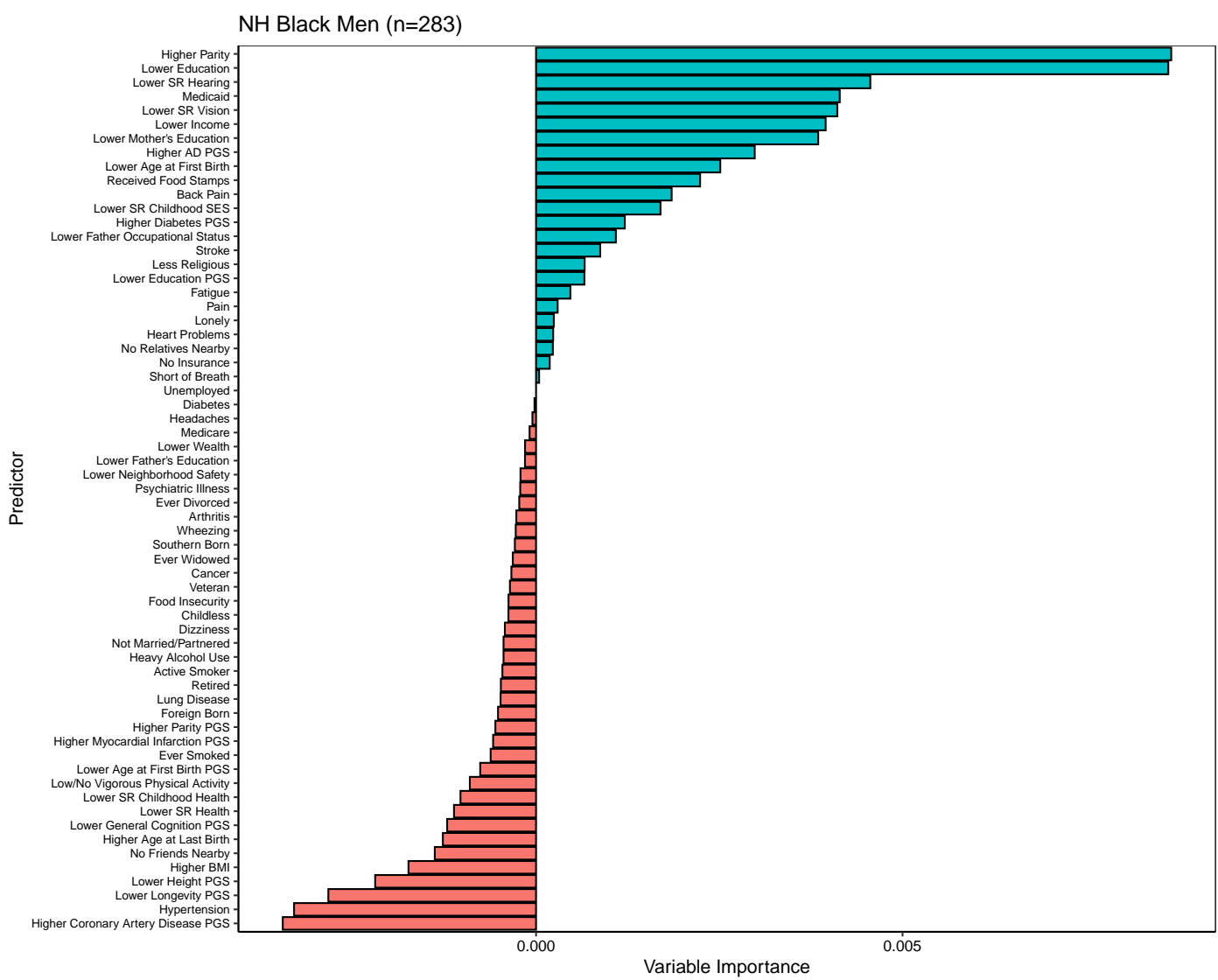

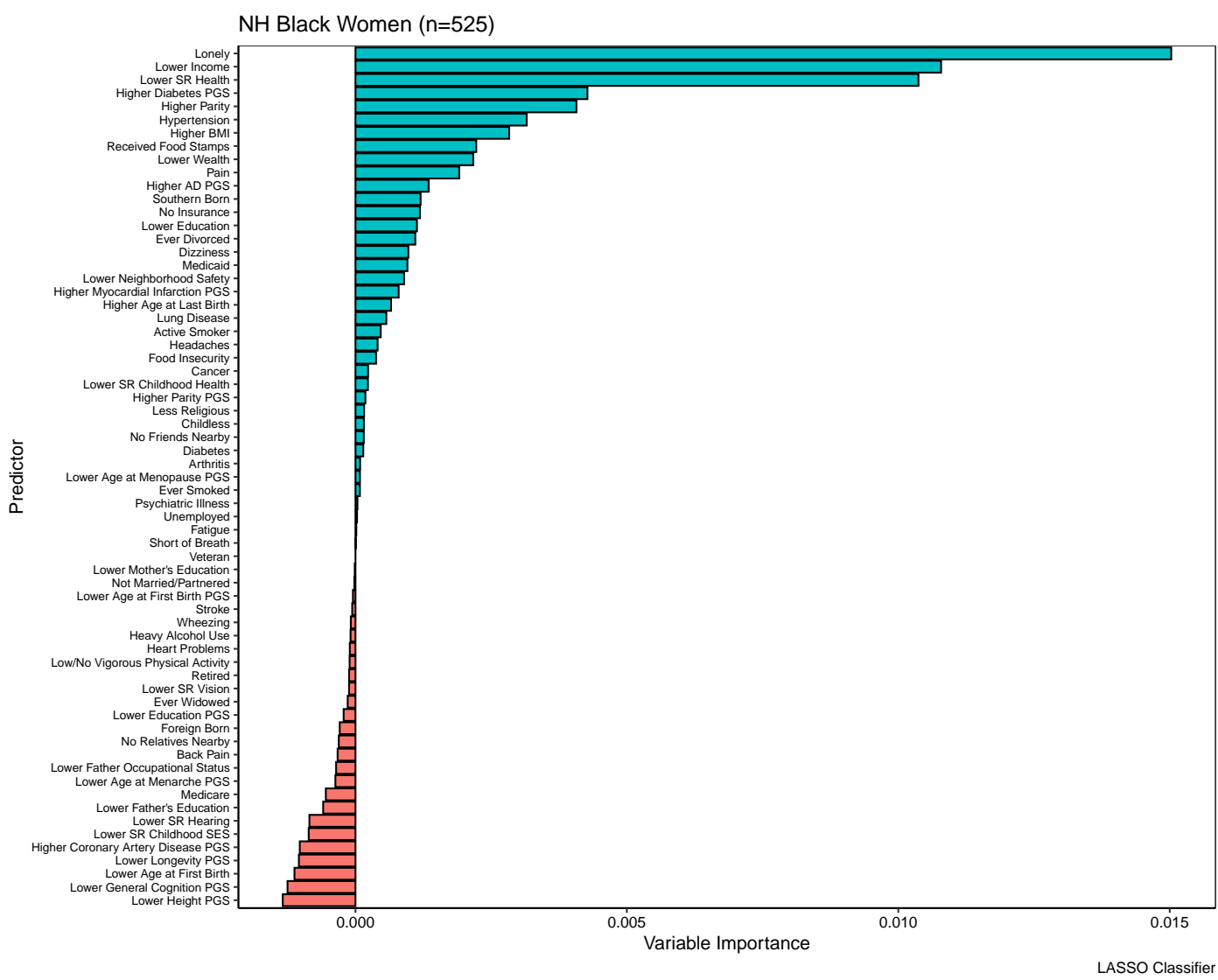

LASSO Classifier

## Fig S11

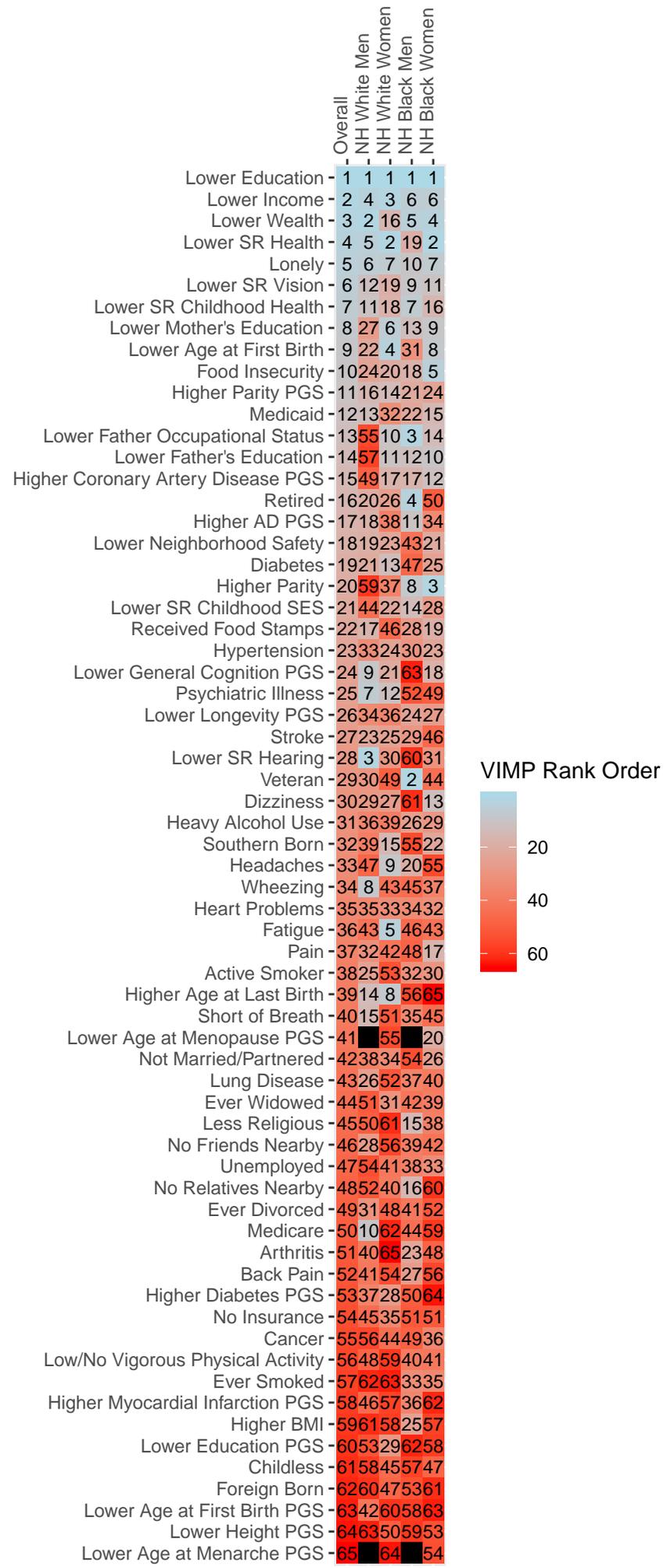

Langa−Weir Classifier

Fig S12

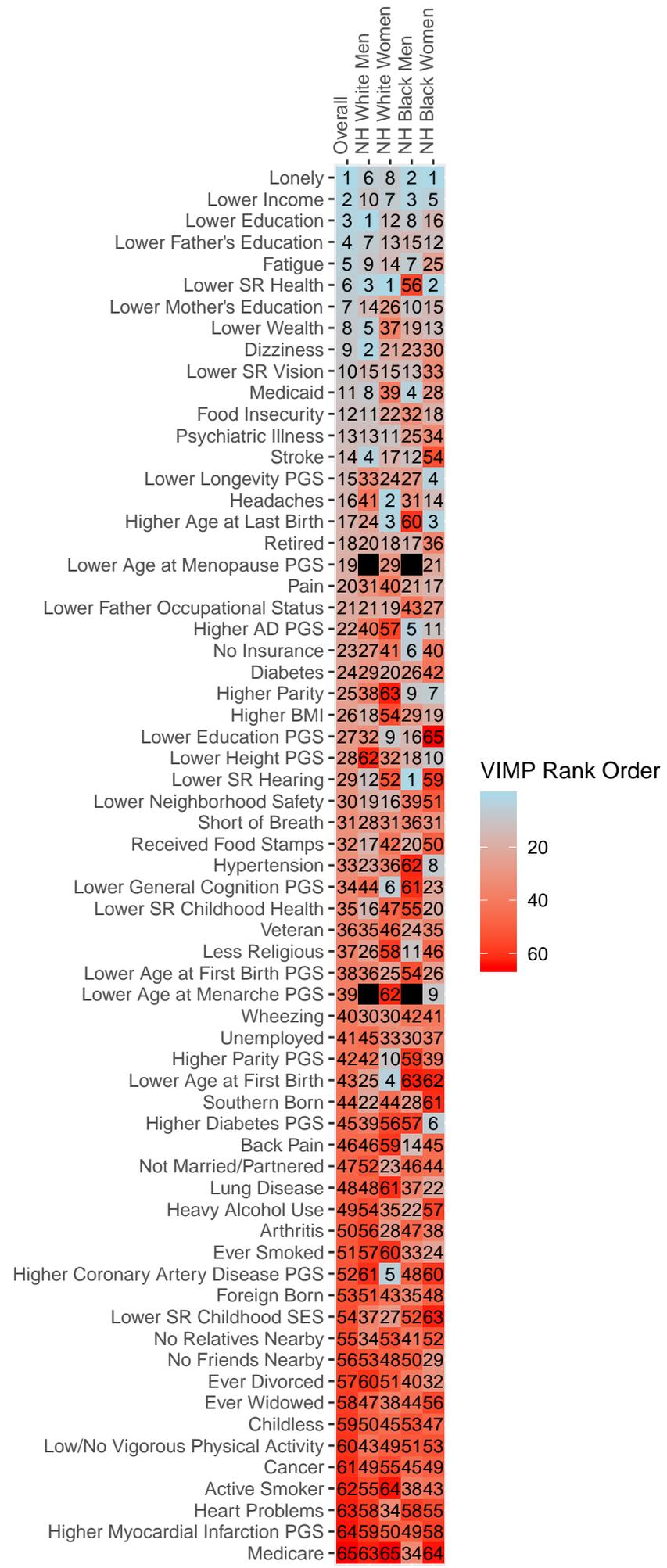

Hurd Classifier

Fig S13

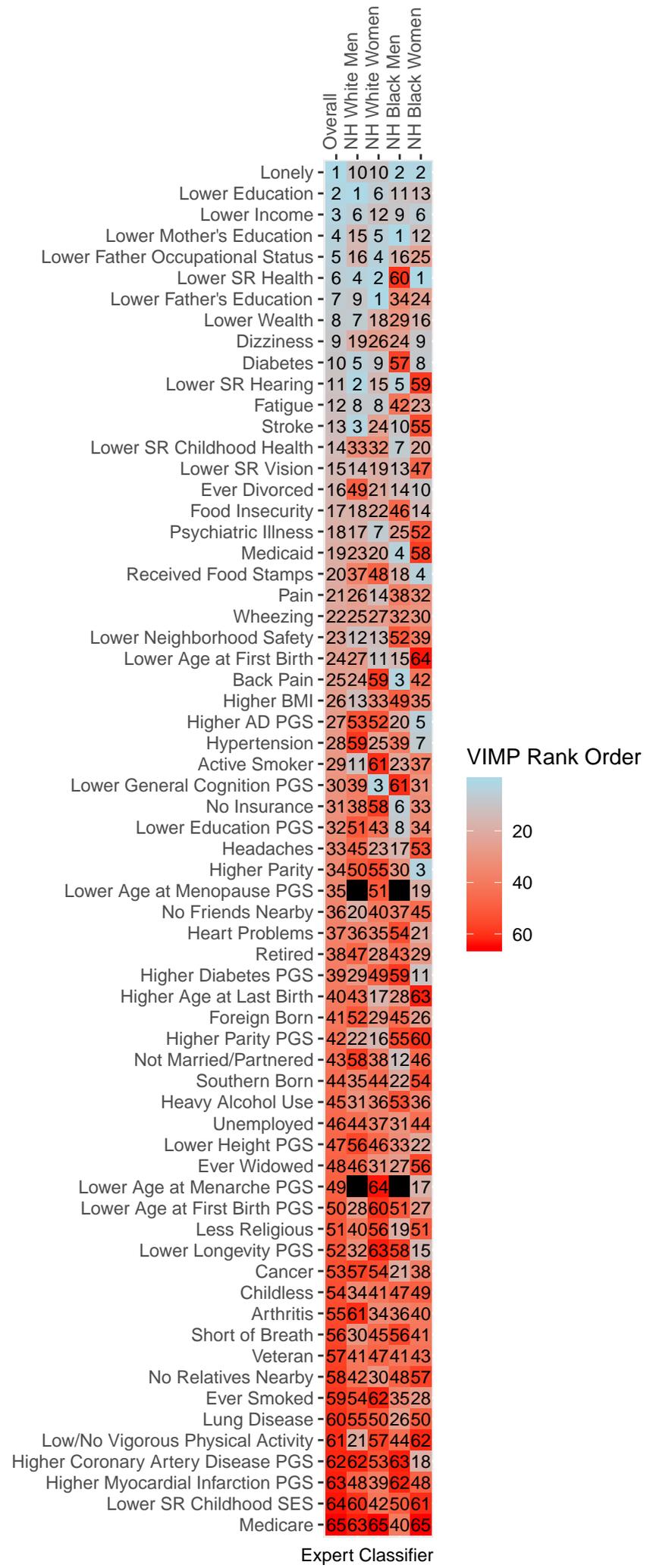

Expert Classifier

# Fig S14

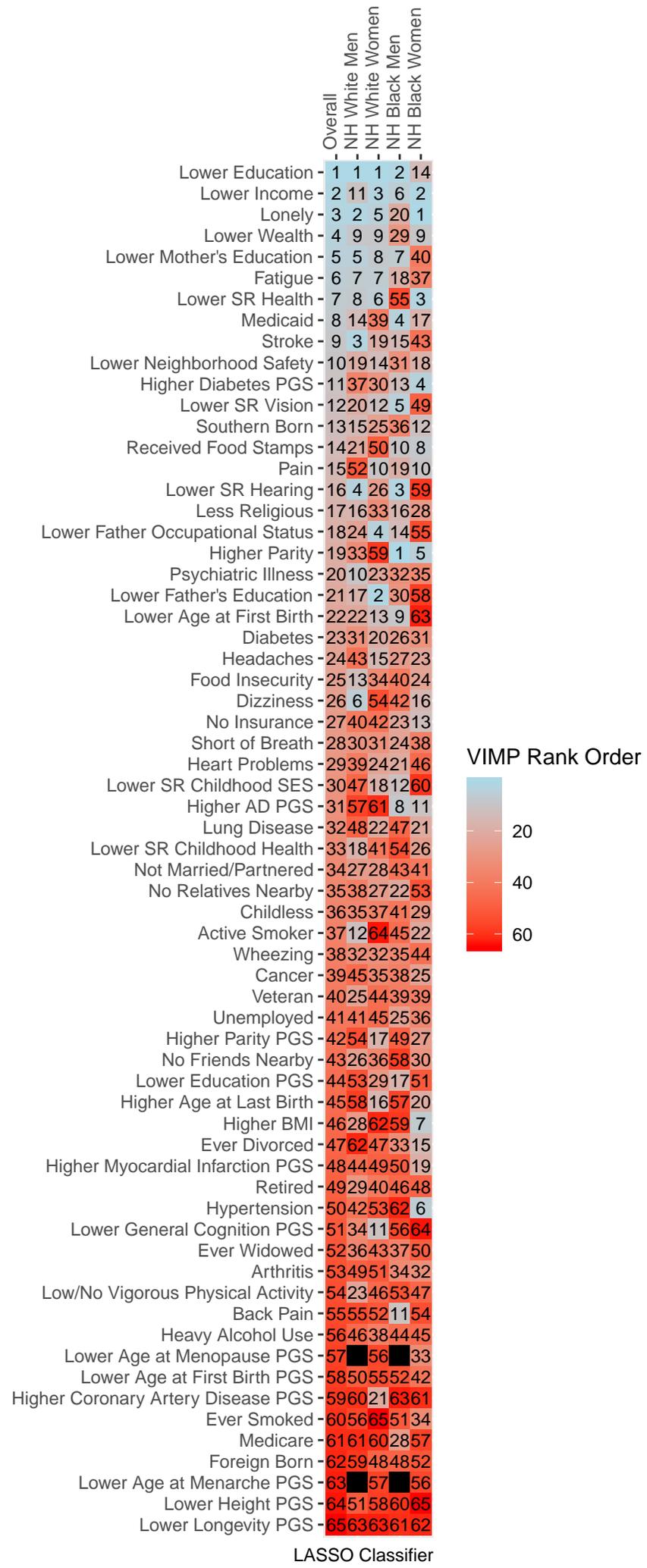